\documentclass[twocolumn]{aastex631}

\usepackage{amsmath}
\usepackage{multirow}
\usepackage{enumitem}
\usepackage{xspace}  

\newcommand{\Mone}{$M_{\rm{1kpc}}/M_*$\xspace}
\newcommand{\Mr}{$M_{\rm{R}}/M_*$\xspace}
\newcommand{\Se}{$\Sigma_e$\xspace}
\newcommand{\Sone}{$\Sigma_{\rm{1kpc}}$\xspace}

\newcommand{\re}{$R_{\rm{e}}$\xspace}
\newcommand{\n}{$n$\xspace}

\newcommand{\zf}{$z_{\rm{form}}$\xspace}
\newcommand{\zp}{$z_{\rm{p}}$\xspace}
\newcommand{\zobs}{$z_{\rm{obs}}$\xspace}
\newcommand{\tage}{$t_{\rm{age}}$\xspace}
\newcommand{\tautot}{$\tau_{\rm{tot}}$\xspace}
\newcommand{\tausf}{$\tau_{\rm{SF}}$\xspace}
\newcommand{\tauq}{$\tau_{\rm{Q}}$\xspace}

\def\B{\ensuremath{B_{435}}\xspace}
\def\V{\ensuremath{V_{606}}\xspace}
\def\I{\ensuremath{I_{814}}\xspace}
\def\Y{\ensuremath{Y_{105}}\xspace}
\def\J{\ensuremath{J_{125}}\xspace}

\def\H{\ensuremath{H_{160}}\xspace}

\newcommand{\Ha}{\textrm{H}\ensuremath{\alpha}\xspace}

\newcommand{\jwst}{\textit{JWST}\xspace}

\newcommand{\hst}{\textit{HST}\xspace}

\newcommand{\galfit}{{\sc Galfit}\xspace}
\newcommand{\sextractor}{{\sc SExtractor}\xspace}
\newcommand{\prospector}{{\sc Prospector}\xspace}

\newcommand{\sersic}{S\'{e}rsic\xspace}

\begin{document}

\title{Reconstructing the Assembly of Massive Galaxies. I. The Importance of the Progenitor Effect in the Observed Properties of Quiescent Galaxies at $z\approx 2$} 

\correspondingauthor{Zhiyuan Ji}
\email{zhiyuanji@astro.umass.edu}

\author[0000-0001-7673-2257]{Zhiyuan Ji}
\affiliation{University of Massachusetts Amherst, 710 North Pleasant Street, Amherst, MA 01003-9305, USA}
\author[0000-0002-7831-8751]{Mauro Giavalisco}
\affiliation{University of Massachusetts Amherst, 710 North Pleasant Street, Amherst, MA 01003-9305, USA}

\begin{abstract}
We study the relationship between the morphology and star formation history (SFH) of 361 quiescent galaxies (QGs) at redshift $\langle z_{\rm{obs}}\rangle\approx 2$, with stellar mass $\log M_*\ge10.3$, selected with the UVJ technique. Taking advantage of panchromatic photometry covering the rest-frame UV-to-NIR spectral range ($\approx40$ bands), we reconstruct the non-parametric SFH of the galaxies with the fully Bayesian SED fitting code \prospector. We find that the half-light radius \re, observed at \zobs, depends on the formation redshift of the galaxies, \zf, and that this relationship depends on $M_*$. At $\log M_*<11$, the relationship is consistent with $R_e\propto(1+z_{\rm form})^{-1}$, in line with the expectation that the galaxies' central density depends on the cosmic density at the time of their formation, i.e. the ``progenitor effect". At $\log M_*>11$, the relationship between \re and \zf flattens, suggesting that mergers become increasingly important for the size growth of more massive galaxies after they quenched. We also find that the relationship between \zf and galaxy compactness similarly depends on $M_*$. While no clear trend is observed for QGs with $\log M_*>11$, lower-mass QGs that formed earlier, i.e. with larger \zf, have larger central stellar mass surface densities, both within the \re (\Se) and central 1 kpc (\Sone), and also larger \Mone, the fractional mass within the central 1 kpc. These trends between \zf and compactness, however, essentially disappear, if the progenitor effect is removed by normalizing the stellar density with the cosmic density at \zf. Our findings highlight the importance of reconstructing the SFH of galaxies before attempting to infer their \textit{intrinsic} structural evolution.

\end{abstract}

\section{Introduction} \label{sec:intro}

The physical process, or processes, that control the transition from widespread star formation to quiescence in massive galaxies at the Cosmic Noon epoch, i.e. $1<z<4$, remain empirically unconstrained (see a recent review by \citealt{Forster2020}). Equally poorly understood at this epoch are the physical reasons behind the apparent differences in morphological, and presumably, dynamical properties between star-forming and quiescent galaxies (QGs) of similar stellar mass \citep[e.g.][]{Daddi2005,Franx2008,Wuyts2011,vanderwel2012,vanderwel2014,Barro2017}. Generally speaking, the former are on average relatively extended systems, with half-light radii up to several kiloparsec (kpc), disk-like light profiles with generally low Sersic index, e.g. $n< 2$, and well-defined rotation curves \citep{Price2016,Wuyts2016,Genzel2017,Lang2017,Tiley2019,Price2020,Genzel2020}. The latter are often very compact, with half-light radii as small as $<0.5$ kpc and steep light profiles, e.g. $n>2$. While the dynamical properties of these galaxies remain largely unexplored, the few cases, where strong gravitational lensing has made spatially resolved dynamical measurements possible \citep{Toft2017,Newman2018b}, have shown to be fast rotators characterized by significant rotation and large velocity dispersion. 

Tracking the structural properties of galaxies at different redshifts clearly reveals their  evolutionary patterns. As time evolves, star-forming galaxies grow in sizes \citep[e.g.,][]{Ferguson2004,Trujillo2007,Buitrago2008,Mosleh2012,Conselice2014,Shibuya2015,Ribeiro2016,Mowla2019}, and apparently change their dynamical states by increasing their rotational velocity and decreasing velocity dispersion \citep[e.g.][]{Simons2017} and re-distribute dark matters \citep{Genzel2020} as they are still undergoing active mass assemblies. On average, this evolution is slow and gradual and takes place as long as the star formation continues. Quiescent galaxies, on the other hand, because by the time of observations very little star formation remains inside them and rejuvenation of star formation seems to be very rare  \citep[e.g.][]{Chauke2019,Tacchella2021}, appear to undergo a rapid structural transformation during quenching, becoming much more compact than star-forming ones, followed by a secular evolution towards larger size  \citep[e.g.][]{Trujillo2006,vanDokkum2008,Williams2010, Newman2012,Cassata2013, vanderwel2014}, and it seems very difficult to reconcile this apparently remarkable structural evolution with the rather tight and homogeneous distributions of the structure and stellar age of QGs at $z\sim0$ (see a review by \citealt{Renzini2006} and references therein).

One possible way to interpret the apparent structural evolution is the late-time, i.e. after quenching, growth of individual QGs via processes such as minor mergers \citep[e.g.][]{Bezanson2009,Naab2009,vanDokkum2010,Oser2012} and/or adiabatic expansions caused by galaxy mass losses either due to the death of stars \citep[e.g.][]{Damjanov2009,vanDokkum2014} or due to the powerful feedback effect from quasar activities \citep{Fan2008}. Alternatively, QGs' structural evolution can also be explained by the progenitor effect \citep{vanDokkum2001, Carollo2013, Williams2017}. In an expanding universe, bound systems that have formed earlier should keep memory of the higher cosmic density at the time of halo collapse, at least when they are observed at high redshift \citep{Lilly2016}. In parallel, studies of the evolution of galaxy stellar-mass function have shown a rapid buildup of the QG population since $z\sim4$ \citep[e.g.][]{Ilbert2010,Ilbert2013,Muzzin2013,Tomczak2014}, adding support to the notion that the increase of the average size of QGs with cosmic time is mostly driven by the addition of freshly quenched ones at lower redshift which, because they have formed later, have larger sizes and lower central densities and become progressively more and more important in terms of number density among the population of QGs at progressively lower redshift \citep{Carollo2013}.

Quantifying empirically the relative importance of late-time growth and progenitor effect in the apparent structural evolution of galaxies is critical for understanding not only of how galaxies grow and, ultimately, of the interplay between baryonic and dark matters, but also of the relationship between galaxy structural transformations and quenching. Currently, the relative time sequence of the quenching process and of the putative morphological and dynamical transformations remains substantially unconstrained. Suggestions that quenching comes first and the structure  changes later have been made based on the observation that at least some high-redshift QGs appear to be disks \citep{Bundy2010,vanderwel2011,Toft2017,Bezanson2018b,Newman2018b}, even if substantially thicker, with steeper light profiles and larger velocity dispersion than modern disk galaxies. 

Another morphological property that has prompted investigations  because of its possible causal link with quenching is stellar mass surface density ($\Sigma$). A number of studies at different redshifts have consistently found what appears to be a threshold in the distribution of $\Sigma$, around $10^{9.5-10}$ M$_{\odot}/$kpc$^2$,  above which galaxies are predominately found to be quiescent \citep[e.g.][just to name a few]{Kauffmann2003,Franx2008,Cheung2012,Fang2013,Lang2014,Whitaker2017}. While many theoretical studies suggest that the so-called wet compaction provides a viable mechanism for this phenomenology, because it can drive the quenching of the galaxies' central regions at $2<z<4$ \citep{Dekel2009,Dekel2014,Zolotov2015,Tacchella2016}, others have shown that the apparent $\Sigma$ threshold can also be explained purely by the progenitor effect \citep{Lilly2016}. Early attempts to measure the stellar age of high-redshift galaxies have indeed found that, in general, more compact galaxies also have older stellar populations \citep[e.g.][]{Saracco2011,Fagioli2016,Williams2017}, meaning that galaxies that have formed and quenched earlier might also have intrinsically higher $\Sigma$, namely the mechanism at the core of the progenitor effect. An important question yet to be answered by observers is: does the well-established relationship between $\Sigma$ and specific star-formation rate imply a causal link between the two? Or is it the product of another physical relationship yet to be identified?

A common approach to investigating this question is to study the redshift evolution of the relationships between stellar mass and size, and between stellar mass and $\Sigma$, which contain, in {\it an integral form}, information of structural evolutionary mechanisms. This means that a certain galaxy population observed at redshift \zobs includes galaxies formed at all different epochs of $z>$ \zobs. The progenitor effect thus, if relevant, is bound to affect any correlation between the size, $\Sigma$ and other galaxy properties measured at \zobs. Unless we know precisely relative contributions from the  different mechanisms, e.g. progenitor effect vs. late-time growth, that govern the assembly of the galaxy population, the interpretation of observations will be highly model dependent (e.g., see section 3.3 of \citealt{Barro2017}). To overcome this problem, alternative approaches have been proposed, including selecting galaxies with a constant number density \citep{vanDokkum2010} or studying the evolution of the number density of QGs of a given size \citep{Carollo2013}.

With Spectral Energy Distribution (SED) modeling growing in sophistication and accuracy, efforts have been devoted to the reconstruction of the star formation history (SFH) of high-redshift galaxies \citep[e.g.][]{Maraston2010,Papovich2011,Ciesla2016,Lee2018,Carnall2018}. Taking advantage of the panchromatic data accumulated in legacy fields that densely samples the rest-frame UV-optical-NIR SED of galaxies at Cosmic Noon, a significant step forward has been made in recent years to model the SFHs in non-parametric forms \citep[e.g.][]{Ocvirk2006,Tojeiro2007,Pacifici2012,Leja2017,Belli2019,Iyer2019}, which has been demonstrated to be critical for the unbiased inference of physical parameters such as stellar mass and stellar age \citep{Leja2019}. 

Because massive galaxies, especially QGs, have assembled most of their stellar mass prior to \zobs, utilizing their reconstructed SFHs provides a  powerful tool to circumvent the progenitor effect, as the SFHs contain key information required to separate galaxies formed at different epochs. This motivates this and upcoming papers in this series, where we use the fully Bayesian SED fitting code \prospector \citep{Leja2017,Johnson2021} to reconstruct the non-parametric SFH of galaxies at $1<z<4$  and investigate the relationships between the star-formation and morphological properties of galaxies and their assembly history. In this first paper of the series, our primary goal is to quantitatively constrain the physical mechanisms behind the apparent structural evolution of QGs at Cosmic Noon. We thus focus on a sample of massive QGs at \zobs$\sim2$, and study the relationships between their morphological properties and SFHs. In an upcoming paper \citep{Ji2022b}, we will include star-forming galaxies to the analysis, and investigate causal links, if any, between galaxy structural transformations and quenching. Throughout the paper, we adopt the AB magnitude system and the $\Lambda$CDM cosmology with $\Omega_m = 0.3$, $\Omega_\Lambda = 0.7$ and $\rm{h = H_0/(100 kms^{-1}Mpc^{-1}) = 0.7}$.

\section{Method}

The goal of this work is to investigate possible correlations of the morphological properties of QGs, such as size and central stellar mass surface density, with their assembly history and stellar mass in an attempt to characterize and quantify the role of the progenitor effect, and to constrain the after-quenching morphological evolution of the galaxies up to the redshift of observation, \zobs. In section \ref{sec:sample}, we describe the sample selection of QGs used in this study, and in section \ref{sec:phot} we detail the photometric data used in this work. In section \ref{sec:SED}, we describe our measurements of the galaxies' assembly history using the SED modeling package {\sc Prospector} and in section \ref{sec:sfh_shape_def}, we describe how we characterize the reconstructed SFHs. Finally, in section \ref{sec:mor}, we describe the morphological parameters used in this work.

\subsection{Sample overview}\label{sec:sample}
\begin{figure}
    \centering
    \includegraphics[width=0.47\textwidth]{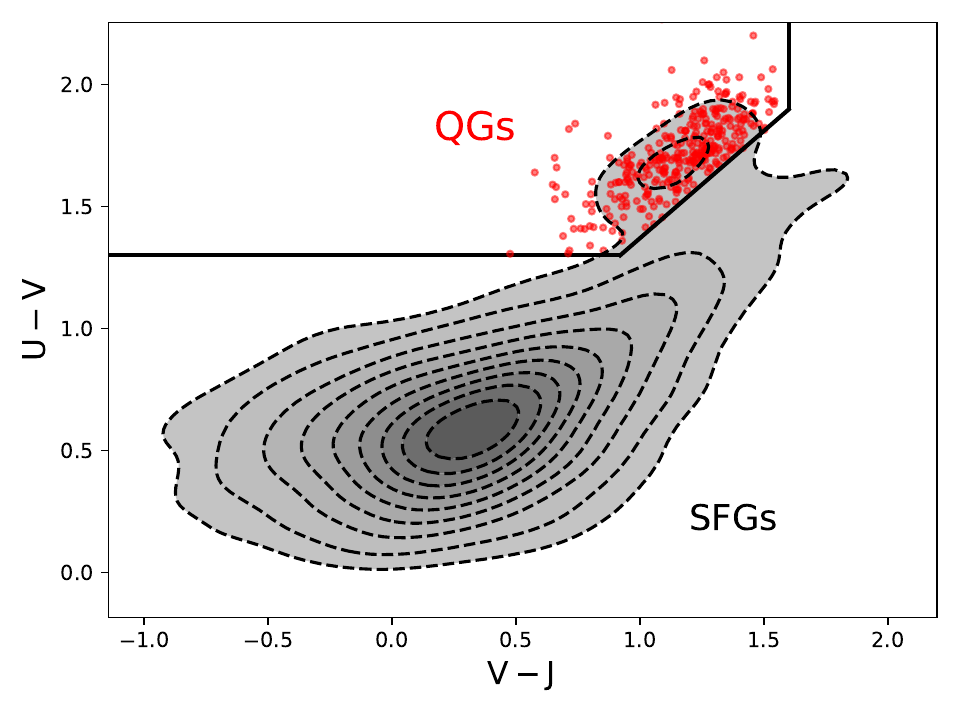}
    \caption{Rest-frame UVJ-color diagram. Red dots are the selected QGs in this study (section \ref{sec:sample}). Background grey contours show the density distribution of $1.2<z<4$ galaxies in the CANDELS/COSMOS. Black solid lines show the SFG-QG separation boundaries used in this work, where QGs are selected to be in the region enclosed by $U - V > 0.88 (V - J) + 0.49$, $U - V > 1.3$ and $V - J < 1.6$.} \label{fig:uvj}
\end{figure}

The sample of QGs considered in this paper is selected using the rest-frame UVJ color diagram \citep{Williams2009} constructed with the CANDELS \citep{Grogin2011,Koekemoer2011} multi-band photometric catalogs in the GOODS-South, GOODS-North and COSMOS fields. As Figure \ref{fig:uvj} shows, we use the selection window proposed by \citet{Schreiber2015}, which has been demonstrated to be sensitive up to $z\sim4$, to identify star-forming galaxies and QGs. In principle, we could use our new SED fitting results from \prospector (section \ref{sec:SED}) to select QGs from the upstart, with no need to first go through this UVJ selection. In practice, however, running \prospector is computationally rather expensive\footnote{In our case, typical running time for a single galaxy is $\approx10-20$ hours}, making it impractical to run it on large catalogs such as the CANDELS ones for the primary selection. Thus, we resort to first select QGs using the very well tested UVJ selection technique and then run \prospector to derive the SFH of a carefully culled sample of QGs.

The parent QG sample is selected to be in the redshift range between $1.2<$ \zobs$<4$ using either photometric, or spectroscopic redshift when available, and to have integrated isophotal signal-to-noise ratio SNR $>10$ in the \hst/WFC3 \H band to ensure good photometric and imaging qualities. We also impose a low-end stellar mass cut, $\log_{10} (M_*/M_\sun)=9$, to ensure good $M_*$ completeness in all fields ($\gtrsim 80\%$, see \citealt{Ji2018} for CANDELS/GOODS and \citealt{Nayyeri2017} for CANDELS/COSMOS). In addition, galaxies with clear evidence of AGN activity, i.e. those flagged using the \texttt{AGNFlag} in the photometric catalogs (section \ref{sec:phot}), have been removed from the sample. These criteria yield a total sample of 677 QGs, 288, 206 and 183 from CANDELS/COSMOS, GOODS-South and GOODS-North, respectively.

\begin{figure*}
    \centering
    \includegraphics[width=1\textwidth]{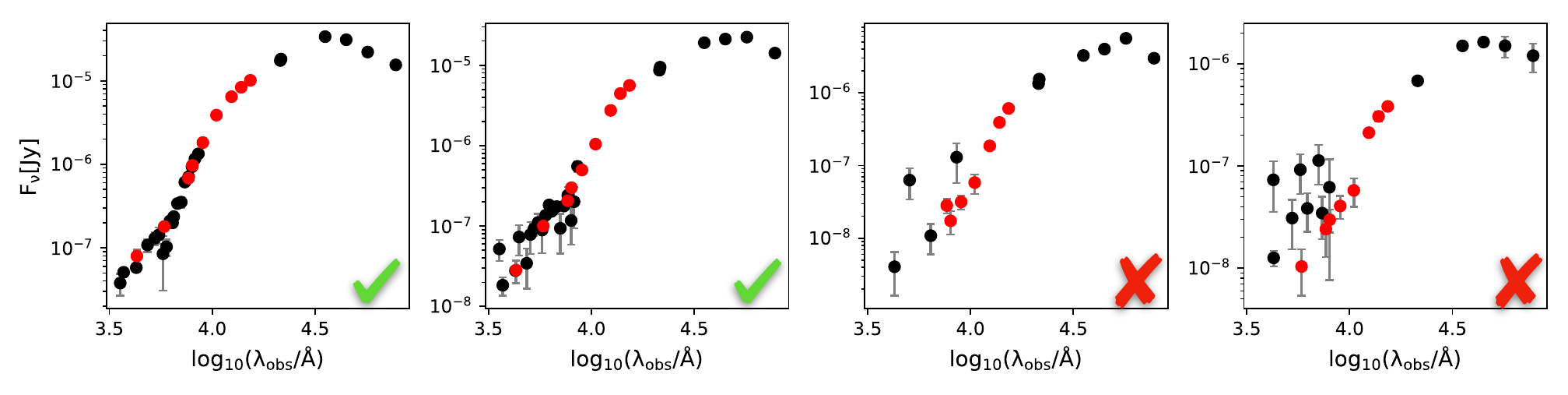}
    \caption{Examples of the photometry of individual UVJ-selected QGs that illustrate our procedure to visually inspect and clean our sample. Only photometry with SNR $>1$ is plotted here. Photometry from \hst is marked in red. The left-most panel shows an example of the best case, where the photometry has high SNR and densely samples the galaxy's SED. The second panel illustrates the next-to-best case, where increased photometric scatter is observed. The third panel shows an example where the galaxy's SED is {\it not} as densely sampled as the first two cases. The right-most panel shows an example where the sampling of the SED is good but the photometric quality is much poorer than the previous cases, primarily due to the shallower depth and lower angular resolution of ground-based observations. To secure high-quality SFH reconstructions, we have limited our final sample only to the QGs with photometry like the first two examples.}
    \label{fig:phot_example}
\end{figure*}

High-quality morphological measurements are crucial for this study. As we will detail in section \ref{sec:mor}, we use the \texttt{GALFITFlag} from \citet{vanderwel2012} to select QGs that have reliable single \sersic\ \galfit fitting results. Equally crucial to this study is to have high-quality photometric data that densely samples the rest-frame UV-Optical-NIR SED of the galaxies, since this is essential to make the high-fidelity SFH reconstructions. Because the adopted catalogs contain photometry from different telescope/instrument combinations, whose depths and angular resolutions are different from one to another, the reliability of the photometry depends critically on the quality of the adopted de-blending procedures of low-resolution images. Given the relatively small sample size, we have visually inspected the galaxies of our sample in all bands to select QGs with high-quality photometry. Examples of galaxies following our visual inspections are illustrated in Figure \ref{fig:phot_example}. After removing the galaxies with either unreliable fitting results of \galfit or poor quality of photometry, the final sample consists 361 QGs, specifically 187, 123 and 51 from the CANDELS/COSMOS, GOODS-South and GOODS-North fields,  respectively\footnote{We notice that the photometric quality in the GOODS-North is not as good as the two other fields, primarily due to the uncertain SHARDS' photometry. This results in a lower rate passing our visual inspection in the GOODS-North.}.

\subsection{Photometric data}\label{sec:phot}
Densely sampled ($\approx$ 40 bands) photometry, obtained from  ground-based and space-borne instruments and covering the rest-frame UV-to-NIR spectral range for galaxies at $z\sim2$, is available in all three fields. 

In this work, we use the photometric catalog from \citet{Nayyeri2017} for the CANDELS/COSMOS field. For the GOODS-North field, we use the photometric catalog from \citet{Barro2019}. In addition to the \hst data taken during the GOODS \citep{Giavalisco2004} and CANDELS surveys and other ground- and space-based ancillary data from UV to FIR, the 25 medium-band photometry at the optical wavelengths acquired during the SHARDS survey \citep{Perez2013}, taken by the OSIRIS instrument at the Spanish 10.4-m telescope Gran Telescopio Canarias, is newly added to the catalog. Given the special characteristic of the OSIRIS, i.e. the varying passband seen by different parts of the detector, each galaxy has its unique set of passbands. As a result, for each galaxy, this requires us to shift the nominal bandpass of every filter to the {\it actual} central wavelength measured by \citet{Barro2019} prior to SED fitting (section \ref{sec:SED}). 

Finally, for the GOODS-South field, we use the latest photometric catalog from the ASTRODEEP project (ASTRODEEP-GS43, \citealt{Merlin2021}). Compared with the old \citealt{Guo2013} catalog in the GOODS-South, this new one uses the latest and improved version of the template-fitting software T-PHOT \citep{Merlin2016} to de-blend low angular-resolution images based on positional priors from high angular-resolution ones to derive aperture matched fluxes. In addition, this new catalog adds 18 new medium-band photometric data observed with the Subaru SuprimeCAM \citep{Cardamone2010}, and another 5 photometric bands observed with the Magellan Baade FourStar \citep{Straatman2016}.

\subsection{SED fitting with \prospector}\label{sec:SED}

We use the panchromatic photometry described above to perform SED modeling for the sample QGs using \prospector \citep{Leja2017,Johnson2021}, with specific emphasis on the robust reconstruction of their SFHs and also to obtain an independent measure of $M_*$. 

\prospector is an SED fitting code built upon a fully Bayesian inference framework, and importantly, it allows flexible parameterizations of SFHs, which has been extensively shown to be critical for characterizing the diversity of the assembly history of galaxies and for unbiasedly measuring the physical properties such as the age of stellar populations \citep[e.g.][]{Lee2018, Carnall2018, Iyer2019,Leja2019, Tacchella2021}. Regarding the basic setup of \prospector, we adopt the Flexible Stellar Population Synthesis (FSPS) code \citep{Conroy2009,Conroy2010} where the stellar isochrone libraries MIST \citep{Choi2016,Dotter2016} and the stellar spectral libraries MILES \citep{Falcon-Barroso2011} are used. During the modeling, we use the sampling code {\sc dynesty} \citep{Speagle2020} which adopts the nested sampling procedure \citep{Skilling2004} that simultaneously estimates the Bayesian evidence and the posterior. 

In this work, we assume the \citealt{Kroupa2001}'s initial mass function (IMF) for  consistency with the IMF used in the \citealt{Byler2017}'s nebular continuum and line emission model, which we adopt in our SED modeling. We assume the \citealt{Calzetti2000}'s dust attenuation law and fit the V-band dust optical depth with an uniform prior $\tau_V\in(0,2)$. We also set redshift as a free parameter, but with a strong prior: a normal distribution centers at the best redshift value from the CANDELS catalogs (either photometric or secure spectroscopic redshift whenever available), with a width equal to the corresponding redshift uncertainty.

Similar to what has been extensively assumed in literature, we leave the stellar metallicity ($Z_*$) of galaxies as \textit{a} free parameter during the SED modeling, meaning that we assume all stars in a galaxy have the same $Z_*$. In this way the detailed history of metal enrichment is ignored. We use a uniform prior in the logarithmic space $\log_{10}(Z_*/Z_\sun)\in(-2,0.19)$, where $Z_\sun=0.0142$ is the solar metallicity, and the upper limit of the prior is chosen because it is the highest value of metallicity that the MILES library has. Measuring accurate $Z_*$ is beyond the scope of this work, as it requires deep spectroscopic data. Nevertheless, we have checked the measurements by fixing $Z_*=Z_\sun$ during the SED fitting (see Appendix \ref{app:metal}), finding no substantial changes in the parameters of our interests. 

Our fiducial model assumes a non-parametric piece-wise SFH composed of $\rm{N_{SFH}=9}$ lookback (i.e. the time prior to the time of observations) time bins, where the value of the star formation rate (SFR) is constant within each bin. Among the intervals of the nine bins, the first two bins are fixed to be $0-30$ and $30-100$ Myr to capture, with relatively higher time resolution, recent episodes of star formation; the last bin is assumed to be $\rm{0.9t_H-t_H}$ where $\rm{t_H}$ is the age of the Universe at \zobs; the remaining six bins are evenly spaced in the logarithmic lookback time between 100 Myr and $\rm{0.9t_H}$. As already extensively explored by \citet{Leja2019} using mock observations generated from cosmological simulations (see their section 2.1 and Figure 15), the recovered physical properties, including the age of stellar populations, generally are insensitive to $\rm{N_{SFH}}$ once it is greater than five.

As highlighted by \citet{Leja2019}, because the procedure above to model non-parametric SFHs includes ``more bins than the data warrant'', it potentially suffers from overfitting problems which are caused by the excessive model flexibility and can lead to overestimated uncertainties. This issue can be effectively mitigated by choosing a prior to weight for physically plausible forms of SFHs (e.g. \citealt{Carnall2019a, Leja2019}). Following this suggestion, we use the well-tested Dirichlet prior \citep{Leja2017} as our fiducial model. We refer readers to those references for the mathematical details of the Dirichlet prior. Here we only outline several key features. The Dirichlet distribution describes a distribution for N parameters $f_i\in(0,1)$ whose summation satisfies the constraint $\sum_{i=1}^Nf_i=1$. Instead of directly modeling the $M_{*,i}$ formed in each lookback time bin, the Dirichlet prior parameterizes a non-parametric SFH as the normalization of the SFH using the total $M_*$ formed by the time of \zobs, and the shape of the SFH using the $(\rm{N_{SFH}}-1)$ fractional mass $m_i$ formed in each time bin, 
\begin{equation}
    m_i = \frac{\delta t_if_i}{\sum_{i=1}^{\rm{N_{SFH}}}\delta t_if_i}
\end{equation}
where $f_i$ is the Dirichlet parameter and $\delta t_i$ is the width of each time bin. In our case, an uniform prior is assumed for logarithmic stellar mass $\log_{10}(M_*/M_\sun)\in(9,12)$. An additional concentration parameter $\alpha_D$ is required for the Dirichlet distribution. The case of $\alpha_D\ge1$ means distributing weight more evenly to all time bins (a smoother SFH) while the opposite case means distributing all weights to a single time bin (a burstier SFH). We decide to fix $\alpha_D=1$ as it has been demonstrated to work very well across various forms of SFHs \citep{Leja2019}. Finally, the Dirichlet prior returns a symmetric prior on stellar age and specific SFR (sSFR=SFR$/M_*$), and a constant SFH prior with $\rm{SFR(t)=M_*/t_H}$. Meanwhile, because the Dirichlet prior does not enforce a smooth transition in SFR across adjacent time bins, this means it allows events such as fast quenching and rejuvenation to happen inside galaxies. 

We have also experimented using the other non-parametric SFH prior, namely the Continuity prior, which {\it enforces} smooth changes of SFR in adjacent time bins (see \citealt{Leja2019} for details) and has been used in some recent studies \citep[e.g.][]{Leja2019b,Tacchella2021}. We first use synthetic galaxies from the IllustrisTNG cosmological simulation \citep{Nelson2018,Springel2018,Marinacci2018,Pillepich2018,Naiman2018} to validate our fiducial \prospector fitting procedure, and then we compare the recovered physical properties derived from the two different priors of non-parameteric SFHs with the intrinsic properties obtained from the simulations (see Appendix \ref{app:tng}). From this comparison we find that the Dirichlet prior works better than the Continuity prior given the typical wavelength coverage and photometric quality of our sample. While tests with simulations provide very valuable guidance to assess the performance of the SED fitting procedure, they are also very limited because it is hard to assess the accuracy of how the simulations model the SFH of real galaxies. We thus decide to additionally do an \textit{empirical} test where we run a new set of \prospector SED fittings for our QGs using the Continuity prior, with the specific goal to quantify the dependence of the physical parameters of our interests on the adopted prior (see Appendix \ref{app:conti}). While systematic offsets are observed, strong and tight correlations are observed among the same parameters derived using the two different priors, meaning that switching to the Continuity prior will not qualitatively change any conclusion of this work. We have ultimately decided to adopt the measurements from the Dirichlet prior as the fiducial ones, because this prior works better according to the IllustrisTNG-based tests that we have conducted. 

\begin{figure*}
    \centering
    \includegraphics[width=1\textwidth]{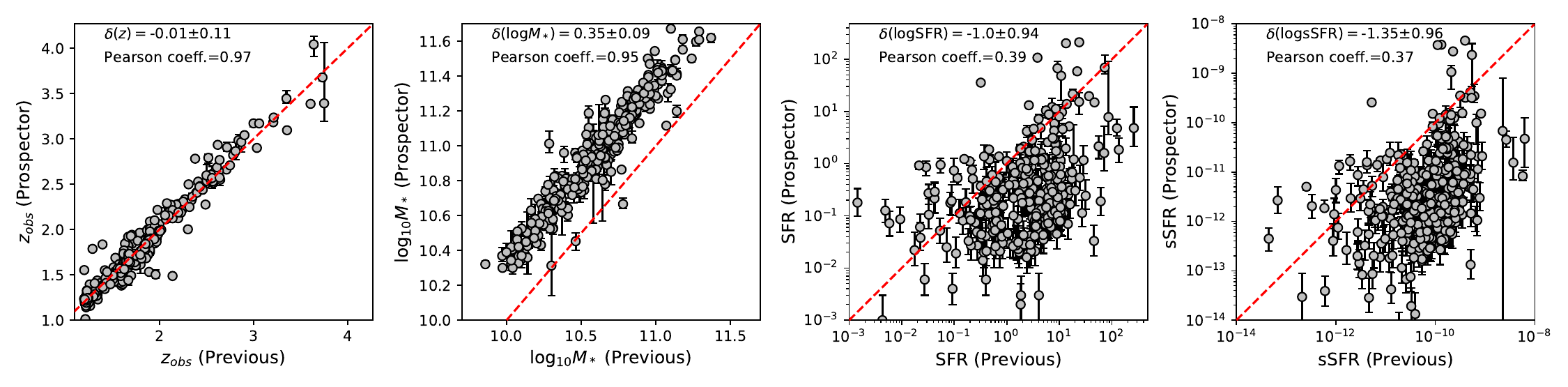}
    \caption{Comparisons between parameters derived from our \prospector fitting (y-axis) and from previous (x-axis) measurements (see section \ref{sec:SED} for details). Also labelled in each panel are the median and standard deviation of the difference between the two measurements, and the Pearson correlation test coefficient between the two measurements. The red dashed line marks the one-to-one relation.}
    \label{fig:compare_sed_param}
\end{figure*}

We now move to check the consistency of our new \prospector SED fitting results with  existing previous measures. We use the previous measurements from \citet{Lee2018} for the CANDELS/GOODS fields, and from \citet{Nayyeri2017} for the CANDELS/COSMOS field. In Figure \ref{fig:compare_sed_param} we compare the \prospector-derived \zobs, $M_*$, SFR and sSFR to the previous measures. We also run Pearson correlation tests between the two measurements. In all cases we find strong correlations, showing the consistency between the \prospector and the previous measurements.

Specifically, excellent agreement is seen for \zobs. A strong and rather tight correlation is also seen for $M_*$, although a +0.35 dex offset, $\delta(\log M_*)=\log M_*(\rm{Prospector})-\log M_*(\rm{Previous})$, is found. We note that the previous measurements above have been obtained assuming the \citet{Chabrier2003}'s IMF which is different from the \citet{Kroupa2001}'s IMF adopted by our \prospector fitting. This however is insufficient to account for the observed offset, as the difference in $M_*$ caused by the two IMFs is typically $\approx10\%$ only \citep[e.g.][]{Speagle2014}. The real cause of the systematic shift in the $M_*$ measurements comes from the different SFHs assumed during SED modeling. Unlike the non-parametric form of SFHs assumed in our \prospector modeling, the adopted \citeauthor{Nayyeri2017} measurements assumed an exponential declining SFH (SFR$(t)\propto e^{-t/\tau}$). Although the \citeauthor{Lee2018} measurements left the form of SFH as a free parameter during their SED fitting, this could only vary among five parametric functional forms, and they found that the majority of QGs prefer an exponential declining SFH as the best-fit solution (see their Figure 2). Using non-parametric SFHs returns older stellar ages, and hence larger stellar masses, than using parametric ones (see \citealt{Carnall2019a,Leja2019,Leja2021} for details). Evidence that the larger stellar masses are actually more accurate has been found. Using synthetic galaxies generated from cosmological simulations, \citet{Lower2020} showed that the $M_*$ derived from non-parametric SFHs is closer to the intrinsic value and it can be as much as $0.3-0.5$ dex larger than that derived from parametric SFHs, which is quantitatively similar to the offset of $M_*$ measurements seen here. \citet{Leja2020} showed that the SFHs inferred from the non-parametric method are more consistent with the observed evolution of galaxy stellar-mass function over $0.5<z<3$. In addition, \citet{Leja2021} showed that using the SFR and $M_*$ derived from non-parametric SFHs can solve a long-standing systematic offset of the star-forming main sequence between observations and predictions from cosmological simulations (see their section 7.1). 

Accurately measuring SFR is very challenging for QGs \citep{Conroy2013} and the results critically depend on the assumed form of SFHs. The adoption of non-parametric forms by \prospector appears to provide substantial improvements if adequate wavelength coverage of galaxy's SED is available \citep{Leja2019b}. Comparing our new results with previous ones shows that although a correlation with a Pearson correlation coefficient $\approx0.4$ is seen between the two sets of SFR measurements, the scatter, unsurprisingly, is much larger than that of \zobs and $M_*$. A similar behavior is also observed when comparing the measurements of sSFR. In the Figure we compare the \prospector-derived {\it instantaneous} SFR, which in our case is essentially the SFR averaged over the past 30 Myr (the first time bin of the assumed SFHs), to the previous SFR measures. These include a mix of measures from the SED fitting and also from $\rm{SFR_{UV}^{obs}+SFR_{IR}}$ diagnostics (see \citealt{Lee2018} for details). We find that in addition to the increased scatter of the correlation, the former is smaller than the latter by $\approx1$ dex. In part, this reflects the different methodology adopted for the SED fitting, as discussed before. In part, this also reflects the fact that, especially for QGs, the $\rm{SFR_{UV}^{obs}+SFR_{IR}}$ diagnostics are affected by significant contributions to the IR flux from AGB stars and/or dust heated by old stars or AGN \citep[e.g.][]{Salim2009,Fumagalli2014,Hayward2014}, rather than on-going star formation, making the latter SFR only an upper limit. Regardless of the offset and scatter seen in the SFR and sSFR comparisons, our SED fitting with \prospector consistently shows that the QGs pre-selected through the UVJ technique have low sSFR $\lesssim10^{-10}\rm{yr}^{-1}$, confirming the quiescent nature of the sample galaxies.

\subsection{Characterizing the shape of SFHs} \label{sec:sfh_shape_def}

In this section we introduce the parameters, which are derived from the reconstructed non-parametric SFHs and take into account the information contained in their shapes, to quantitatively describe the epochs and timescales of SFHs. Similar to what has been done in recent literature, the first set of parameters are characteristic redshifts, including
\begin{itemize}
    \item \zf (formation redshift): the redshift when the time interval between \zf and \zobs equals to the mass-weighted age (\tage) of a galaxy
    \item \zp (p$=50,70,90$): the redshift when p-percent $M_*$ seen by the time of \zobs have already formed.  
\end{itemize}
We point out that the definition of \zf is very similar to that of $z_{50}$, and the difference between the two is only 5\% on average and 21\% at most. In addition, the uncertainty of \zp increases as p becomes smaller because of the large uncertainty in estimating the contributions from older stellar populations given the data coverage we currently have \citep{Papovich2001}. \citet{Tacchella2021} used the time intervals between two \zp as proxies for the timescales of star formation and quenching of galaxies. While this is able to describe the general shape of SFHs, the choice of the values of p is somewhat arbitrary, and it is unable to describe other possible features of SFHs such as star-formation rejuvenation. Instead of using arbitrary time epochs  (e.g. \zp), we propose a new set of parameters that utilize the information of the \textit{full} shape of SFHs. In this way the aforementioned drawbacks can be mitigated. The new parameters are detailed as follows.

\begin{figure*}
    \centering
    \includegraphics[width=1\textwidth]{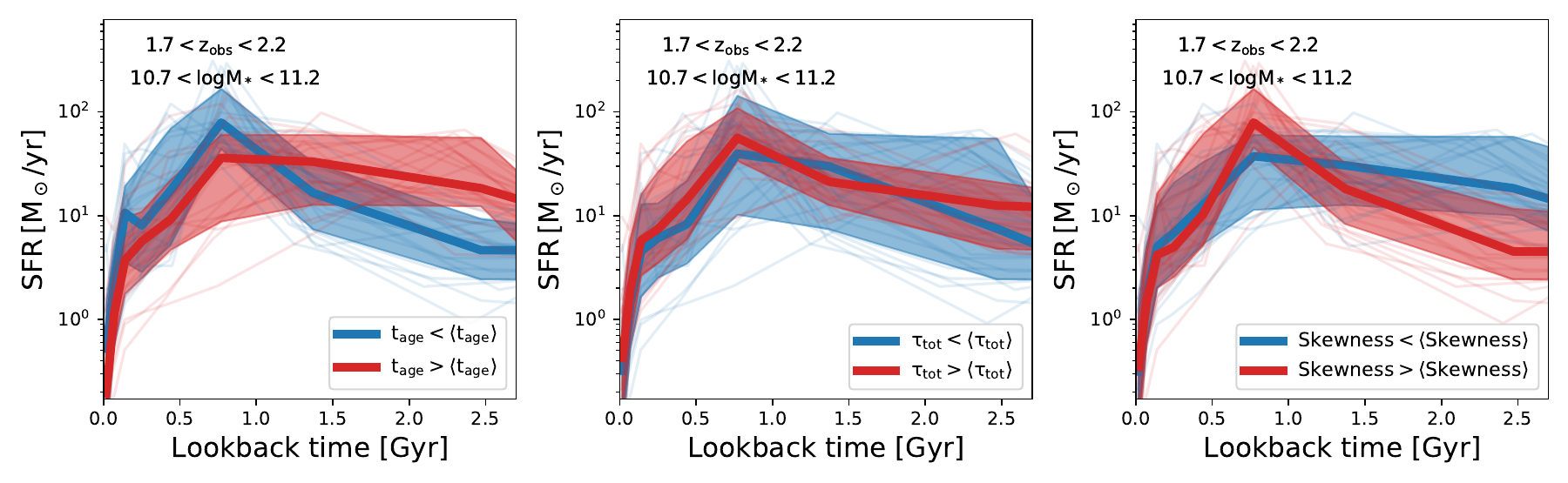}
    \caption{ Demonstration of the effectiveness of the metrics introduced in section \ref{sec:sfh_shape_def} to quantitatively characterize the shape of reconstructed SFHs. In the plots we only show the SFHs, expressed as SFR versus lookback time from $z_{\rm{obs}}$, for the QGs within narrow ranges of redshift and stellar mass, namely $1.7<z_{\rm{obs}} <2.2$ and $10.7<\log_{10}(M_*/M_\sun)<11.2$. A full collection of reconstructed SFHs is shown in Figure \ref{fig:sfh}. From left to right the QGs are divided into two groups using the sample median value of $\langle t_{\rm{age}}\rangle$, $\langle\tau_{\rm{tot}}\rangle$ and $\langle$Skewness$\rangle$ (check section \ref{sec:sfh_shape_def} for parameter definitions). The thin dashed line shows the individual SFHs. The SFH averaged over each group is shown as a thick solid line (median) with a shaded region (16-84th percentile range).}
    \label{fig:SFH_shapes_rep}
\end{figure*}

The first parameter \tautot measures the width (i.e. duration) of the main star-formation episode of an SFH,
\begin{equation}
    \tau_{\rm{tot}} =2\times\sqrt{\frac{\int_{0}^{t_{\rm{H}}}(t-t_{\rm{age}})^2\cdot\rm{SFR}(t)dt}{\int_{0}^{t_{\rm{H}}}\rm{SFR(t)}dt}}.
    \label{equ:tautot}
\end{equation}
This definition is essentially the square root of the mass-weighted second central moment of stellar ages, times the factor two to account for the full (two-side) width. In our case, since the SFH has a constant piece-wise shape with nine time bins, the equation above can be re-written as
\begin{equation}
    \tau_{\rm{tot}} =2\times\sqrt{\frac{\sum_{i=1}^{9}\,(t_i-t_{\rm{age}})^2\cdot \rm{SFR}_i\delta t_i}{M_*}}.
\end{equation}
In the middle panel of Figure \ref{fig:SFH_shapes_rep}, we plot the reconstructed SFHs for the subsample of QGs within a narrow redshift ($z_{\rm{obs}}=1.7-2.2$) and stellar mass ($\log_{10}M_*=10.7-11.2$) ranges. We divide the subsample into two groups using its median \tautot. It can be seen in Figure \ref{fig:SFH_shapes_rep} that the group having larger \tautot also has a more extended period of star-formation episode.

Another important characteristic is the overall asymmetry of SFHs which contains key information of the relative timescale between the period of actively forming stars and the period of the quenching of star formation. We introduce two ways to quantify this property. The first one is the Skewness, i.e. the third standardized moment of stellar ages,
\begin{equation}
    \begin{split}
    \rm{Skewness}= \frac{1}{(0.5\tau_{\rm{tot}})^3}\cdot \frac{\int_{0}^{t_{\rm{H}}}(t-t_{\rm{age}})^3\cdot\rm{SFR}(t)dt}{\int_{0}^{t_{\rm{obs}}}\rm{SFR(t)}dt} \\ 
    = \frac{8}{\tau_{\rm{tot}}^3}\cdot\frac{\sum_{i=1}^{9}\,(t_i-t_{\rm{age}})^3\cdot \rm{SFR}_i\delta t_i}{M_*}
    \end{split}
\end{equation}
A more positive Skewness means that an SFH has a longer tail towards $t>t_{\rm{age}}$, which is demonstrated in the right-most panel of Figure \ref{fig:SFH_shapes_rep}. The other way is to first divide an SFH into two parts, namely an actively star-forming period when $t_{\rm{age}}<t<t_{\rm{H}}$ and a quenching period when $0<t<t_{\rm{age}}$, and then directly compare the time widths between the two periods, i.e. $\tau_{\rm{SF}}/\tau_{\rm{Q}}$, where
\begin{equation}
    \begin{split}
    \tau_{\rm{SF}}=\sqrt{\frac{\int_{t_{\rm{age}}}^{t_{\rm{H}}}(t-t_{\rm{age}})^2\cdot\rm{SFR}(t)dt}{\int_{t_{\rm{age}}}^{t_{\rm{H}}}\rm{SFR(t)}dt}}\\
    \tau_{\rm{Q}}=\sqrt{\frac{\int_{0}^{t_{\rm{age}}}(t-t_{\rm{age}})^2\cdot\rm{SFR}(t)dt}{\int_{0}^{t_{\rm{age}}}\rm{SFR(t)}dt}}
    \end{split}.
\end{equation}
For our measurements, we replace both integrals with discrete summations in the same way as we did for \tautot and the Skewness. We have compared the Skewness and $\tau_{\rm{SF}}/\tau_{\rm{Q}}$ in Appendix \ref{app:skew_ratio}, where we find that these two metrics are highly consistent with each other in quantifying the asymmetry of SFHs. We also show that combining the two metrics can effectively identify the subtle events such as star formation rejuvenation in SFHs.
 
\subsection{Morphological properties}\label{sec:mor}

Morphological measurements of individual QGs are taken from \citet{vanderwel2012}, where the morphology of galaxies is described in terms of their light profiles modeled with a single \sersic function using \galfit \citep{galfit2010}. To secure robust measurements, we only use those with the best quality flag, namely \texttt{GALFITFlag = 0}, following the recommendation of \citeauthor{vanderwel2012}. This means that we exclude the galaxies whose single \sersic  fitting (1) has a suspicious result where the \galfit magnitude significantly deviates from the expected magnitude derived from \sextractor \citep{Bertin1996}, or (2) has a bad result where at least one parameter reaches the constraint value (e.g. $0.2<n<8$, $b/a>0.1$) set to \galfit, or (3) is unable to converge.

We use the measurements in the \H band because it better probes the stellar morphology for $z\sim2$ galaxies as it is the reddest \hst imaging data available in the three fields. Throughout this work, all morphological parameters are measured in the \H band unless specified otherwise. For each QG, we obtain the half-light radius \re (circularized, $=R_{e,maj}\sqrt{b/a}$) and the \sersic index $n$. Together with $M_*$ derived from the SED fitting, we obtain the projected stellar mass surface density inside the effective radius \Se($=M_*/(2\pi R_e^2)$) and inside the central radius of 1 kpc \citep{Cheung2012},
\begin{equation}
    \Sigma_{\rm{1kpc}} =\frac{M_{\rm{1kpc}}}{\pi \cdot1\rm{kpc}^2} = \frac{M_*}{\pi \cdot1\rm{kpc}^2}\frac{\gamma(2n,\,x)}{\Gamma(2n)}
\end{equation}
where $x=b_n(\frac{1\rm{kpc}}{R_e})^{1/n}$, $\gamma/\Gamma$ is the ratio of incomplete gamma function divided by complete gamma function, and $b_n$ is calculated by numerically solving $\Gamma(2n)=2\gamma(2n,b_n)$ \citep{Graham2005}. 

The measurements of \re and \n are usually highly covariant, making the determination of best-fit values of the individual parameters (\n in particular) highly uncertain \citep[e.g.][]{Ji2020}. As discussed in \citet{Ji2022}, because \Sone is derived as the combination of \re and \n, this helps to reduce the uncertainty stemming from the strong covariance between the two parameters, making \Sone significantly better constrained than the two parameters individually. In this sense, \Sone is a more robust outcome of the light profile modeling procedure than \n. 

In addition, because \Se and \Sone strongly correlate with $M_*$, any correlation observed with $\Sigma$ then can in principle be driven by $M_*$ rather than galaxy morphology. This motivates the use of the fractional mass within the central radius of 1kpc, 
\begin{equation}
    \frac{M_{1\rm{kpc}}}{M_*}=\frac{\pi (\rm{1kpc})^2\Sigma_{\rm{1kpc}}}{M_*}=\frac{\gamma(2n,\,x)}{\Gamma(2n)}, \label{equ:mone}
\end{equation} 
which was introduced by \citet{Ji2022}, and has been demonstrated to be able to quantify the compactness of galaxies while removing most of the dependence on $M_*$.  

\section{Results}

We use the reconstructed SFHs of the QGs to study the relationships of their assembly history with $M_*$, size and other morphological properties. In section \ref{sec:diverse}, we show the diverse assembly history of QGs. In section \ref{sec:mass_sfh}, we study the dependence of the assembly history of QGs on $M_*$. In section \ref{sec:re_zform}, we explore the relationship between the assembly history of QGs and \re. In section \ref{sec:re_zform_stacking}, we carry out the stacking analysis to further study the relationship between the multi-band light distribution of QGs and their assembly history. Finally, in section \ref{sec:zf_sigma}, we explore the relationships between the compactness of QGs and their assembly history.

\subsection{The Diverse Assembly History of QGs}\label{sec:diverse}

\begin{figure}
    \centering
    \includegraphics[width=0.47\textwidth]{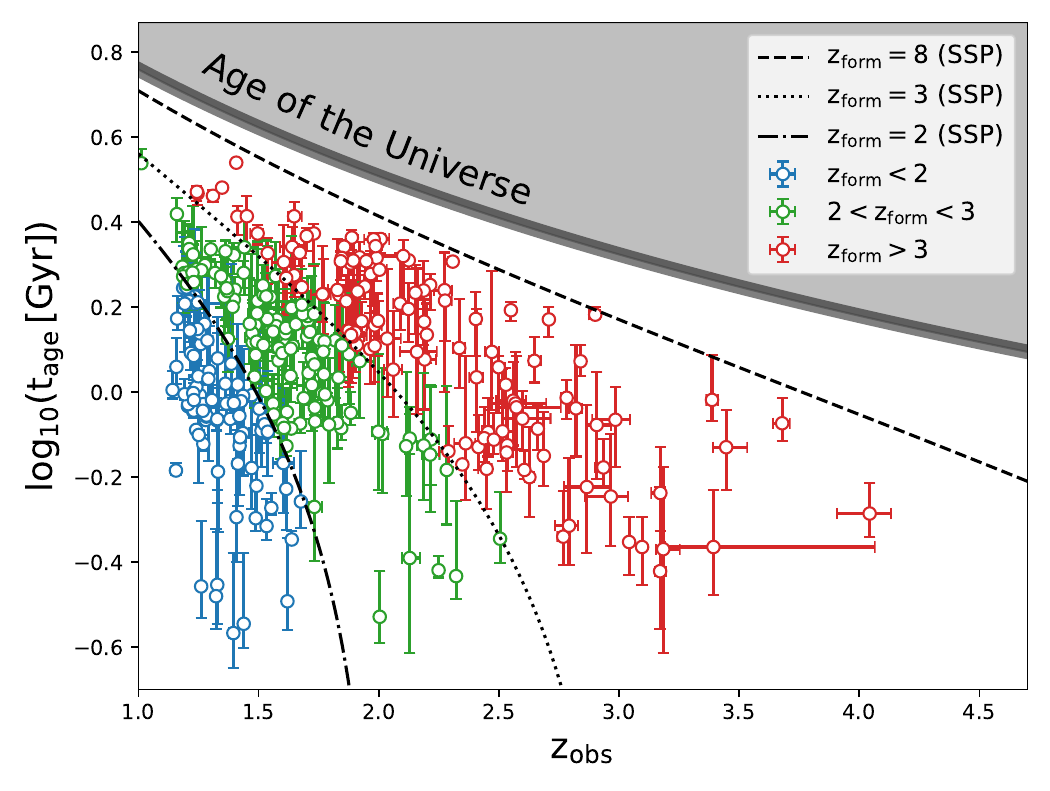}
    \caption{The relationship between mass-weighted age (\tage, vertical axis) and observed redshift (\zobs, horizontal axis). Individual QGs are color-coded according to their \zf. The thick grey line marks the age of the Universe. The dotted, dashed and dot-dashed lines present the expected age for a single stellar population (SSP) formed at $z=8$, $z=3$ and $z=2$ respectively.} \label{fig:age_z}
\end{figure}

To begin, we show the diverse assembly history of QGs by plotting \tage as a function of \zobs. Figure \ref{fig:age_z} shows that at fixed \zobs QGs are a mixture of those that have been quenched for a long time and those that have just quenched. The span of \tage is $\approx0.3\sim2.5$ Gyr at fixed \zobs, being very similar to the range of stellar ages measured by \citet{Belli2019} for a sample of 24 QGs at $z\sim1.6$ using the rest-frame optical spectra taken by the Keck/MOSFIRE, and by \citet{Carnall2019} for a sample of 75 QGs at $z\sim1.1$ using the rest-frame UV/Optical spectra taken by the VLT/VIMOS, as well as by \citet{Estrada2020} for a sample of QGs at $0.7<z<2.5$ using low-resolution spectra taken by the \hst/G102 and G141 grisms. If we plot the expected stellar age of single stellar populations (SSPs) formed at different redshifts, we find that most of our QGs fall between the SSPs formed between $z=1.5$ and $z=8$. This is consistent with the measurements of \citet{Tacchella2021} for a sample of lower-redshift ($z\sim0.8$) QGs using the deep rest-frame optical spectra taken by the Keck/DEIMOS. We note that the \tage of some QGs in our sample suggests that they have formed at $z\sim8$ when the Universe was only $\approx0.6$ Gyr old. This means that they have assembled most of their mass within a very short period of time since the Big Bang. These QGs provide an excellent test bed of the history of metal enrichment in galaxies in the early Universe and the formation of the First Galaxies. This idea has been exploited by  \citet{Kriek2016}, who carried out ultradeep rest-frame optical spectroscopy with the Keck/MOSFIRE for a $z=2.1$ QG and reported a [Mg/Fe] $\approx$ 0.6 which is consistent with that the galaxy has formed at $z\approx10$.

\begin{figure*}
    \centering
    \includegraphics[width=1\textwidth]{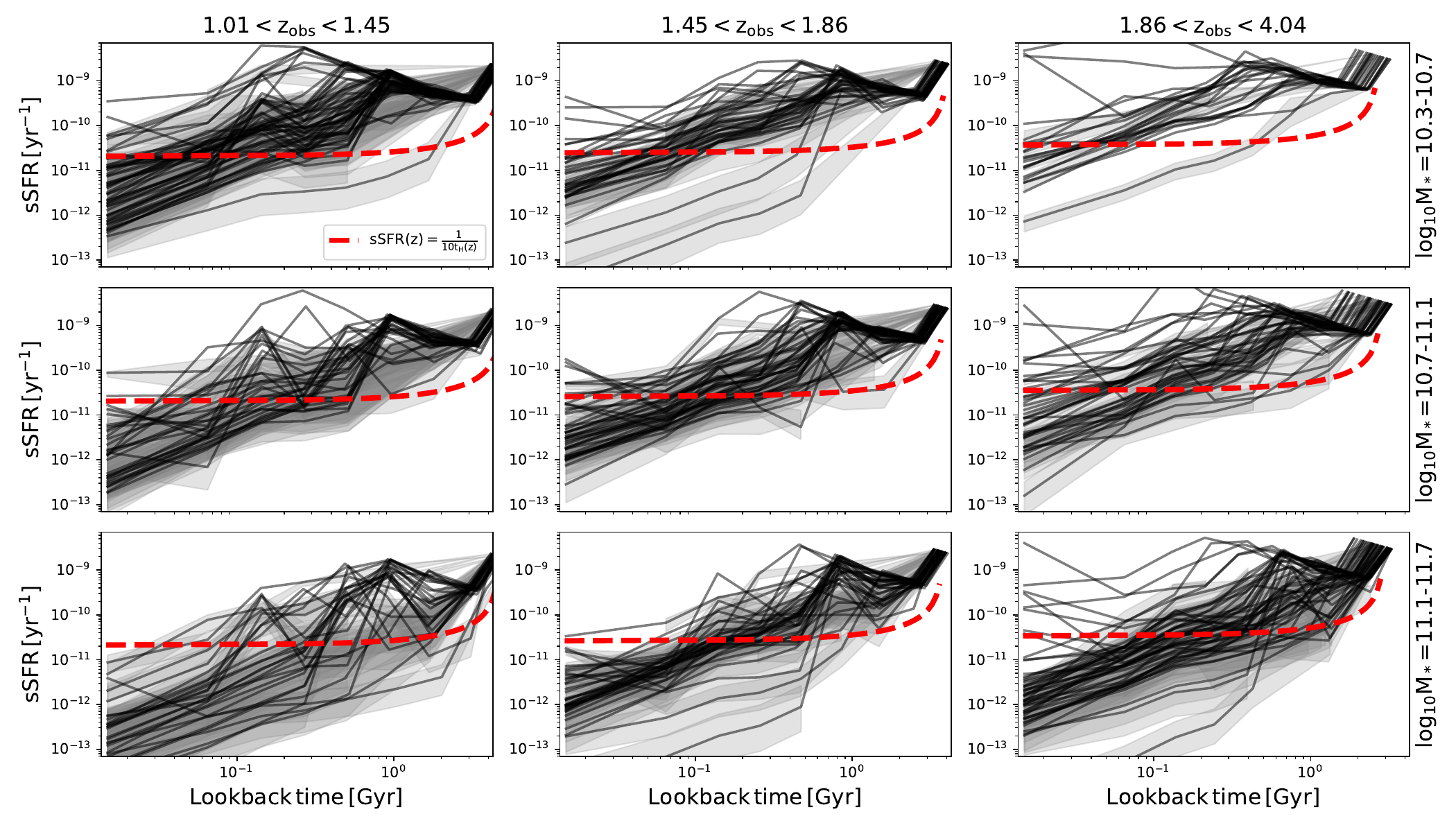}
    \caption{The collection of the reconstructed SFHs. The sample is divided in intervals of \zobs and $M_*$, where \zobs increases from left to right and $M_*$ increases from top to bottom. The y-axis is plotted as sSFR. The red dashed line marks the relation of $1/(10t_{\rm{H}} (z))$. sSFR below the red dashed line indicates a very low level of star formation. } \label{fig:sfh}
\end{figure*}

In Figure \ref{fig:sfh} we present the full collection of the reconstructed SFHs of our sample galaxies. Also marked in Figure \ref{fig:sfh} is the relation of $1/(10t_{\rm{H}} (z))$. Because the inverse of sSFR is the mass-doubling time of galaxies assuming that they continue forming stars with the current SFR, when the sSFR is below the marked relation,  this means the mass-doubling time is longer than even 10$\times$ the age of the Universe, implying a very low level of star formation. At fixed \zobs, QGs with larger $M_*$ have lower sSFR at the time of observations. Meanwhile, at fixed $M_*$, QGs observed at higher \zobs have higher sSFR. This is consistent with the downsizing effect of galaxy formation \citep{Cowie1996}. A closer inspection of the variety of shapes of the individual SFHs reveals a broad diversity of assembly history. Looking at QGs with similar $M_*$ and \zobs shows that some have quickly become quiescent after an initial, relatively rapid burst of star formation, while others have experienced a more extended assembly history, with some showing multiple episodes of enhanced star formation activity before \zobs. 

\begin{figure*}
    \centering
    \includegraphics[width=1\textwidth]{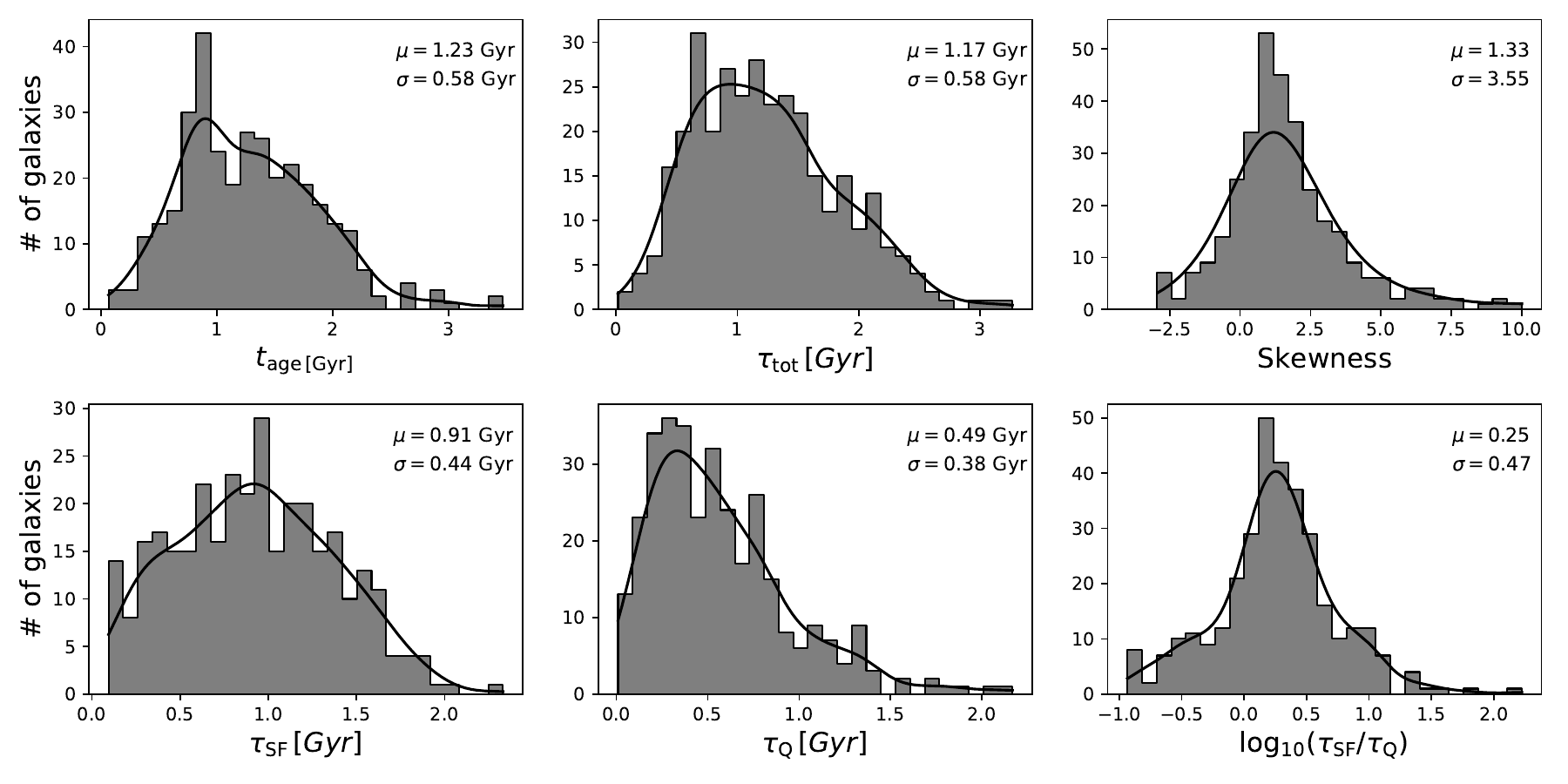}    
    \caption{Histograms of the key parameters used to quantify the shape of SFHs. We refer readers to section \ref{sec:sfh_shape_def} for the definition of individual parameters in detail. Overplotted black solid curve show the underlying density distributions estimated using a Gaussian kernel. The median and standard deviation are also labelled for each distribution.}
    \label{fig:SFH_shapes_dist}
\end{figure*}

In Figure \ref{fig:SFH_shapes_dist} we plot the distributions of the individual characteristic parameters of SFHs introduced in section \ref{sec:sfh_shape_def}. All parameters span quite wide ranges, again illustrating the diversity of the assembly history of QGs. Using \tautot as a proxy, the median timescale of the main star-forming episode inside our $z\sim2$ QG sample is 1.17 Gyr with a 0.58 Gyr standard deviation of the distribution. This is in quantitative agreement with spectroscopic studies of QGs at $z>1$. For example, \citet{Kriek2019} used elemental abundances [Fe/H] and [Mg/Fe], measured using the spectra taken by the Keck/LRIS and MOSFIRE for a sample of five massive QGs, to constrain the star-formation timescale to be 0.2$-$1 Gyr at $z\sim1.4$. The agreement is tantalizing, especially since the data sets and methodologies used are dramatically different. The histograms of \tausf and \tauq also reflect the relatively longer duration of the star-forming phase, i.e. the time span before the peak of the SFH, relative to the quenching phase, namely the period past the peak. The former is about 1 Gyr long on average, while the latter is half as long, reinforcing the conclusions of a large body of research that consistently suggests that quenching is generally a comparatively shorter process than star formation \citep[e.g.][]{Barro2013,Renzini2015, Estrada2020, Chen2020, Tacchella2021}. The diversity of the assembly history of QGs is also reflected in the very broad distributions of the Skewness and of \tausf/\tauq. Both parameters describe the relative timescales of the star formation and quenching of high-redshift QGs. While the focus of this paper is on the progenitor effect, rather than on quenching, in an another, upcoming paper of this series we will address the relationship between the structure of galaxies and the Skewness and \tausf/\tauq of their SFHs. 

\begin{figure}
    \centering
    \includegraphics[width=0.47\textwidth]{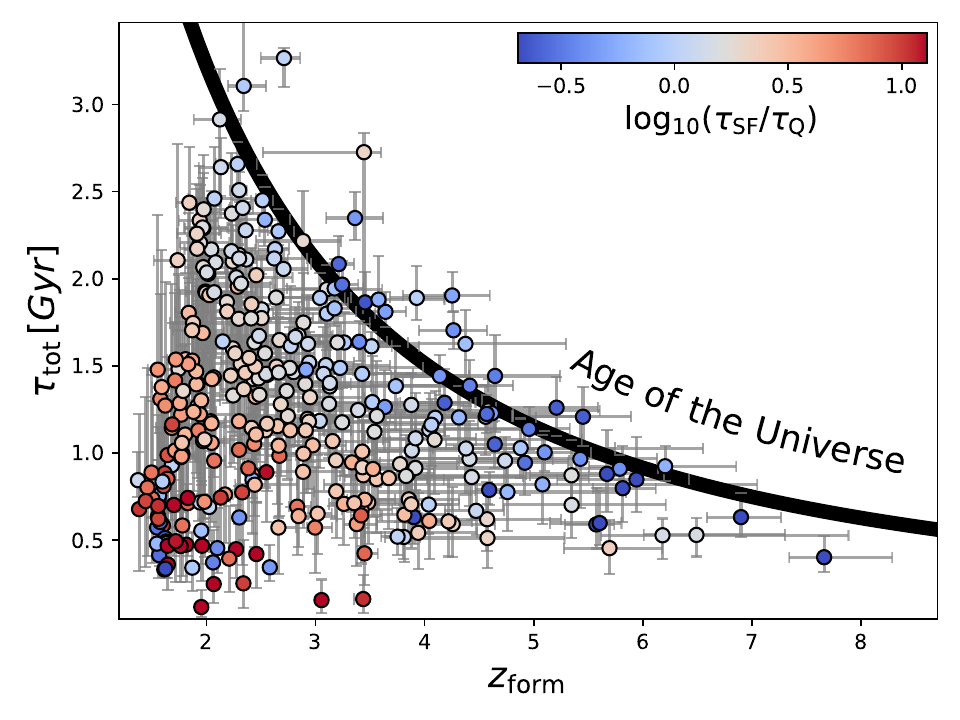}
    \caption{The relationship between \zf and \tautot. Each point is color-coded using \tausf/\tauq. The age of the Universe at \zf is also shown as the black solid line.}
    \label{fig:zf_tau_skew}
\end{figure}

Finally, we combine \zf, \tautot and \tausf/\tauq, and plot them together as a single scatter plot which is shown in Figure \ref{fig:zf_tau_skew}. Again, the diverse assembly history of QGs can be clearly recognized. The \tautot spans the entire range from $\sim0.1$ Gyr to the age of the Universe at \zf, meaning that for the QGs formed at the similar epoch of the Universe, the timescale of their mass assemblies can be dramatically different. While the reconstructed SFHs suggest that some of the QGs have experienced a rapid monolithic assembly history (small \tautot), namely that they have assembled most of their mass within a very short period of time, other QGs have assembled their mass following a much more smoother and extended process (large \tautot). 

\subsection{The Stellar Mass of QGs and Their Assembly History}\label{sec:mass_sfh}

\begin{figure*}
    \centering
    \includegraphics[width=0.7\textwidth]{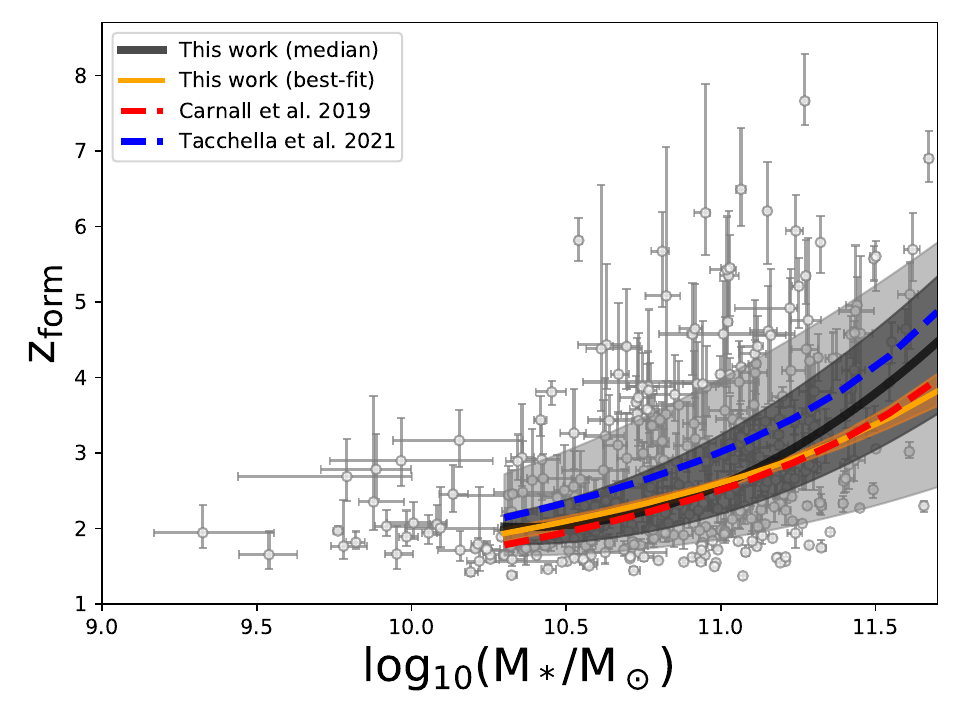}
    \caption{The relationship between \zf and $M_*$. While we only study the QGs with $\log_{10}(M_*/M_\sun)\ge10.3$ in this work, the plots here show all UVJ-selected QGs in the CANDELS/COSMOS, GOODS-South and GOODS-North. We use the COnstrained B-Splines package in R ({\sc Cobs}) to carry out the non-parametric inter-quantile regressions. The median (black solid line), 31th-69 percentiles (0.5$\sigma$, dark grey shaded region) and 16th-84th percentiles (1$\sigma$, light grey shaded region) relations are plotted in the panel. The orange solid line and the shaded region show the best-fit and 1$\sigma$ range of the linear relation between \tage and $M_*$ (see Section \ref{sec:mass_sfh} for details). Also shown as the red and blue dashed lines are the best-fit relations from \citet{Carnall2019} and \citet{Tacchella2021}, respectively. }
    \label{fig:mass_zp}
\end{figure*}

In this section we study the relationship between $M_*$ and the assembly history of QGs. In Figure \ref{fig:mass_zp}, $M_*$ is plotted against \zf. While the entire discussions of this work focus only on $\ge 10^{10.3}M_\sun$ QGs where the environmental effects should only play a relatively minor role, we also include $<10^{10.3}M_\sun$ QGs to the plot here, just to illustrate the entire UVJ-selected QG sample in the CANDELS/COSMOS and GOODS fields. 

To begin, \zf spans a wide range from 1.5 to 6  at fixed $M_*$, which is very similar to the ranges found by other studies of high-redshift QGs \citep{Saracco2019,Carnall2019,Estrada2020,Tacchella2021}. The range of \zf also becomes wider as QGs are more massive. For a more quantitative comparison, we fit a linear relation between $M_*$ and \tage. The best-fit relation (the orange solid line in Figure \ref{fig:mass_zp}) is found to be
\begin{equation}
    \frac{t_{\rm{age}}}{Gyr} = 2.46^{+0.07}_{-0.04} - 1.23^{+0.14}_{-0.21}\cdot \log_{10}(\frac{M_*}{10^{11}M_\sun})
\end{equation}
where the uncertainties of best-fit parameters are estimated as follows. We first bootstrap the sample, during which we also resample the value of each measurement using a normal distribution with the uncertainties of \tage and $M_*$ derived from {\sc Prospector} fitting results. We then repeat the process 1000 times, and get the 1$\sigma$ uncertainty for the best-fit relation. In this way both sample random errors and measurement uncertainties are taken into account.

\citet{Carnall2019} and \citet{Tacchella2021} also performed the same linear fit using their measurements\footnote{For reference, the best-fit relation of \citet{Carnall2019} is $t_{\rm{age}}/Gyr=2.56^{+0.12}_{-0.1}-1.48^{+0.34}_{-0.39}\log_{10}(M_*/10^{11}M_\sun)$, while \citet{Tacchella2021} found it to be $t_{\rm{age}}/Gyr=2.1\pm0.1-1.3\pm0.4\log_{10}(M_*/10^{11}M_\sun)$.}, which are shown as the red and blue dashed lines in Figure \ref{fig:mass_zp}. Compared with our QGs, their samples are at lower redshift $0.5<z<1.3$ where rest-frame UV/optical spectroscopy is still feasible for statistically significant QG samples using current instrumentations. Both studies included spectra to their SED modellings for the SFH reconstructions. We find very good agreement between our best-fit relation and that of \citet{Carnall2019}, suggesting that the SFH of QGs can be robustly reconstructed even without spectroscopy, if sensitive, panchromatic photometry is available (also see Appendix \ref{app:tng} for our tests with simulated galaxies). In fact, given that in both \citeauthor{Carnall2019} and \citeauthor{Tacchella2021} cases the wavelength range of spectral coverage is very narrow, and because the accurate absolute flux calibration for the ground-based spectroscopic data is challenging, the continuum shape is modelled using the polynomials rather than a pure stellar synthesis model. Unless the spectra have multiple detections of stellar spectral features (e.g. absorptions) with high SNRs, the real information gained from the spectra is limited.

A $\approx0.4$ Gyr offset is seen between our best-fit relation and that of \citet{Tacchella2021}. This is unlikely due to the sample selection effects, because all QGs are selected using the UVJ technique and the $M_*$ distributions are very similar. While \citet{Tacchella2021} also used \prospector, they assumed the Continuity prior while we are using the Dirichlet prior. As one can see in detail in Appendix \ref{app:tng}, compared with the Dirichlet prior, using the Continuity prior over-estimates \tage by $\approx0.5$ Gyr for the Illustris-TNG QGs at $z\sim2$, which is quantitatively consistent with the offset seen here. We therefore believe the offset is primarily caused by different assumptions on the prior of non-parametric SFHs. Although the magnitude of the offset is small (within 0.5$\sigma$) compared to the dispersion of the relation (next paragraph), nevertheless, it immediately shows the potentially significant systematics introduced by priors for the similar studies, and also highlights the importance in understanding the imprint of the priors on the reconstructed SFHs.

We also measure the percentile trends between \zf and $M_*$. We adopt the non-parametric quantile regression method, in which case no arbitrarily defined bins are required. We use the COnstrained B-Splines ({\sc cobs}, see \citealt{Ng2007,Ng2020} for details) package in R where the total number of knots required for the regressions B-spline method is determined using the Akaike-type information criterion. The resulting median (black solid line), 31th-69th percentiles ($\approx0.5\sigma$, dark grey shaded region) and 16th-84th percentiles trends ($1\sigma$, light grey shaded region) are plotted in Figure \ref{fig:mass_zp}. We see that the measurements of \citet{Tacchella2021} fall well within the 0.5$\sigma$ range of our measurements. Compared with the best-fit linear relation, a clear uptick is seen at the high-mass end ($>10^{11.5}M_\sun$). Although the significance of this uptick suffers from small number statistics, \citet{Tacchella2021} indeed also found very high \zf for their $>10^{11.5}M_\sun$ galaxies (see their Figure 11). 

Finally, while in the discussions above only \zf is considered, in Appendix \ref{app:results_w_zp} we also show the relationships of $M_*$ with other characteristic redshifts, including $z_{50}$, $z_{70}$ and $z_{90}$. Very similar trends, namely that more massive QGs formed earlier, are observed, confirming the findings above when we use \zf.

\subsection{The Size Evolution of QGs and Their Assembly History} \label{sec:re_zform}

\begin{figure*}
    \centering
    \includegraphics[width=1\textwidth]{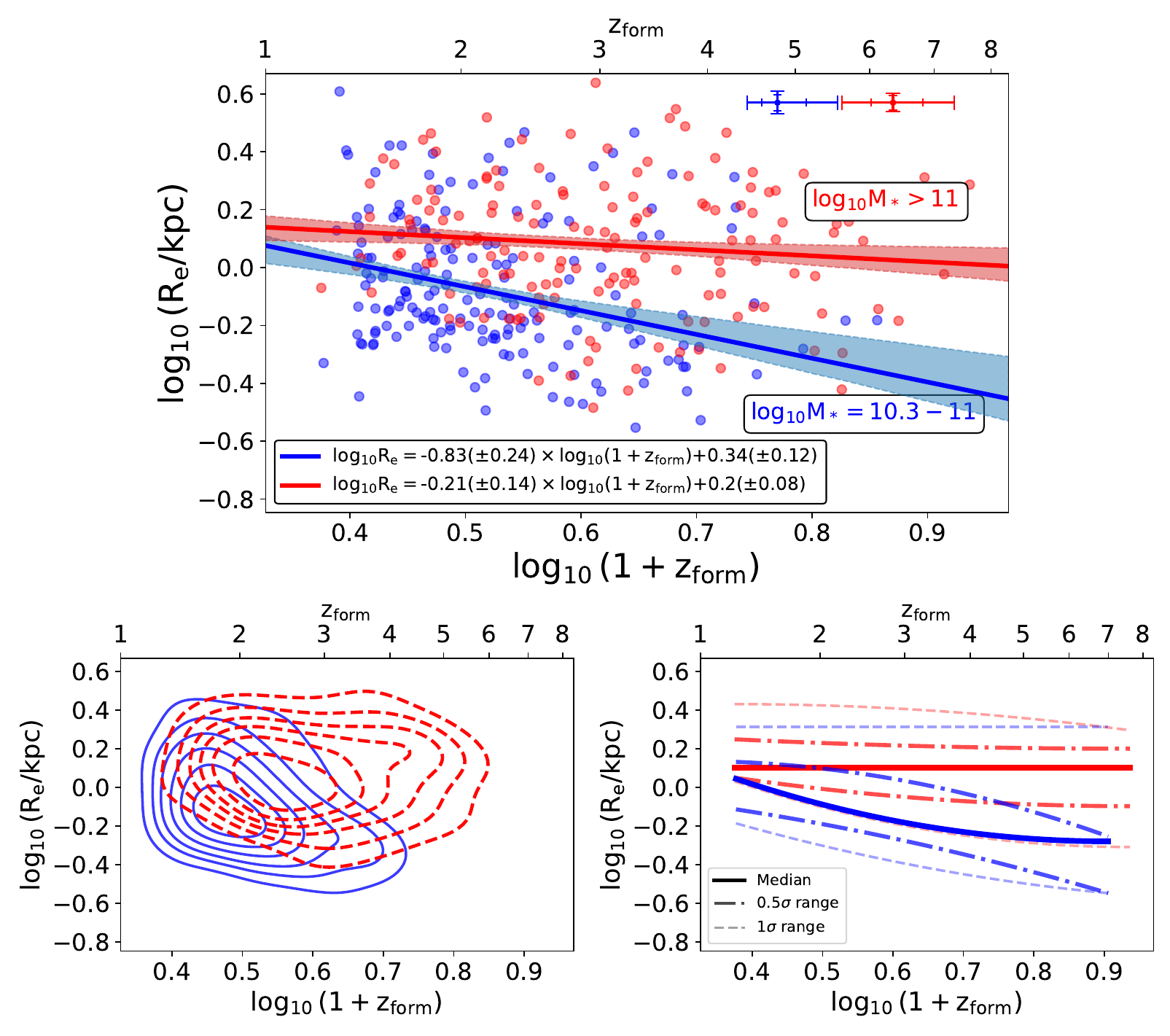}
    \caption{The relationship between \re and \zf. {\bf Top:} The dots represent individual QGs with $10^{10.3}M_\sun<M_*<10^{11}M_\sun$ (lower-mass, shown in blue) and $M_*>10^{11}M_\sun$ (higher-mass, shown in red). The typical uncertainty of individual measurements is plotted in the upper-right corner, where the inner and outer error bars show the median and 84-th percentile of the individual uncertainties, respectively. The thick solid lines show the least squares linear regressions, and the light shaded regions show the 1$\sigma$ uncertainties. The best-fit relations and their uncertainties are labelled in the figure legend. {\bf Bottom left:} The density distributions for the lower-mass and higher-mass QGs estimated using a Gaussian kernel. {\bf Bottom right:} The median, 0.5$\sigma$ and 1$\sigma$ quantile trends derived using the {\sc Cobs}. } 
    \label{fig:size_zform}
\end{figure*}

In this section we study the relationship between the size evolution of QGs and their assembly history. In Figure \ref{fig:size_zform}, the effective radius \re is plotted against \zf. A linear fit is conducted in the logarithmic space, i.e.
\begin{equation}
    \log_{10} R_{\rm{e}} = -\beta \log_{10}(1+z_{\rm{form}})+C.
\end{equation}
The uncertainty of the best-fit relation is derived in the same way as we did in section \ref{sec:diverse} for the linear relation between \tage and $M_*$.

The slope of the relation contains key information about the physics of galaxy size evolution. One should note that the Virial radius $R_{\rm{vir}}$ of a dark matter halo is set by the Virial density which by construction is the density contrast relative to the mean density of the Universe at the formation epoch of the halo and that it has a very weak time dependence, i.e. is essentially constant. This means that at fixed mass the size of halos formed at different redshifts should scale following the expansion of the Universe, i.e. $R_{\rm{vir}}\propto(1+z_{\rm{form}}^{\rm{h}})^{-1}$ \citep{Mo2010}. Note that to differentiate from \zf of a galaxy we use $z_{\rm{form}}^{\rm{h}}$ as the formation redshift of a halo\footnote{This corresponds to the time of halo collapse, which is sometimes also referred to as the halo virialization time.}. If the ratio between the galaxy \re and the halo $R_{\rm{vir}}$ is roughly constant, which in fact is supported both by theories \cite[e.g.][]{Mo1998} and by observations \citep[e.g.][]{Kravtsov2013,Somerville2018}, then \re should also scale as $(1+z_{\rm{form}}^{\rm{h}})^{-1}$. However, because galaxy morphology is only known at \zobs, one key condition to make the above argument stand is that galaxies \textit{cannot} have significant size growths after they formed. Or, in other words, the morphological properties measured at \zobs still need to have the memory of the time when galaxies formed. If the observed slope of the relation were to significantly deviate from $-1$, this would be strong empirical evidence of late-time size growth of QGs.

\begin{figure}
    \centering
    \includegraphics[width=0.47\textwidth]{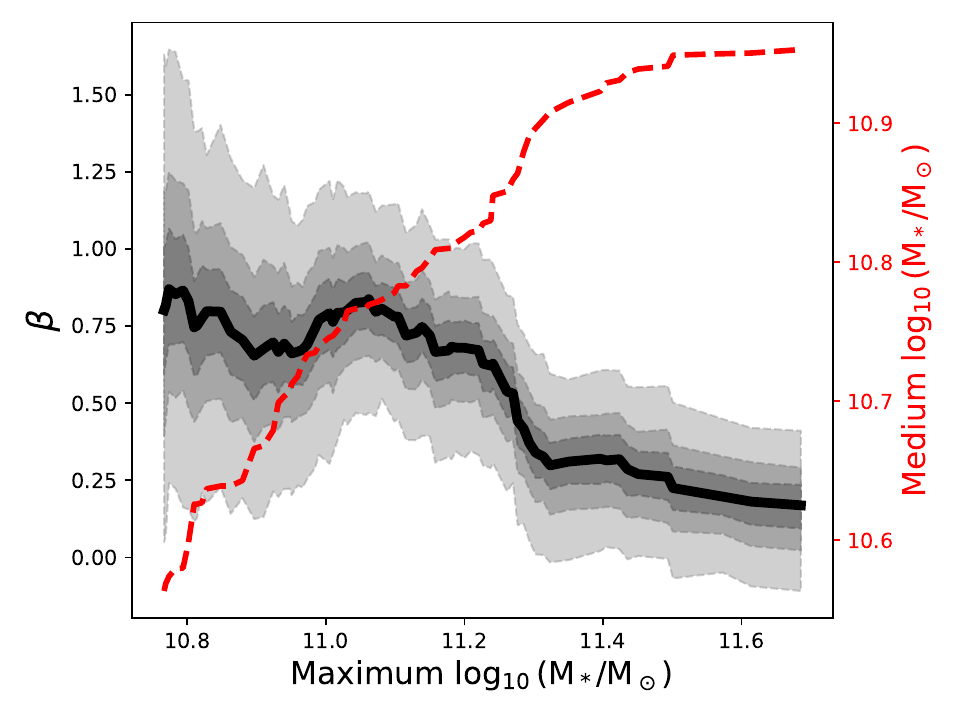}
    \caption{The change of the best-fit $\beta$ as more massive QGs are included to the fit (black solid line). The x-axis shows the maximum $M_*$ of the QGs included to the fit (see section \ref{sec:re_zform} for details). The grey shaded regions mark the 0.5, 1 and 2$\sigma$ uncertainties. The red dashed line shows the medium $M_*$ of the QGs included to the fit.}
    \label{fig:beta}
\end{figure}

The dependence on $M_*$ is found in Figure \ref{fig:size_zform}, which shows that the evolution of \re with \zf becomes flatter for more massive QGs. In other words, more extended galaxies, i.e. those with larger $R_e$, of smaller mass generally formed later than more massive ones with similar $R_e$. This is also seen through the kernel density estimations (the bottom left panel) and the non-parametric quantile trends (the bottom right panel). As it will be discussed in section \ref{sec:galfit_stack}, this finding is further confirmed by our stacking analysis. For illustration, the entire QG sample is only divided into two $M_*$ bins in Figure \ref{fig:size_zform}. We have checked the result by dividing the sample into more $M_*$ bins. In particular, considering the relatively large scatter of the \re-\zf relation, instead of binning the QGs using arbitrary $M_*$ bins, we first sort $M_*$ of individual QGs into an increasing order. Then, starting from the first 30\% of the sorted sample, we keep adding more massive QGs into the linear fit and study the change of $\beta$. The uncertainty of $\beta$ is derived in the same way as described before. We plot in Figure \ref{fig:beta} the best-fit $\beta$ against the maximum $M_*$ of the QGs included to the fit. The $\beta$ generally decreases with $M_*$.  While $\beta$ is consistent with $\approx 1$ for the QGs with $\log_{10}(M_*/M_\sun)\lesssim 11$, i.e. $R_e\propto(1+z_{\rm{form}})^{-1}$, it quickly decreases to $\sim0.2$ as more massive QGs are included to the fit. Using a different approach, similar results were obtained by \citet{Fagioli2016}, although at a somewhat lower redshift $0.2<z<0.8$. They stacked the rest-frame optical spectra for a sample of $\approx$ twenty thousand QGs and found that the relationship between stellar age and size depends on $M_*$. Specifically, at fixed \zobs, on the one hand, among QGs in the mass range $10.5< \log_{10}(M_*/M_\sun) < 11$, those with larger sizes also tend to be younger (i.e. smaller \zf). On the other hand, for QGs with mass in the range $\log_{10}(M_*/M_\sun)>11$, no clear trend between stellar age and size is observed. Similar conclusions have also been reached by \citet{Carollo2013} who found the $M_*$ dependence of the evolution of the number density of QGs {\it of a given size} over $0.2<z<1$ in the COSMOS field.

Before we proceed to discuss the physical implications of the observed relationship between \re and \zf, we caution about two possibly important systematics. First, although the dependence on $M_*$ still remains, we find that the best-fit slope becomes flattened if we use the Continuity prior to reconstruct the non-parametric SFHs. Specifically, for the lower-mass ($\log_{10}M_*=10.3-11$) QGs $\beta$ changes from $\approx0.8\pm0.2$ (Dirichlet) to $\approx0.5\pm0.2$ (Continuity), and for the higher-mass ($\log_{10}M_*>11$) QGs it changes from $\approx0.2\pm0.1$ to $\approx0.1\pm0.1$. Second, the slope of the relation depends on which definition of formation redshift is adopted. Because the goal is to use SFHs to separate the progenitor effect from the late-time growth of galaxies, as we already discussed in the beginning of this section, the ideal choice then is $z_{\rm{form}}^{\rm{h}}$ which unfortunately is unknown. Using \zf thus is purely from an empirical stand point. Other proxies for $z_{\rm{form}}^{\rm{h}}$ are also considered, including the redshift $z_{\rm{form}}^{-0.5\tau_{\rm{tot}}}$ of lookback time (\tage$-$ \tautot/2), and the redshift $z_{\rm{form}}^{-\tau_{\rm{SF}}}$ of lookback time (\tage$-$ \tausf). No substantial changes to the best-fit relations are found. Quantitatively, the best-fit slope for the lower-mass QGs is $\beta=0.7\pm0.2$ ($0.6\pm0.2$) if $z_{\rm{form}}^{-\tau_{\rm{SF}}}$ ($z_{\rm{form}}^{-0.5\tau_{\rm{tot}}}$) is used. For the higher-mass QGs, the best-fit slope is $\beta=0.2\pm0.1$ ($0.1\pm0.1$) if $z_{\rm{form}}^{-\tau_{\rm{SF}}}$ ($z_{\rm{form}}^{-0.5\tau_{\rm{tot}}}$) is used. 

We have also attempted to empirically evaluate the performances of individual proxies for $z_{\rm{form}}^{\rm{h}}$. In particular, assuming that the central regions of galaxies are much less affected by processes such as mergers relative to their outskirts, which in fact is supported by simulations of galaxy mergers \citep[e.g. see Figure 3 in][]{Hilz2013}, we can compare the distributions of \Sone for galaxy samples selected to have formed at the same epoch, using different proxies for the formation redshift. If one proxy is better than others in separating the progenitor effect, then the galaxy sample selected using that proxy should have a narrower distribution of \Sone. Perhaps because of the combination of relatively small sample size and large scatter (both intrinsic and introduced by our measures), however, we do not find any statistically significant difference among different proxies. This needs to be further investigated with larger samples in the future.

\begin{figure}
    \centering
    \includegraphics[width=0.47\textwidth]{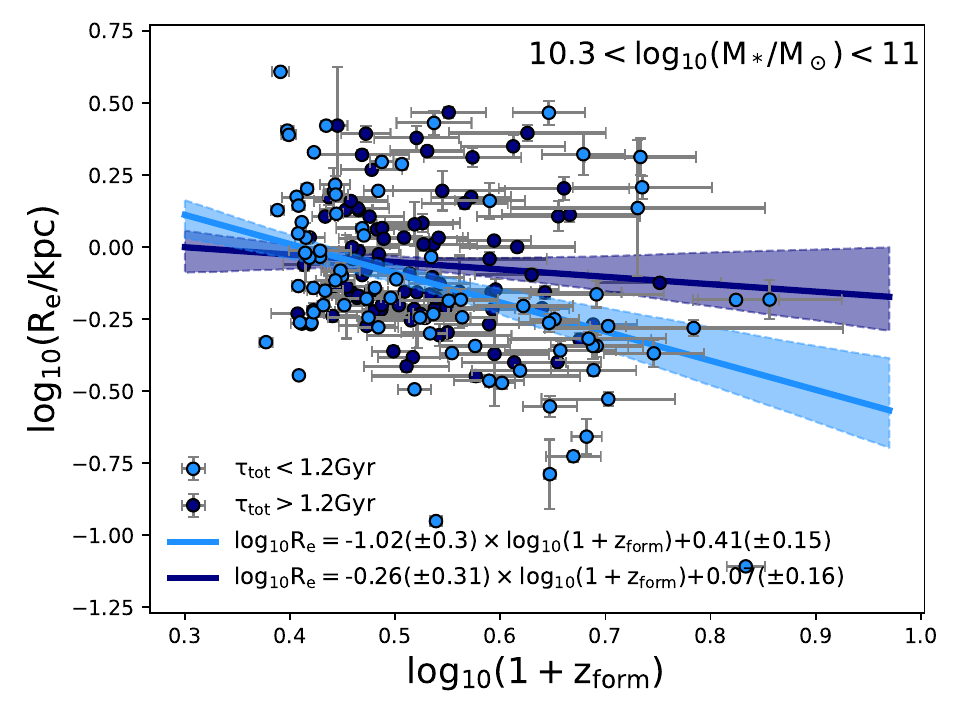}
    \caption{The relationship between \re and \zf for the sample of lower-mass QGs. The galaxies are binned using the median value of \tautot. The solid lines show the least squares linear regressions, and the light shaded regions show the 1$\sigma$ uncertainties. Best-fit relations and uncertainties are labelled in the figure legend.}
    \label{fig:size_zform_tau}
\end{figure}

Setting aside the aforementioned systematics, for the lower-mass QGs the slope of the \re-\zf relation is always found to be steeper and closer to the value $\beta=1$ compared to the higher-mass QGs, regardless of which non-parametric SFH prior and proxy for formation redshift are used. This suggests that the progenitor effect becomes increasingly important in driving the apparent size evolution of QGs as they become less massive. In addition to \zf, other characteristics of SFHs can also affect the apparent size evolution. As we already showed in section \ref{sec:diverse}, galaxies can have very similar \zf, yet, the timescale to assemble most of their mass can be dramatically different (Figure \ref{fig:zf_tau_skew}). In Figure \ref{fig:size_zform_tau}, we further divide the lower-mass QG sample into two subsamples using \tautot. The relatively small sample size and large scatter notwithstanding, two interesting things are noticed. Compared to the lower-mass QGs with larger \tautot, we find likely evidence that those with smaller \tautot have a steeper evolution of \re with \zf. Meanwhile, at fixed \zf, they also seemingly to have smaller \re. These suggest that the progenitor effect is more prominently observed in the lower-mass QGs with a more rapid and monolithic assembly history, as expected. 

For the higher-mass QGs, the relationship between \re and \zf is in general nearly flat. A natural interpretation would be the increasing role that mechanisms of post-quenching size and mass growths play, i.e. from right after quenching to the time of observations. Note that this time lapse also tends to be larger for more massive galaxies, since these systems complete their formation and quenching at an earlier time (see Figure \ref{fig:mass_zp}). One very plausible mechanism for such post-quenching growth is galaxy mergers, and minor mergers in particular. Assuming energy conservation of a merging system, as the distribution of orbital parameters in hydrodynamical simulations suggests that most merging halos are on parabolic orbits \citep{Khochfar2006}, and using the Virial theorem, \citealt{Naab2009} showed that minor mergers are much more efficient in growing galaxy sizes compared to major mergers, given a fixed amount of mass to be added to the central galaxy. Because the merger rate is expected to increase with stellar mass and redshift \citep[e.g.][]{Hopkins2010,Rodriguez2015,OLeary2021}, galaxy mergers can effectively grow galaxy sizes by adding mass to the outskirts of compact QGs that formed relatively early. This scenario is also supported by the continuity analysis of the galaxy stellar-mass function, as discussed by \citet{Peng2010}, who argued that QGs with $M_*>10^{11}M_\sun$ should have significant growths in mass after the quenching, whilst a similar situation is very unlikely for lower-mass QGs. As we will discuss in section \ref{sec:re_zform_stacking}, our stacking analysis reaches similar conclusions, namely that the late-time size growth of very massive QGs is very likely driven by merging and accretion of small galaxies primarily to their outskirts. Finally, we have also verified that if we further divide the higher-mass QGs into subsamples using their \tautot we do not find statistically significant differences among the \re-\zf relationship as a function of \tautot. This again suggests that, for these higher-mass QGs, by the time of observation, \zobs, post-quenching size and mass growths have erased the memory of the structural properties that these massive galaxies had at \zf, i.e. the time when they formed.

\subsection{The Morphology of QGs in the Stacked \hst Images}\label{sec:re_zform_stacking}

In this section, we further study the morphology of QGs by stacking their multi-band \hst images. We pay particular attention to the low surface brightness flux which is too faint to be studied for individual QGs. \hst bands included to the analysis are ACS/\I, WFC3/\J and WFC3/\H, because they are available in all three fields simultaneously. Motivated by the $M_*$ dependence of the relationship between \re and \zf (Figure \ref{fig:size_zform}), the QGs are first divided into a lower-mass sample with $10^{10.3}M_\sun<M_*<10^{11}M_\sun$ and a higher-mass sample with $M_*>10^{11}M_\sun$. Each sample is then divided into three subsamples using the 3-quantiles of \zf. These add up to 6 subsamples in total. 

Before stacking, for each QG we first center on its \sextractor-derived light centroid in the \H band to make 10''$\times$10'' cutouts of the \hst images and the corresponding sigma maps. Each image is then rotated using the position angle in the \H band from \citet{vanderwel2012}, which is measured by fitting the light profile with a single-\sersic profile using \galfit. After this, the semi-major axis of all QGs is aligned along the North direction. We also use the segmentation map of the \H band provided by the CANDELS team to mask out any intervening objects detected in the cutouts. Finally, we stack the reprocessed images with inverse-variance weights.

\subsubsection{Surface brightness profiles of the stacked QGs}\label{sec:SB_stack}

Here we compare the multi-band surface brightness profiles of the stacked QGs. All stacked images are first PSF-matched to that of the \H band. Surface brightness profiles are then measured in the PSF-matched images using the elliptical annular  apertures whose axis ratio (b/a) is fixed to the best-fit value in the \H band measured by \galfit assuming a single-\sersic profile. 

Two types of uncertainties are taken into account for the measurements. The first comes from sample random errors, which can be particularly important for the stacking analysis because sometimes a small number of galaxies can dominate the stacked signal. To quantify this uncertainty, each subsample has been bootstrapped 100 times. The other type of uncertainty comes from image noise. During each bootstrapping we thus use the corresponding stacked sigma map to Monte Carlo resample the pixel values of the stacked image 30 times with a normal distribution. We repeat the surface brightness measurements 100$\times$30$=$3000 times for each subsample and for each \hst band. Finally, we use the range between 16th and 84th percentiles as 1$\sigma$ uncertainties of the surface brightness profiles.

\begin{figure*}
    \includegraphics[width=0.327\textwidth]{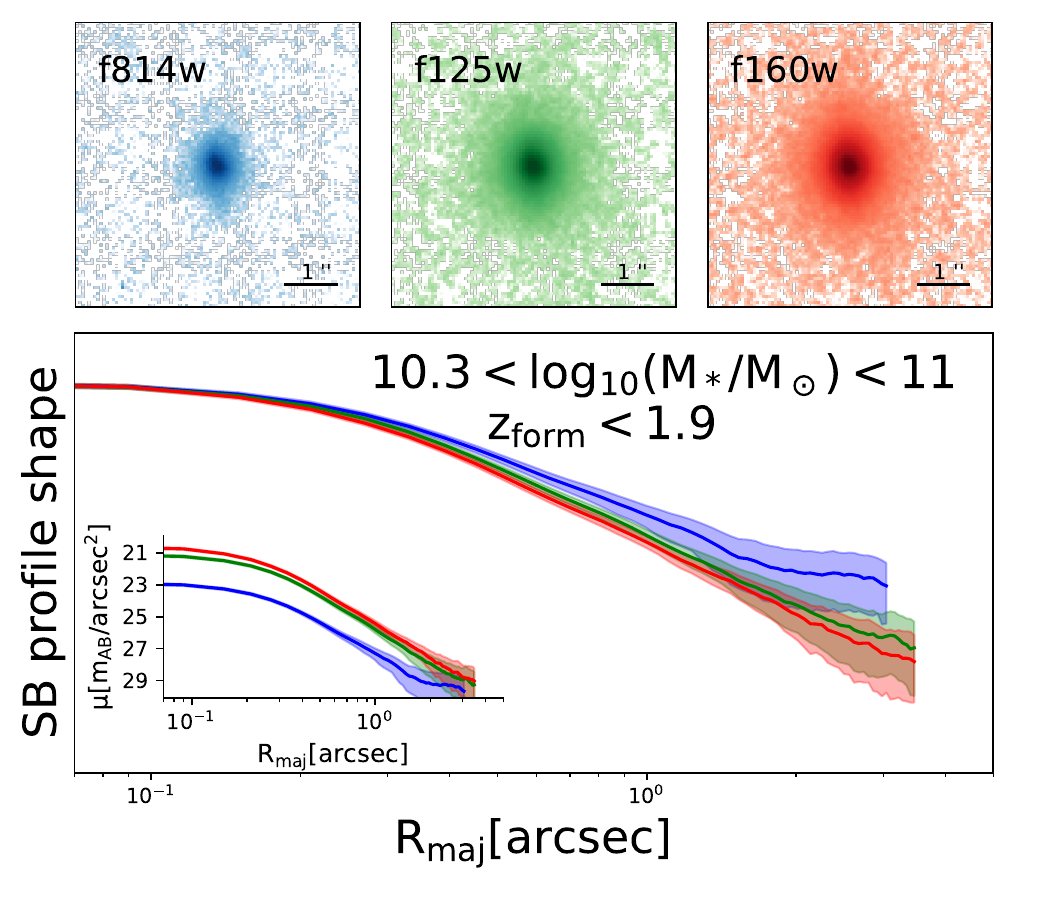}
    \includegraphics[width=0.327\textwidth]{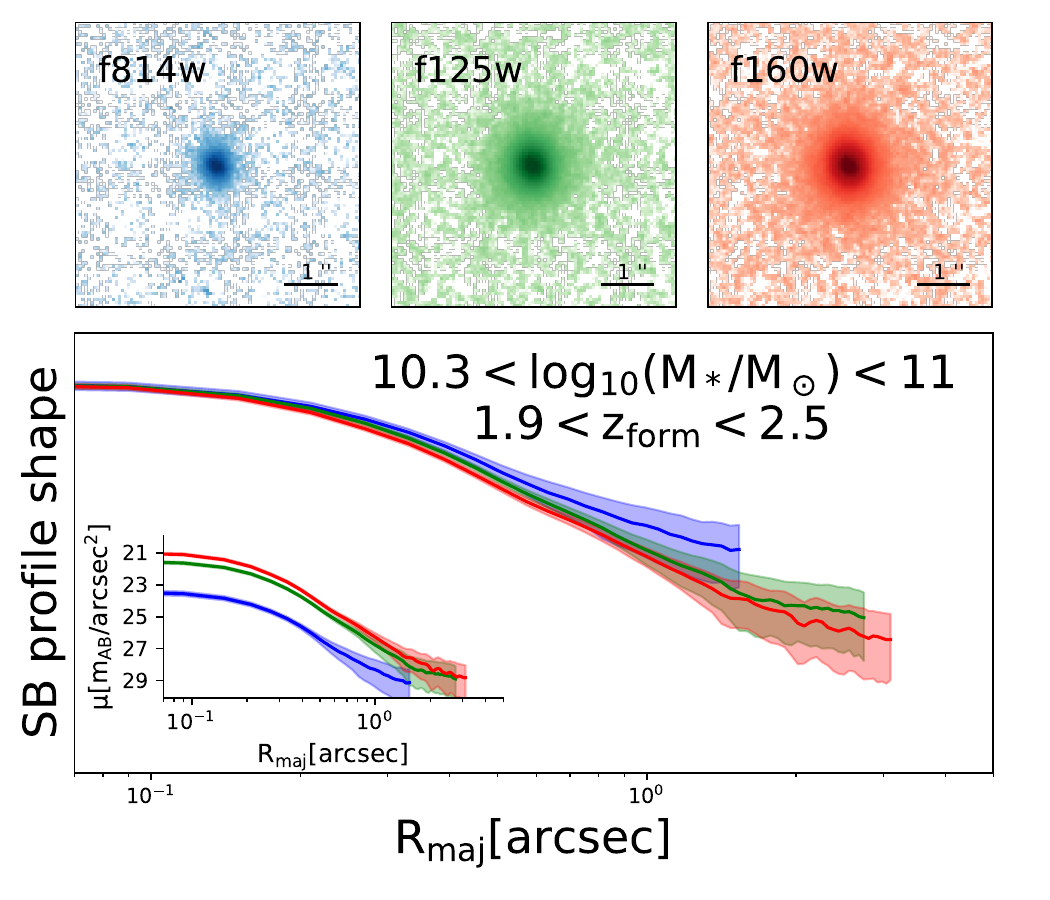}
    \includegraphics[width=0.327\textwidth]{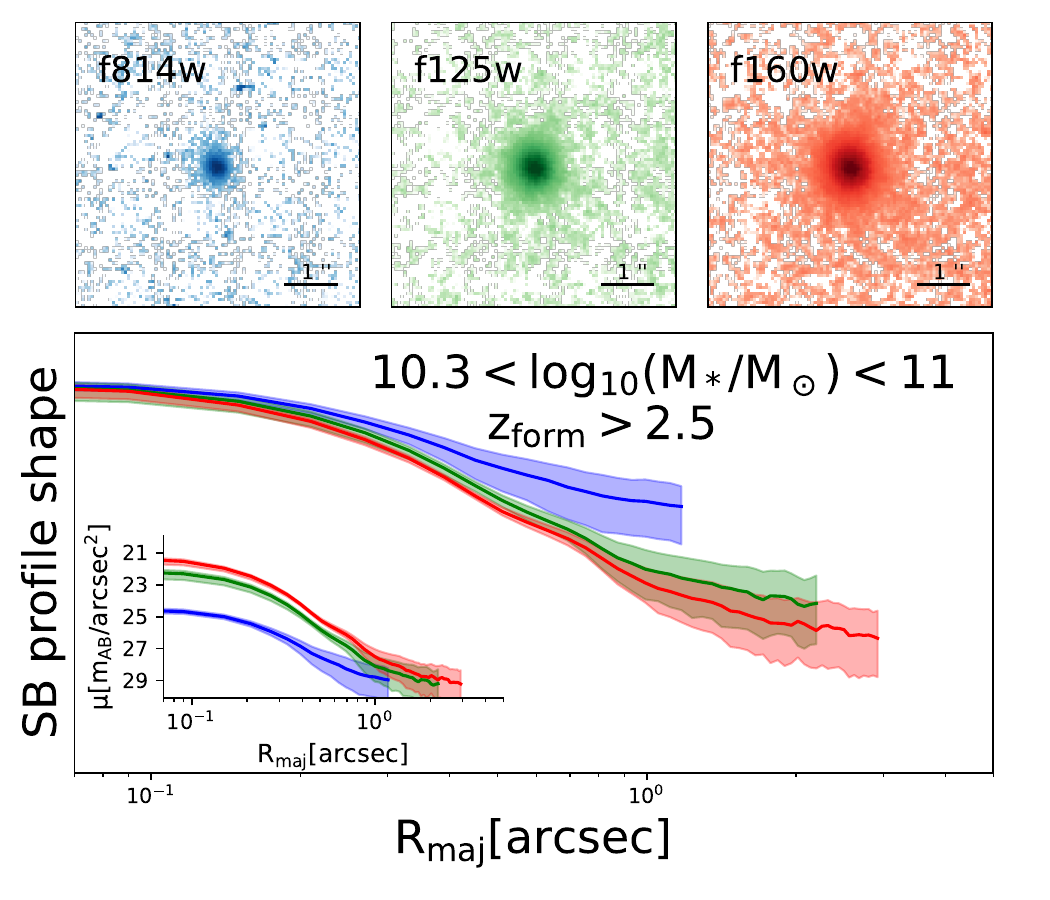}
    \includegraphics[width=0.327\textwidth]{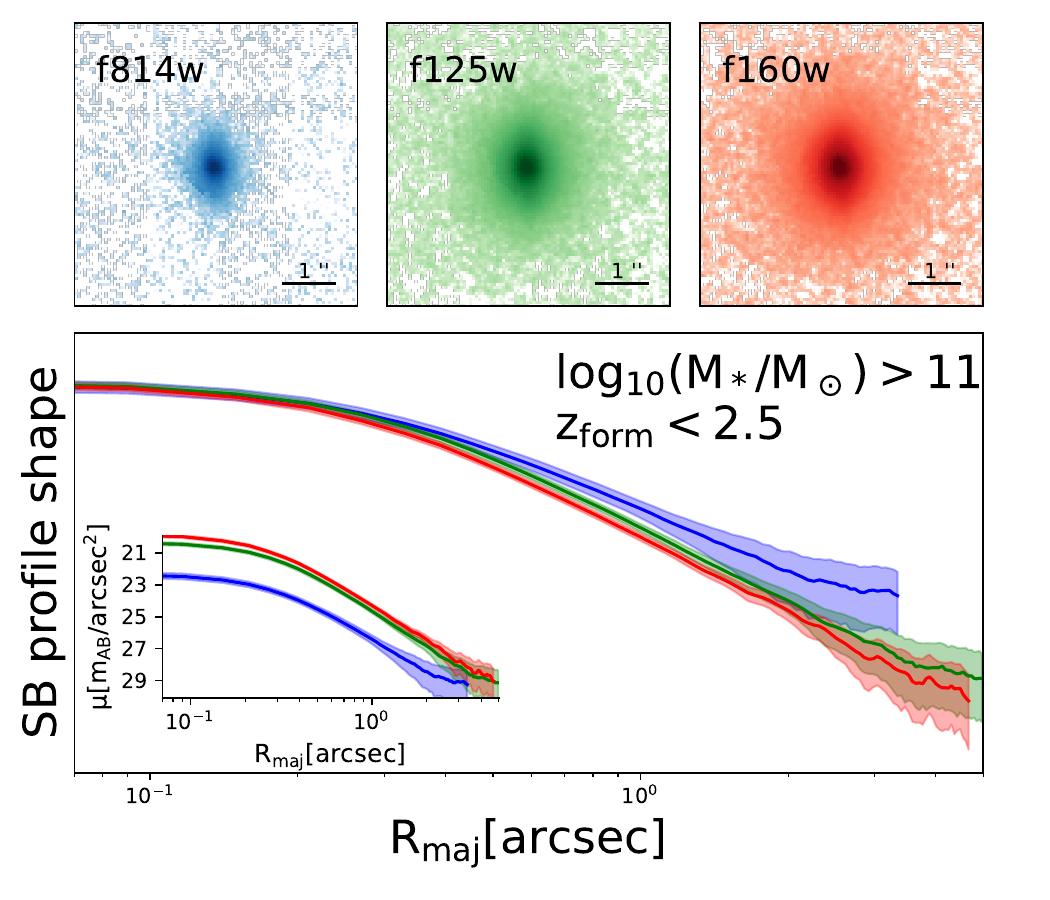}
    \includegraphics[width=0.327\textwidth]{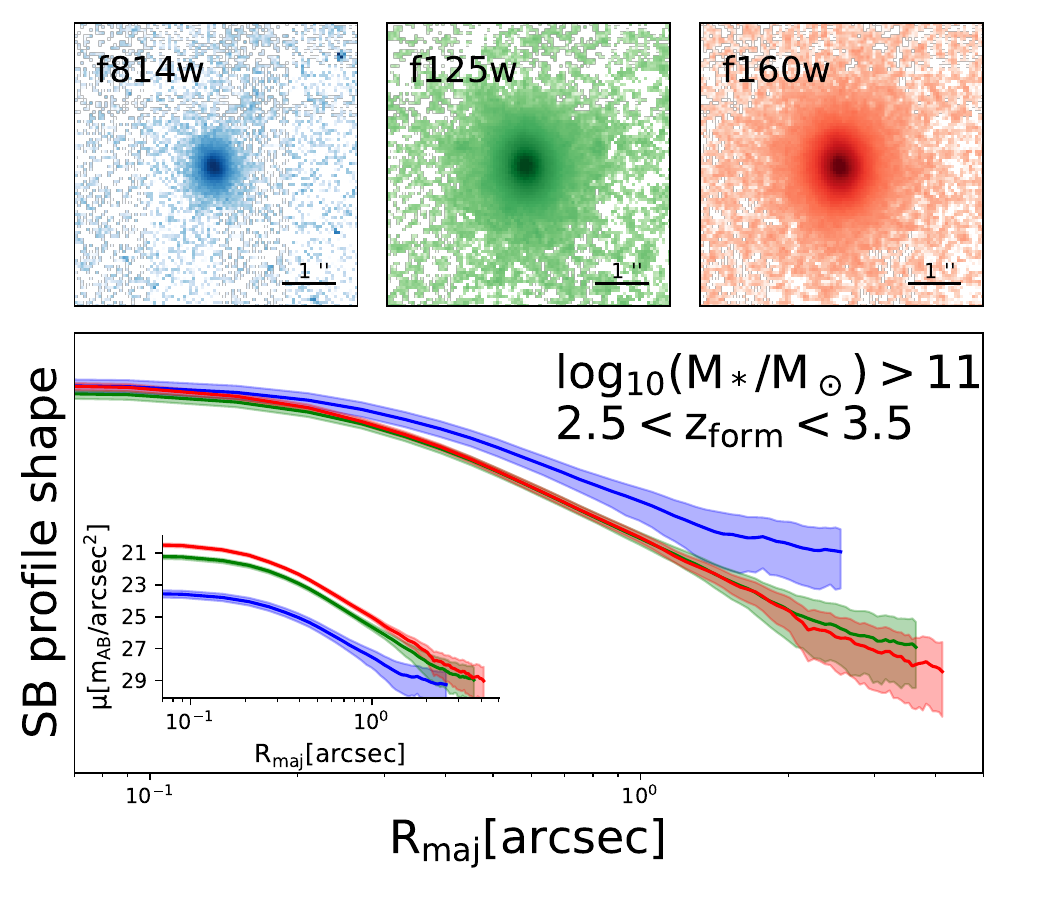}
    \includegraphics[width=0.327\textwidth]{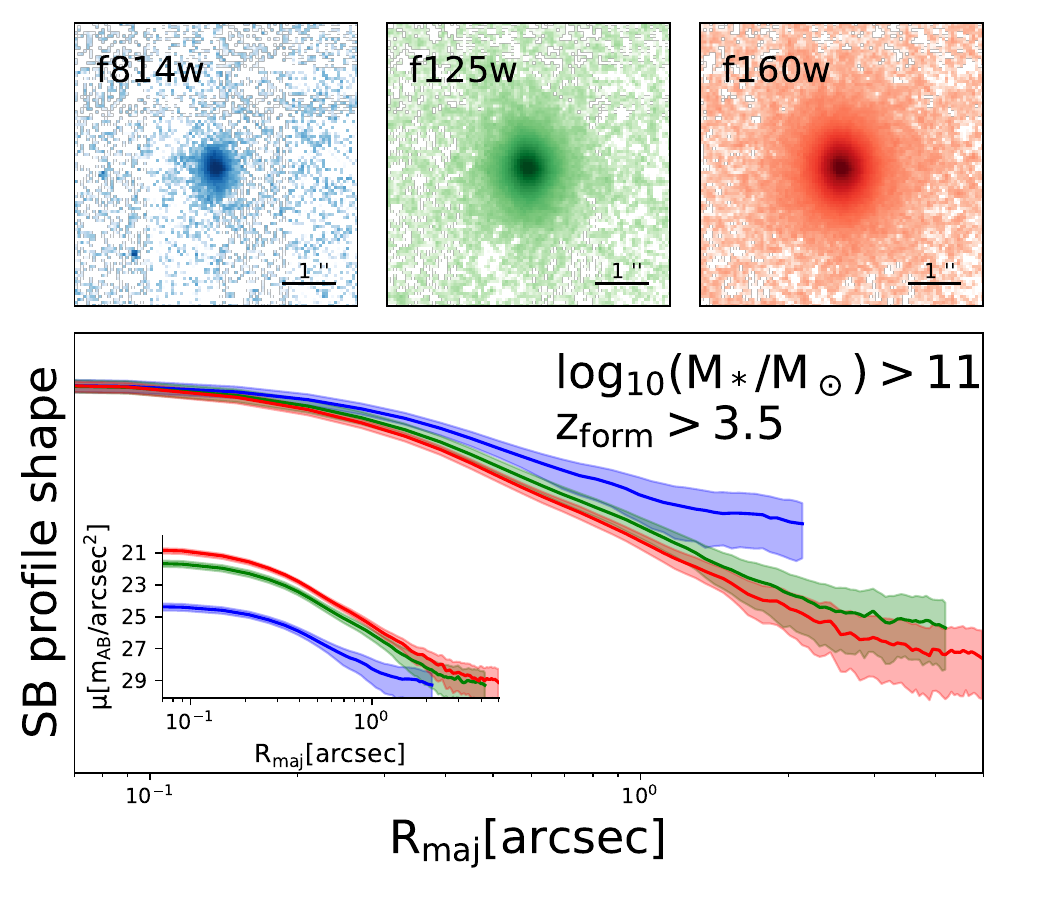}
    \centering
    \caption{Surface brightness profile analysis for the stacked \hst images of the six QG subsamples. In the panel of individual subsamples, the subpanels on the top show the three-band stacked \hst images (PSF-unmatched). The {\it shape} of surface brightness profiles (in logarithmic scale) measured in the PSF-matched images are shown in the bottom subpanel, while the inset of it shows the surface brightness profiles in the physical unit of $\rm{mag/arcsec^2}$. The shaded region marks the 1$\sigma$ range of a measurement. Every surface brightness measurement stops at the radius corresponding to the 1$\sigma$ detection limit of the stacked image. } 
    \label{fig:SB_stacking}
\end{figure*}

In Figure \ref{fig:SB_stacking} we show the stacked multi-band \hst images and surface brightness profiles of every subsample. The stacking pushes the 1$\sigma$ detection limit\footnote{Note that the sensitivities are derived using the original, i.e. PSF-unmatched, stacked images.} of surface brightness to $\approx29$ $\rm{mag/arcsec^2}$ in all three bands. For all subsamples the surface brightness in the \I band is much fainter than that in the \J and \H bands (see the inset of each panel), illustrating once again the effectiveness of the UVJ selection criteria in culling galaxies with very low levels of on-going star formation activity. We continue to compare the {\it shape} of surface brightness profiles by normalizing each profile with the central surface brightness. While the profile shapes of the \J and \H bands are very similar, we see evidence that the light in the \I band is more extended. Although the relatively small sample size prevents us putting the statistics on a solid ground at the moment, based on the uncertainties derived from our bootstrapping and Monte Carlo resampling, the finding that the light in the \I band is more extended than that in the \J and \H bands seems to be more prominent for the higher-mass QGs, particularly in the outer regions with $R_{maj}>0.5$ arcsec. This finding will soon be reinforced in section \ref{sec:galfit_stack} via our single-\sersic fitting analysis of the stacked images.

\subsubsection{Single-\sersic fitting results of the stacked QGs} \label{sec:galfit_stack}

\begin{table*}[]
    \centering
    \caption{Single-\sersic fitting results for the stacked \hst images of individual subsamples.}
    \begin{tabular}{||c|c|c|c|c|c|c|c|c|c||}
    \hline
     $\rm{M_*/M_\sun}$ & $\rm{z_{form}}$ & $\rm{\#^{(a)}}$ & $\rm{\langle z_{obs}\rangle^{(b)}}$ & $\rm{{R_e^{F814W}}^{(c)}}$ & $\rm{n^{F814W}}$  & $\rm{{R_e^{F125W}}^{(c)}}$ & $\rm{n^{F125W}}$ & $\rm{{R_e^{F160W}}^{(c)}}$ & $\rm{n^{F160W}}$ \\
    \hline
      \multirow{3}{*}{$10^{10.3}\sim10^{11}$} & $<1.9$        & 51 & 1.3 & 1.32$\pm$0.19 & 1.64$\pm$0.38 & 1.27$\pm$0.09 & 2.05$\pm$0.24 & 1.28$\pm$0.09 & 1.93$\pm$0.22  \\
                                    & $1.9\sim2.5$ & 65 & 1.5 & 1.14$\pm$0.11 & 1.28$\pm$0.26 & 1.13$\pm$0.06 & 1.41$\pm$0.19 & 1.16$\pm$0.06 & 1.42$\pm$0.16 \\
                                    & $>2.5$        & 65 & 2.0 & 1.11$\pm$0.83 & 1.54$\pm$1.04 & 0.80$\pm$0.15 & 1.81$\pm$0.37 & 0.81$\pm$0.13 & 1.82$\pm$0.32 \\
    \hline
       \multirow{3}{*}{$>10^{11}$} & $<2.5$        & 47 & 1.4 & 1.59$\pm$0.12 & 2.11$\pm$0.31 & 1.38$\pm$0.08 & 2.13$\pm$0.22 & 1.39$\pm$0.06 & 2.03$\pm$0.18 \\
                              & $2.5\sim3.5$ & 53 & 1.7 & 1.64$\pm$0.17 & 1.53$\pm$0.27 & 1.31$\pm$0.12 & 1.79$\pm$0.22 & 1.30$\pm$0.10 & 1.81$\pm$0.20 \\
                              & $>3.5$        & 53 & 2.2 & 1.75$\pm$0.39 & 1.54$\pm$0.49 & 1.51$\pm$0.17 & 2.06$\pm$0.38 & 1.40$\pm$0.10 & 1.96$\pm$0.23 \\
    \hline
    \end{tabular}
    \label{tab:stack}
    \footnotesize{ {\bf (a)--} Total number of galaxies in the subsample; {\bf (b)--} Median redshift of the subsample; {\bf (c)--} In the unit of kpc.}
\end{table*}

We now study the morphology of individual stacked QG subsamples using \galfit. During the analysis, we adopt the PSFs produced by the CANDELS team and fit each stacked image with a 2-D \sersic light profile, from which we obtain the \sersic index \n and \re.

\begin{figure*}
    \includegraphics[width=1\textwidth]{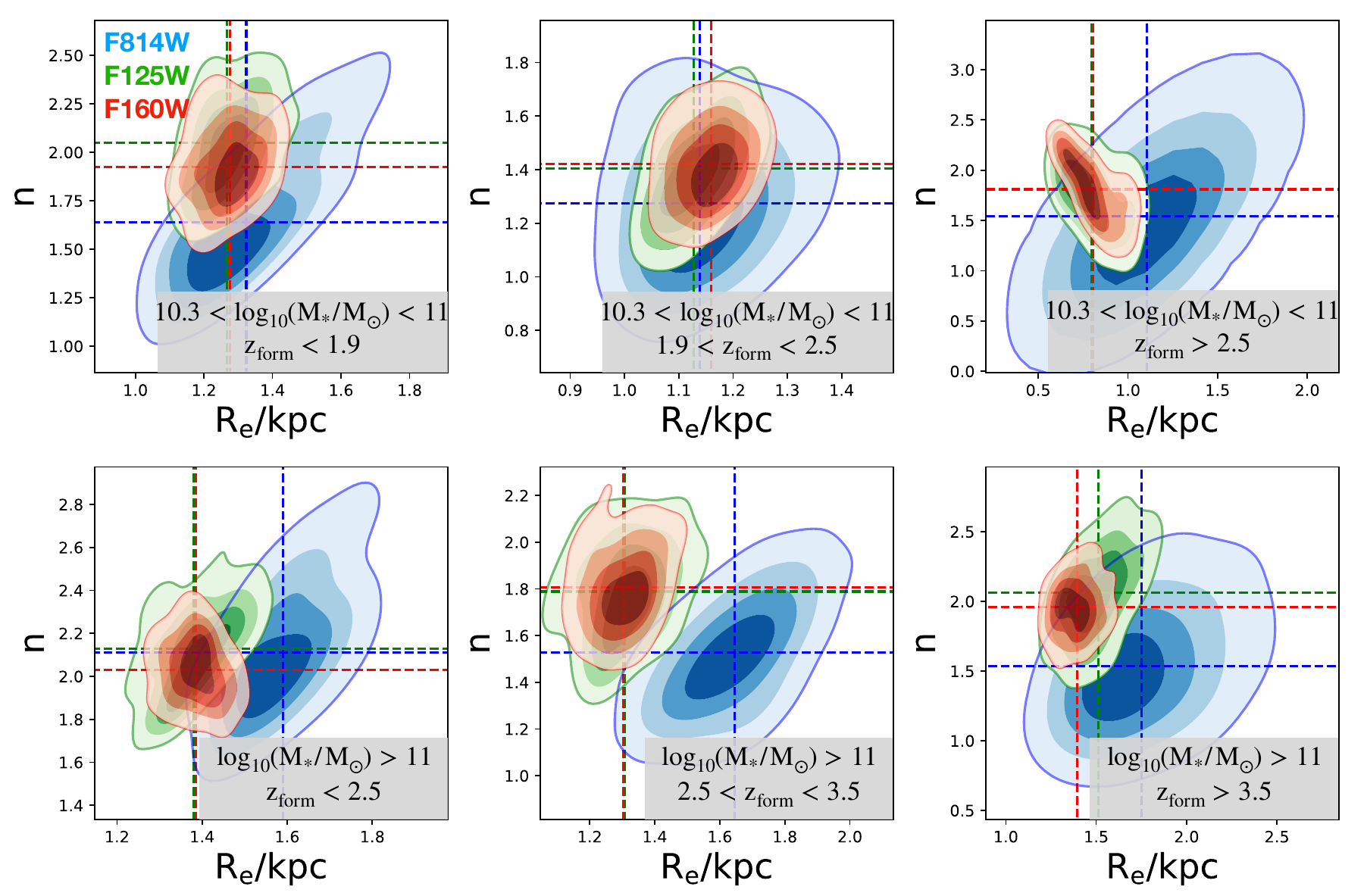}
    \centering
    \caption{Single \sersic fitting analysis with \galfit for the stacked \hst images of the six QG subsamples. The contours show the distributions of best-fit \re and \n derived from our Monte Carlo plus bootstrap resamplings (see section \ref{sec:galfit_stack} for details). The dashed lines mark the medians. }
    \label{fig:galfit_stacking}
\end{figure*}

One complication of this analysis is to properly estimate the uncertainty of the fitted parameters. Similarly to what has already been described in section \ref{sec:SB_stack}, both image noise and sample random errors are taken into account. We follow section \ref{sec:SB_stack} to estimate the uncertainty from the former by Monte Carlo resampling the image pixel values 30 times with the stacked sigma maps and the uncertainty from the latter by bootstrap resampling the stacked sample 100 times. We have compared the two uncertainties and found that sample random errors dominate over the uncertainty introduced by image noise. These add up to 3000 times \galfit fittings for each band and for each subsample. The \galfit single-\sersic fitting results are tabulated in Table \ref{tab:stack} and plotted in Figure \ref{fig:galfit_stacking}.

\begin{figure*}
    \centering
    \includegraphics[width=1\textwidth]{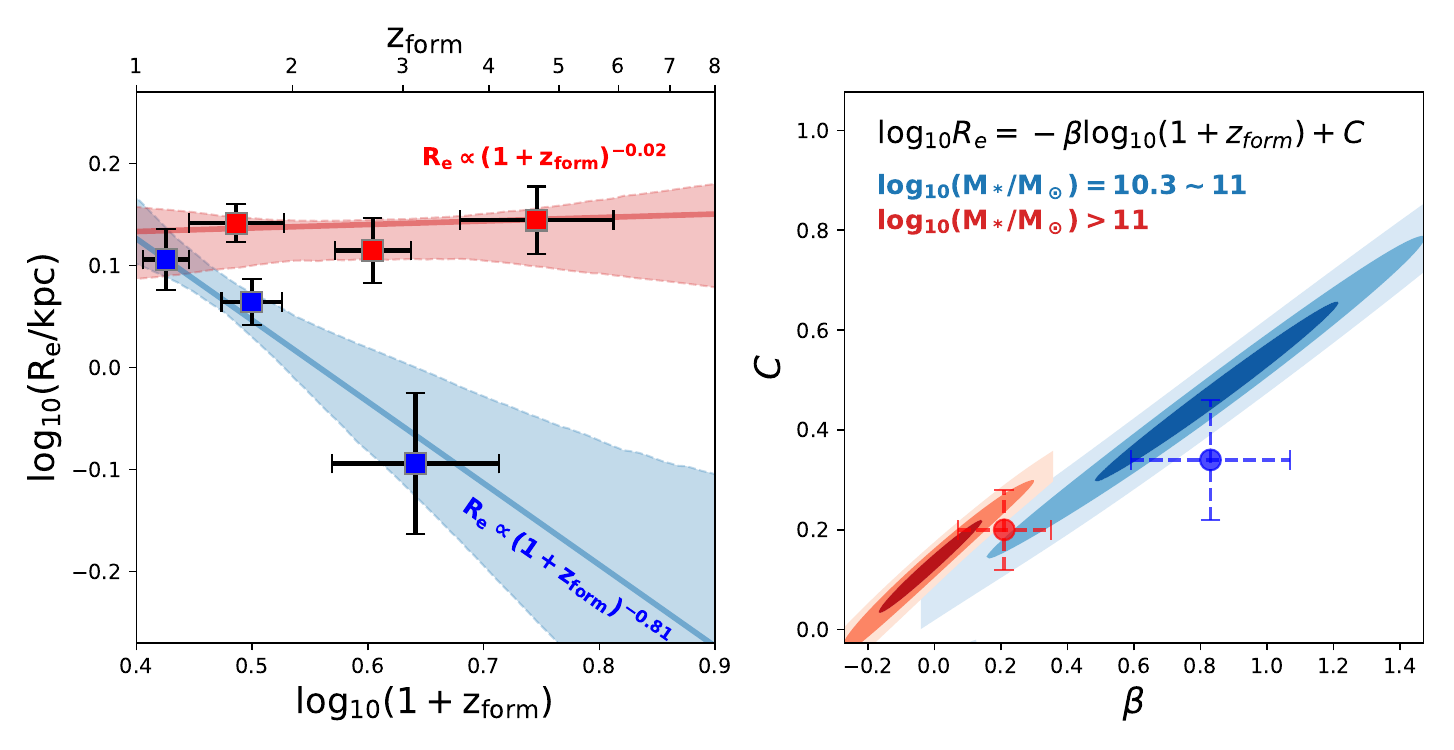}
    \caption{{\bf Left:} The relationship between \re and \zf derived from our stacking analysis (section \ref{sec:galfit_stack}). Blue shows the best-fit relation of the lower-mass QGs and red shows the best-fit relation of the higher-mass QGs. {\bf Right:} The contours show the 0.5, 1 and 2$\sigma$ parameter uncertainties of the best-fit \re-\zf relation. Also marked using the circle with dashed error bars are the best-fit values and 1$\sigma$ uncertainties derived from individual QGs (see section \ref{sec:re_zform} and Figure \ref{fig:size_zform}).}
    \label{fig:re_zform_stacking}
\end{figure*}

To begin, we study the relationship between \zf and \re (\H) derived from the stacking where we use the median \zobs of individual subsamples to convert angular scales to physical scales. We show in Figure \ref{fig:re_zform_stacking} the best-fit relations of the lower-mass (blue) and higher-mass (red) stacked QG samples. The stacking analysis confirms the $M_*$ dependence of the \re-\zf relationship as reported in section \ref{sec:re_zform} based on individual QGs. The relation becomes shallower as the QGs become more massive. Quantitatively speaking, the best-fit values of $\beta$ derived from the stacking analysis are in great agreement ($\sim0.5\sigma$ difference) with those derived from individual QGs. The best-fit intercept $C$ for the lower-mass QGs derived from the stacking seems to be larger than (though still within $\sim 1\sigma$ range) that derived from individual QGs (Figure \ref{fig:size_zform}). Interestingly, we find that this difference in $C$ disappears if we stack the images without pre-aligning the galaxies' semi-major axis. Also noticed is that in all subsamples the best-fit \sersic index $n$ is found to be $\sim1.5-2$, while it becomes to $\sim3-5$ when stacking without aligning the semi-major axis in advance. These suggest that the pre-alignment is critical in retrieving low surface brightness light on the outskirts. More importantly, these indicate that on average the morphology of high-redshift QGs, at least on their outskirts, is more disk-like rather than spheroidal, which is in agreement with recent morphological and dynamical studies of a few cases of $z\sim2$ QGs where strong magnification by gravitational lensing helps reaching sub-kpc resolutions in the source plane \citep{Newman2018}. We defer a detailed study on this finding and its implications on galaxy quenching to a forthcoming paper.

We finally compare the \sersic fitting results in the \H band with the other two bluer \hst bands, i.e. \I and \J (Figure \ref{fig:galfit_stacking}). While the morphologies of the lower-mass QGs in all three bands are very similar, the higher-mass QGs have a more extended morphology, namely having large \re, in the \I band than in the \J and \H bands. These are consistent with the analysis of the surface brightness profiles in section \ref{sec:SB_stack}. Together with the dependence of the relationship between \re and \zf on $M_*$, the findings here are consistent with the picture that very massive QGs after they formed keep growing their sizes via merging with satellite star-forming galaxies, while for the lower-mass QGs, which presumably formed in the regions of lower overdensities compared to the higher-mass QGs, experienced much less mergers and (hence) their morphological properties still have the memory of the time when they formed.

\subsection{The Relationship between the Compactness of QGs and Their Assembly History} \label{sec:zf_sigma}

We now study the relationships between the compactness of QGs and their SFH. While the results presented below are only about the relationships with \zf, in Appendix \ref{app:results_w_zp} we also show the results of using other characteristic \zp where we find no substantial changes in any of our findings.

\begin{figure}
    \centering
    \includegraphics[width=0.47\textwidth]{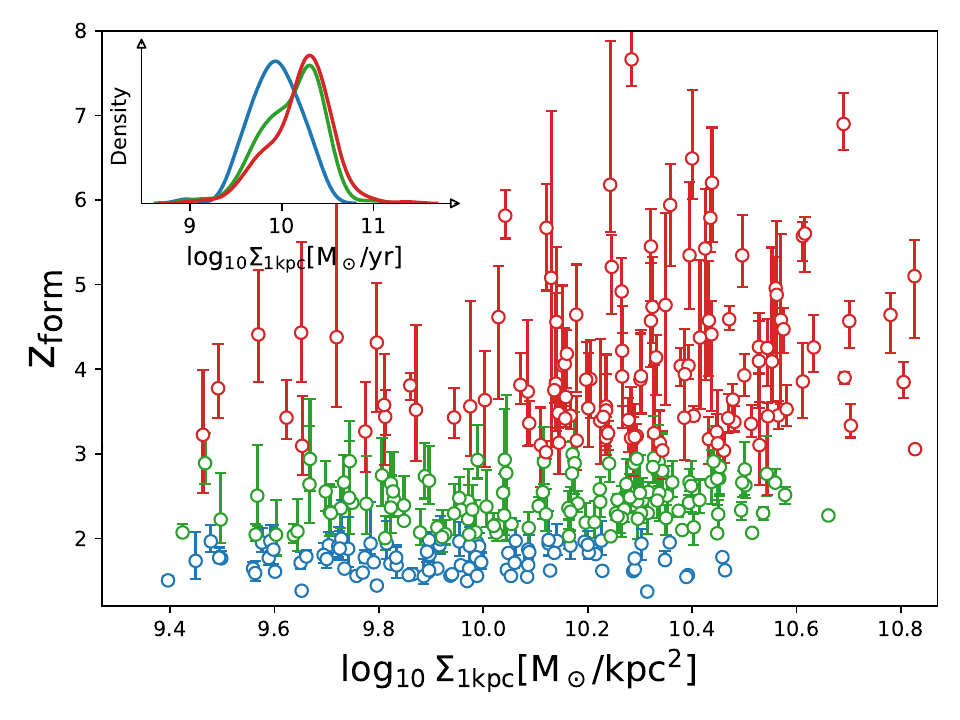}
    \caption{The relationship between \Sone and \zf. Individual QGs are color-coded following \zf$<2$ (blue), $2<$\zf$<3$ (green) and \zf$>3$ (red). The density distributions of \Sone estimated using a Gaussian kernel are plotted in the inset.} \label{fig:S1_zf}
\end{figure}

\begin{figure}
    \centering
    \includegraphics[width=0.47\textwidth]{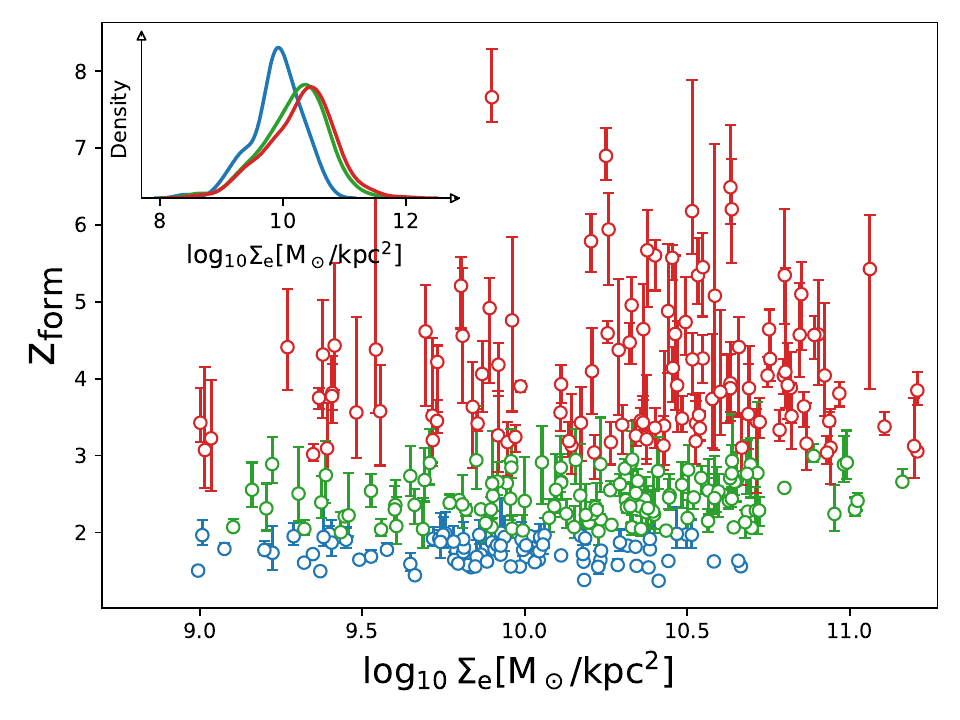}
    \caption{Similar to Figure \ref{fig:S1_zf}, but for the relationship between \Se and \zf.} \label{fig:Se_zf}
\end{figure}

To begin, we present the relationships between stellar mass surface densities and \zf derived from the reconstructed SFHs (section \ref{sec:sfh_shape_def}). Figure \ref{fig:S1_zf} shows the results for \Sone, and Figure \ref{fig:Se_zf} shows the results for \Se. On average, the QGs with larger stellar mass surface densities, regardless if measured within \re or within the central radius of 1 kpc, tend to form earlier, i.e. have larger \zf. This is broadly consistent with other studies \citep[e.g.][]{Tacchella2017,Williams2017} where compact QGs were found to have formed earlier compared to non-compact QGs. The majority of our QGs with \zf$>3$, meaning that they have assembled most of their mass within $<2$ Gyr since the Big Bang, have \Sone$\rm{>10^{10}\,M_\sun/kpc^2}$, being quantitatively consistent with the finding of \citet{Estrada2020}. 

\begin{figure}
    \centering
    \includegraphics[width=0.47\textwidth]{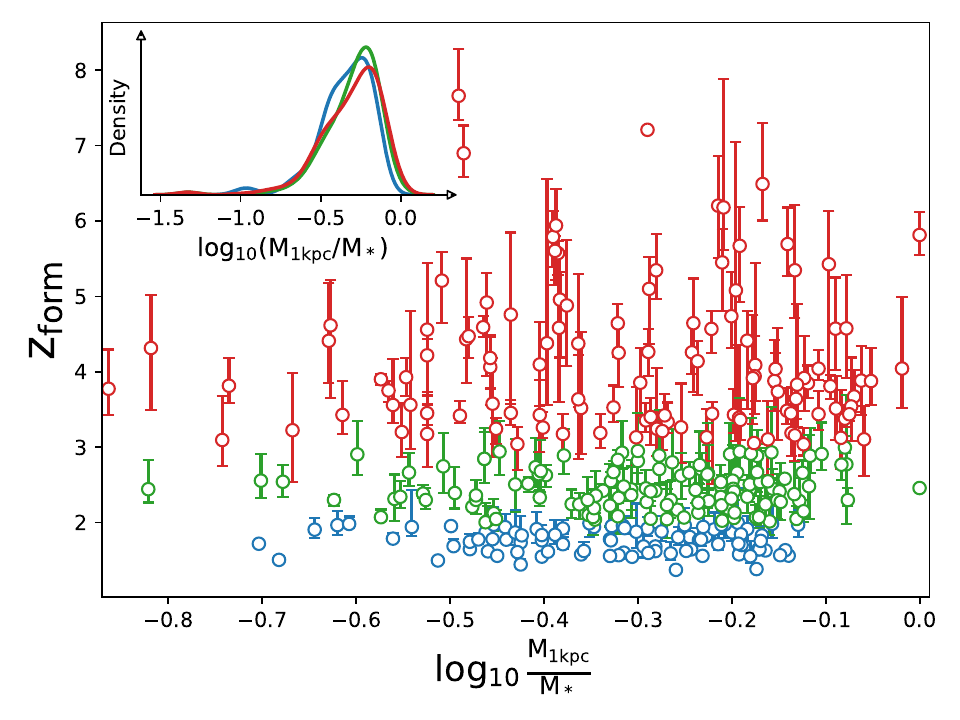}
    \caption{Similar to Figure \ref{fig:S1_zf}, but for the relationship between \Mone and \zf.} 
    \label{fig:M1_zf}
\end{figure}

We continue by studying the relationship between \Mone and SFH. Because $M_*$, $\Sigma$ and \zf are inter-correlated with each other, making it very challenging to identify the real underlying physics driving the observed trend between $\Sigma$ and SFH, we use \Mone to mitigate the strong dependence of $\Sigma$ on $M_*$ (section \ref{sec:mor}, also see \citealt{Ji2022} for more details) in order to have a more direct view on the relationship between galaxy morphology, specifically compactness and SFH. As Figure \ref{fig:M1_zf} shows, we find that the dependence of \zf on \Mone becomes much weaker than the dependence on $\Sigma$. This suggests that the increasing trend of \zf with \Se and \Sone likely is just a reflection of the combined effect of the strong and positive correlation between $\Sigma$ and $M_*$ \citep[e.g.][]{Barro2017,Ji2022}, and the increasing trend of \zf with $M_*$ (section \ref{sec:mass_sfh}).

\begin{figure*}
    \centering
    \includegraphics[width=1\textwidth]{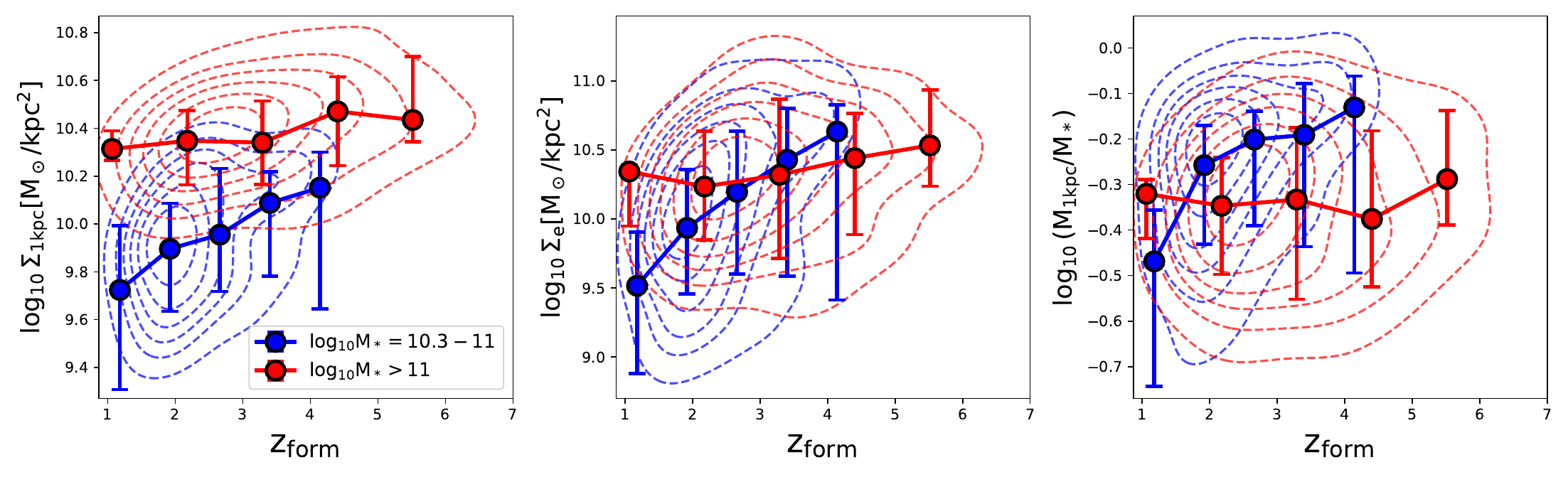}
    \caption{The relationships between \zf (horizontal axis) and galaxy compactness (vertical axis) for QGs with different $M_*$. Blue shows the results of the lower-mass QGs, and red shows the results of the higher-mass QGs. Contours show the density distributions estimated using a Gaussian kernel, while the circles with error bars show the median and 1$\sigma$ range of the compactness of QGs in individual \zf bins. From left to right the adopted compactness metrics are \Sone, \Se and \Mone.} 
    \label{fig:compact_zf_massbin}
\end{figure*}

Motivated by the dependence of the relationship between \re and \zf on $M_*$ (section \ref{sec:re_zform} and \ref{sec:re_zform_stacking}), similarly to what we did in previous sections, we again divide the QGs into the lower-mass and higher-mass samples and study the relationships between the different metrics of galaxy compactness and \zf. The $M_*$ dependence of the relationships is clearly seen in Figure \ref{fig:compact_zf_massbin}. For the lower-mass QGs, we find the increasing trends of \zf with \Sone, and with \Se. For the higher-mass QGs, however, the trends are essentially flat. We also find an increasing trend of \Mone with \zf for the lower-mass QGs, which is interesting given that no clear trend between \zf and \Mone was seen when QGs with all different stellar mass were mixed together (Figure \ref{fig:M1_zf}). This shows the complication, likely introduced by the interplay between different physical processes, of the interpretation of the apparent (null) correlations. Qualitatively speaking, the findings here are fully in line with the picture that the progenitor effect becomes increasingly important as QGs become less massive, as it has already been discussed in section \ref{sec:re_zform} when studying the size evolution. On the one hand, because the density of the Universe increases with redshift, this means that galaxies formed earlier also have larger densities. On the other hand, as is the case for very massive QGs, if significant late-time size growths take place after galaxies have formed and quenched, the morphological properties of galaxies can reduce the memory of the time of their formations, making the relationship between galaxy compactness and \zf flat. As will soon be discussed quantitatively in section \ref{sec:diss_progenitor}, the increasing trends of \zf with \Sone, \Se and \Mone observed for the lower-mass QGs seemingly can be fully explained by the strong progenitor effect.

\section{Discussions}
To summarize, we utilized the fully Bayesian SED fitting code \prospector to reconstruct the non-parametric SFHs for a carefully culled sample of 361 massive QGs at $\langle $\zobs$\rangle\sim2$ selected through the UVJ technique in the CANDELS/COSMOS and GOODS fields. The diverse assembly history is clearly seen in these high-redshift QGs. We have studied in detail the relationships between the evolution of the morphological properties of QGs and their assembly history. The relationships of \zf with \re, and with galaxy compactness (\Sone, \Se and \Mone) show a clear dependence on stellar mass. All these findings converge to a general physical picture of the morphological evolution of QGs, namely that the progenitor effect plays a crucial role in the apparent evolution of the morphological properties of the QGs with $10.3<\log_{10}(M_*/M_\sun)<11$, while the late-time size growths via galaxy mergers become increasingly important as QGs become more and more massive. In the following, we discuss the implications of our findings for the dynamical (section \ref{sec:diss_dyn}) and chemical (section \ref{sec:diss_che}) evolution of QGs. Finally, in section \ref{sec:diss_progenitor}, we provide simple, empirical arguments that the progenitor effect alone can explain the apparent evolution of certain morphological properties of lower-mass QGs.

\subsection{Implications for the dynamical evolution of QGs}\label{sec:diss_dyn}

The dependence of morphological evolution of QGs with \zf on $M_*$ could imply a similar $M_*$ dependence of dynamical evolution. On the one hand, among the more massive galaxies, those with lower \zf were born larger than those with higher \zf, which were more compact at the time of formation. This implies that a substantial mass growth took place via mergers, which also grew the size of QGs with stellar mass $>10^{11}M_\sun$. This suggests that QGs at the high-mass end formed in dense environments, presumably in the regions with large overdensity of the primordial density field. Early on, these massive halos can gravitationally attract large amount of gas from surrounding cosmic webs. The gas is highly dissipative and thus can quickly sink to the center of gravitational potential, fueling  intense star formation until the cumulative mass becomes so large that the infalling gas accreted later on are shock-heated by the halo potential \citep{Dekel2006} which eventually leads to the quenching of star formation. After the formation, these very massive galaxies continue growing their sizes via merging with other galaxies through dynamic friction. We do not expect these galaxies to be fast rotators, since multiple incoherent merging episodes are expected to cause the loss of angular momentum \citep[e.g][]{Emsellem2011}.

On the other hand, lower-mass QGs very likely formed in regions with comparatively smaller overdenties, meaning that they have shallower gravitational well and hence were not able to accrete  gas from cosmic webs as efficiently as those galaxies in larger halos. As a result, lower-mass QGs may assemble in a much more slower and smoother manner. They may initially form as disks with high gas fractions, and then quench via secular processes like the bulge formations through the coalescence of star-forming clumps to the center of galaxies \citep{Dekel2009}. For these galaxies we expect to see significant rotations. Strong observational supports for the scenarios above come from the dynamical studies of Early Type Galaxies (ETGs) in the nearby Universe using ground-based Integral Field Unit spectroscopy. The distribution of the dynamical state of ETGs at $z\sim0$ is bimodal, with one population dubbed ``fast rotator'' having much larger $v_{rot}/\sigma$ than the other one dubbed ``slow rotator''. Interestingly, in the local universe fast and slow rotators can be very well separated using a characteristic $M_*=2\times10^{11}M_\sun$, with the former being less massive (see a review by \citealt{Cappellari2016} and references therein). Therefore,  our finding of the $M_*$ dependence of the morphological evolution of QGs at $z\sim2$ seems to be quantitatively in line with the dynamical studies of ETGs at $z\sim 0$. 

A critical test on the dynamical evolution is to directly measure the kinematics of high-redshift QGs. However, at high redshift, unlike star-forming galaxies whose dynamical properties can still be measured through strong emission lines either from ionized gas such as \Ha or from cold molecular gas such as CO, measuring the kinematics of QGs is very challenging because QGs have very low level of on-going star formation and very little molecular gas \citep[e.g.][]{Bezanson2019,Williams2021,Whitaker2021}, and they are very compact. The robust dynamical measurements of QGs thus require deep and spatially-resolved spectroscopy at the rest-frame optical wavelengths where abundant stellar absorption features exist. Substantial progress has been recently made in understanding the kinematics of QGs at $0.5<z<1$ \citep[e.g.][]{Bezanson2018}. At $z\sim2$ and beyond, while current instrumentation only allows such measurements for a few cases of gravitationally lensed QGs \citep{Toft2017,Newman2018b}, the upcoming \jwst should significantly improve this situation.

\subsection{Implications for the chemical evolution of QGs} \label{sec:diss_che}

The chemical composition of galaxies is another key parameter to constrain the physics of formation and evolution of QGs across cosmic time. The dependence of morphological evolution on $M_*$ could also imply the $M_*$ dependence of the evolution of chemical properties of QGs. Without strong nebular emission lines, the only probe of the chemical properties of QGs is stellar absorption features which require ultradeep NIR spectroscopy with current ground-based 10m telescopes. Some progress has been made recently in measuring the chemical compositions of a handful of QGs at $z>1$ using either individual spectra \citep{Lonoce2015,Kriek2019,Lonoce2020} or stacked spectra \citep{Onodera2015}. Some intriguing, but not yet conclusive, results have already been found. For example, \citet{Kriek2019} successfully measured the abundance ratios [Mg/Fe] and [Fe/H] in three massive ($\log_{10}M_*\sim11$) QGs at $z\sim1.4$, where they found tentative evidence that (1) while the relationships between [Mg/Fe] and $M_*$ are consistent at $z\sim1.4$ and at $z<0.7$, [Fe/H] is lower at $z\sim1.4$ than at $z<0.7$; and (2) relative to QGs at $z\sim0$, the offset of [Mg/Fe] of massive QGs decreases with redshift. Taken together, this evidence suggests that high-redshift massive QGs need to accrete low-mass and less $\alpha$-element enhanced galaxies to explain the observations. We also highlight another work from \citet{Jafariyazani2020} where for the first time they were able to measure abundance ratios of other six elements (apart from Mg and Fe) and their radial gradients inside a very massive QG ($\log_{10}M_*\sim11.7$) at $z=1.98$ thanks to the magnification effect of strong gravitational lensing. This QG was found to be not only Mg-enhanced but also Fe-enhanced compared to the center of nearby ETGs. If this QG is a typical progenitor of ETGs at $z\sim0$, the findings then suggest that significant radial mixing of stars with different chemical abundances have to take place inside very massive QGs, even in their central regions.

Despite the very small sample size of existing measurements of chemical compositions of QGs at $z>1$, all studies seem to consistently suggest that processes like galaxy mergers are needed to explain the chemical evolution of QGs with $\log_{10}M_*>11$, which is fully in line with our conclusions based upon the morphological evolution of QGs. Future \jwst observations will not only significantly enlarge the number of chemical measurements of very massive QGs at high redshift, but also push the measurements to lower masses and eventually enable testing if the chemical evolution of QGs also depends on stellar mass.

\subsection{Can the progenitor effect alone explain the observed relationships between galaxy compactness and \zf for the lower-mass QGs?} \label{sec:diss_progenitor}

Setting aside the potential systematic errors from the assumed prior of non-parametric SFHs and the proxy used for $z_{\rm{form}}^{\rm{h}}$, which we discussed in detail in section \ref{sec:re_zform}, for the lower-mass QGs ($10.3<\log_{10}(M_*/M_\sun)<11$) the slope of the relationship between \re and \zf is consistent with the value $-1$. This can be fully explained in terms of the formation of dark matter halos, indicating that the progenitor effect plays a predominant role in the apparent size evolution of the lower-mass QGs. Obviously, this means that the progenitor effect should also play an important role in the apparent evolution of QGs' compactness because all compactness metrics discussed in section \ref{sec:zf_sigma} directly depend on \re. An important question yet to be answered is, can the progenitor effect alone be enough to explain the apparent evolution of compactness of lower-mass QGs? The answer is critical for understanding not only if there are additional physical links between galaxy compactness and \zf, but also if there is a causal link between galaxy compactness and quenching.

\begin{figure*}
    \centering
    \includegraphics[width=1\textwidth]{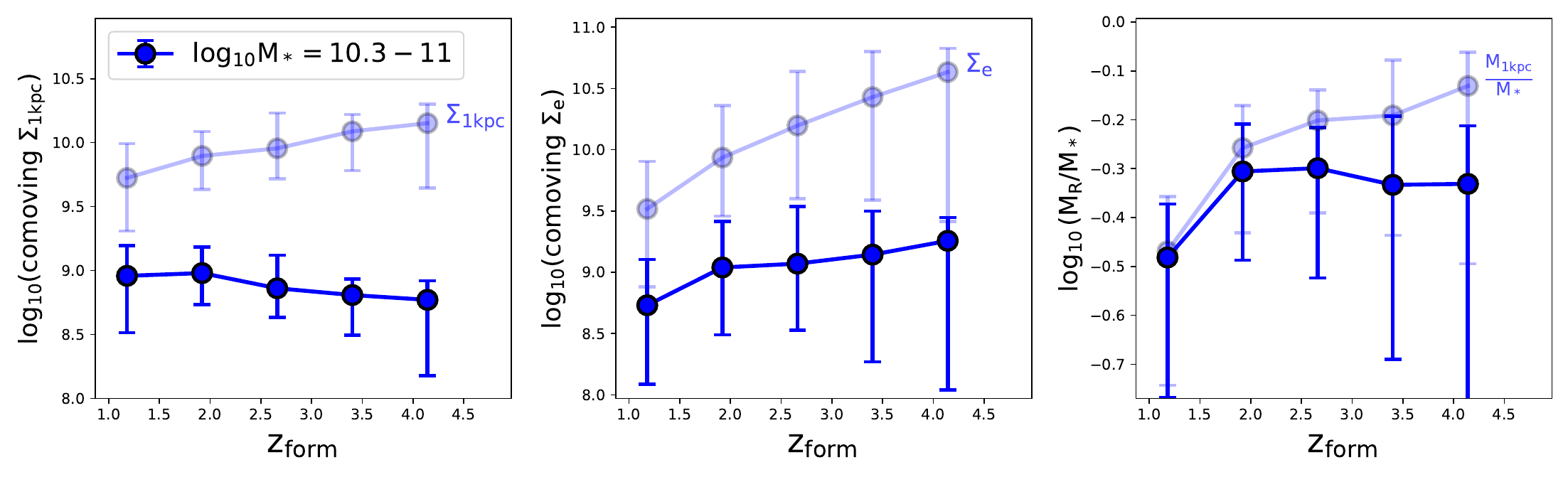}
    \caption{Empirical tests of the progenitor effect on the apparent morphological evolution of the lower-mass QGs. From left to right show the relationships of \zf with comoving-\Sone, comoving-\Se and \Mr (see section \ref{sec:diss_progenitor} for details). For comparison, we also use light blue to plot the relationships already shown in Figure \ref{fig:compact_zf_massbin} for the lower-mass QGs.} 
    \label{fig:comoving_compact}
\end{figure*}

To begin, we test the progenitor effect on the apparent evolution of stellar mass surface densities. We introduce a purely empirical but physically motivated parameter, comoving stellar mass surface density, which is defined as
\begin{equation}
    \rm{comoving\mbox{-}\Sigma = \frac{\Sigma}{(1+z_{\rm{form}})^2}}
\end{equation}
The reason of normalizing $\Sigma$ with \zf, rather than \zobs, has already been explained in section \ref{sec:re_zform}, namely that the morphological properties (e.g. density) of galaxies should reflect the earlier time when they formed, if the progenitor effect is strong which is the case for the lower-mass QGs based on the \zf-\re relationship. The left two panels of Figure \ref{fig:comoving_compact} show the relationships between \zf and comoving-\Sone, and between \zf and comoving-\Se, respectively. For comparison, we also show the relationships of \Sone and \Se in light blue. It is clear that the evolution becomes very weak, if any, once stellar mass surface densities are normalized by the cosmic density at \zf, suggesting that the progenitor effect alone is largely sufficient to explain the apparent correlations between \zf and $\Sigma$.  

We stress however that, depending on the radius within which $\Sigma$ is measured, the characteristic redshift used for the normalization should, strictly speaking, corresponds to the exact epoch when the regions within that radius assembled. This needs the knowledge not only of the galaxies' integrated assembly history but also that of their sub-structures, for example the central regions and the outskirts. This requires the  reconstruction of spatially resolved SFH which unfortunately is not currently feasible for high-redshift QGs. We argue, however, that using \zf to normalize stellar mass surface densities, both for comoving-\Sone and comoving-\Se, provides a reasonable substitute, as we explain in what follows. 

The mass assembly of galaxies \textit{is not} self-similar, namely some parts (e.g. bulge) form earlier while others (e.g. disk) form later. Studies of the growth of massive star-forming galaxies at $z>1$ consistently show that massive galaxies grew their inner regions earlier \citep[e.g.][]{Nelson2012,Nelson2016,Tacchella2017}. Despite being crude, it is therefore not unreasonable for us to assume that high-redshift QGs assembled their mass in a strictly inside-out manner. Under this assumption, \zp derived from the reconstructed SFHs then is the epoch when the central region within the radius containing p-percent total stellar mass has assembled. The redshift used for the normalization of \Se thus should be either $z_{50}$ or \zf. We have checked that the results of using the two redshifts are essentially identical. Similarly, the redshift used for the normalization of \Sone should be \zp where p equals to the fractional mass within the central 1kpc, i.e. \Mone. Because the median \Mone of the sample of lower-mass QGs is $\approx0.5$ (0.59$\pm$0.18, Figure \ref{fig:M1_zf}), using \zf for comoving-\Sone therefore is also reasonable. We have also tested our results by normalizing \Sone using the \zp determined by \Mone for individual QGs, where no substantial changes were found. To this end, we believe that our empirical approach of normalizing galaxy compactness metrics with \zf in general should be able to capture the progenitor effect imprinted on the apparent evolution of galaxy compactness of our lower-mass QG sample. 

We finally discuss the apparent evolution of \Mone. Because galaxies formed earlier tend to have smaller \re, and meantime \Mone is measured using the fixed central 1 kpc no matter what \zf a galaxy has, these mean that the 1 kpc scale probes a {\it relatively} larger area for galaxies having larger \zf than those having smaller \zf. This is how the progenitor effect enters in the apparent evolution of \Mone. To quantify its contribution, we modify \Mone (equation \ref{equ:mone}) to \Mr where, instead of using the fixed 1 kpc, for individual galaxies we adopt
\begin{equation}
    R = 1\,\rm{kpc}\cdot \frac{1+z_{\rm{form}}^{\rm{min}}}{1+z_{\rm{form}}}
\end{equation}
and $z_{\rm{form}}^{\rm{min}}$ is the minimum \zf of the sample. As the right-most panel of Figure \ref{fig:comoving_compact} shows, the increasing trend of \Mone with \zf almost disappears after using \Mr, suggesting that the key driver of the apparent correlation between \zf and \Mone is the progenitor effect.

\section{Summary}

This is the first of a series of papers that aim to investigate the relationships between the star-formation and structural properties of galaxies and their assembly history at the Cosmic Noon epoch. In this work, we focus on a sample of UVJ-selected massive quiescent galaxies (QGs) with stellar mass $M_*>10^{10.3}M_\sun$ and at a median redshift $\langle z_{\rm{obs}}\rangle\approx1.9$ in the CANDELS/COSMOS and GOODS fields (section \ref{sec:sample}). Thanks to the accumulated high-quality panchromatic photometry that densely samples the rest-frame UV-Optical-NIR spectra of the galaxies (section \ref{sec:phot}), we utilize the fully Bayesian SED fitting code \prospector to reconstruct the SFH of QGs (section \ref{sec:SED} and \ref{sec:sfh_shape_def}) and find a broad diversity among their assembly histories (section \ref{sec:diverse}).

Following \citet{Lilly2016}, we refer to the dependence of properties of galaxies observed at \zobs on \zf, the time of formations defined as the redshift when a galaxy has assembled half of the mass it has at \zobs, as the progenitor effect. The key finding of this paper is that the progenitor effect is important, and if not accounted for, it introduces spurious correlations between galaxies' observables. Specifically,
\begin{itemize}
    \item We have studied the progenitor effect in terms of the relationship between the half-light radius \re measured at \zobs, and \zf (section \ref{sec:re_zform}). We have quantified the relationship with the form $R_e\propto (1+z_{\rm{form}})^{-\beta}$. We have found that the relationship becomes steeper, i.e. having larger $\beta$, as QGs become less massive. This dependence on stellar mass has been further confirmed by our stacking analysis (section \ref{sec:re_zform_stacking}).
    
    \item Regardless of the systematics introduced by the specific prior assumed during the non-parametric SFH reconstructions, which appear to be relatively minor, and the usage of \zf (section \ref{sec:re_zform}), using our fiducial SED modeling which has been validated with the IllustrisTNG simulations (Appendix \ref{app:tng}), we have found that, for lower-mass QGs with $10.3<\log_{10}M_*<11$, the relationship is consistent with $R_e\propto (1+z_{\rm{form}})^{-1}$, namely what is expected if the central density of the galaxies keeps memory of the value it had at the initial collapse, suggesting that, after quenching, merging and interactions had a minimal influence in continue shaping the central light (and thus, mass) profile.
    
    For higher-mass QGs with $\log_{10}M_*>11$ the relationship flattens, tending towards the value $\beta=0.2\pm0.1$. This suggests that for very massive galaxies, after quenching, the light profile continues to evolve and converges to a constant shape, which is typical of relaxed structures. In turn, this suggests that merging played a substantial role. We have also stacked the multi-band \hst images for the QGs. Evidence that the higher-mass QGs have a more extended light profile in the \I band than in the \J and \H bands, based on both the analysis of surface brightness profiles and the analysis of single-\sersic fittings with \galfit, has been found, suggesting that after-quenching growths of massive galaxies are due to merging satellite star-forming galaxies.
    
    We have also highlighted that the mass value $\log_{10}M_*\approx11$, which separates the two post-quenching evolutionary regimes, is essentially the same as the one that separates fast rotators from slow rotators in local ETGs, suggesting a direct evolutionary link with the populations of $z\sim 2$ QG. 
    
    \item Similar to the stellar-mass dependence observed for the relationship between \re and \zf, the relationship between the compactness of QGs and \zf also depends on stellar mass (section \ref{sec:zf_sigma}). Increasing trends of \zf are found with \Sone, \Se and \Mone for the lower-mass QGs, while they become very flat for the higher-mass QGs. 
    
    We have empirically evaluated the progenitor effect on the apparent trend between compactness and \zf for the lower-mass QGs by introducing the ``comoving" central surface mass density of galaxies (comoving-$\Sigma$), which is the central surface mass density normalized by the cosmic density at the time of their formations. Unlike what we found for $\Sigma$, comoving-$\Sigma$ does not depend on \zf, meaning that the larger central stellar density and compactness of the lower-mass QGs at higher redshift can be fully explained by the progenitor effect, without requiring any additional physical mechanisms. 
\end{itemize}

As a concluding remark, finally, we observe  that this work highlights the  importance, and the effectiveness, of measuring the SFH of galaxies before attempting any inference into their evolution from observed correlations of properties at any given epoch. Our tests, as well as those done by others \citep[e.g.][]{Leja2019, Tacchella2021}, conducted using cosmological simulations show that the \prospector code appears capable to robustly reconstruct the SFH of galaxies at $z= 1-7$ with sufficient precision and accuracy to unveil trends of galaxy properties, when good-quality and densely sampled panchromatic photometry and/or spectroscopy is available. The reconstruction of SFHs by SED modeling seems to be leaving its infancy and enter a phase where it can be used as a power tool of investigation of galaxy evolution.  

\section{ACKNOWLEDGMENT}

We thank the anonymous referee for useful comments. We are indebted to Alvio Renzini for illuminating discussions and for his comments on an earlier version of the manuscript. We thank Joel Leja for his generosity in providing us with advice and guidance on running \prospector. This work was completed in part with resources provided by the University of Massachusetts' Green High Performance Computing Cluster (GHPCC).  

\software{Prospector \citep{Leja2017,Johnson2021}, FSPS \citep{Conroy2009,Conroy2010}, MIST \citep{Choi2016, Dotter2016}, MILES \citep{Falcon-Barroso2011}, dynesty \citep{Speagle2020}, cobs \citep{Ng2007,Ng2020}, GALFIT \citep{galfit2002,galfit2010}}

\bibliography{ji_2021_QG_SFH}

\appendix
\section{Validation of the \prospector fitting procedure with the IllustrisTNG simulations} \label{app:tng}

\begin{figure*}
    \includegraphics[width=0.47\textwidth]{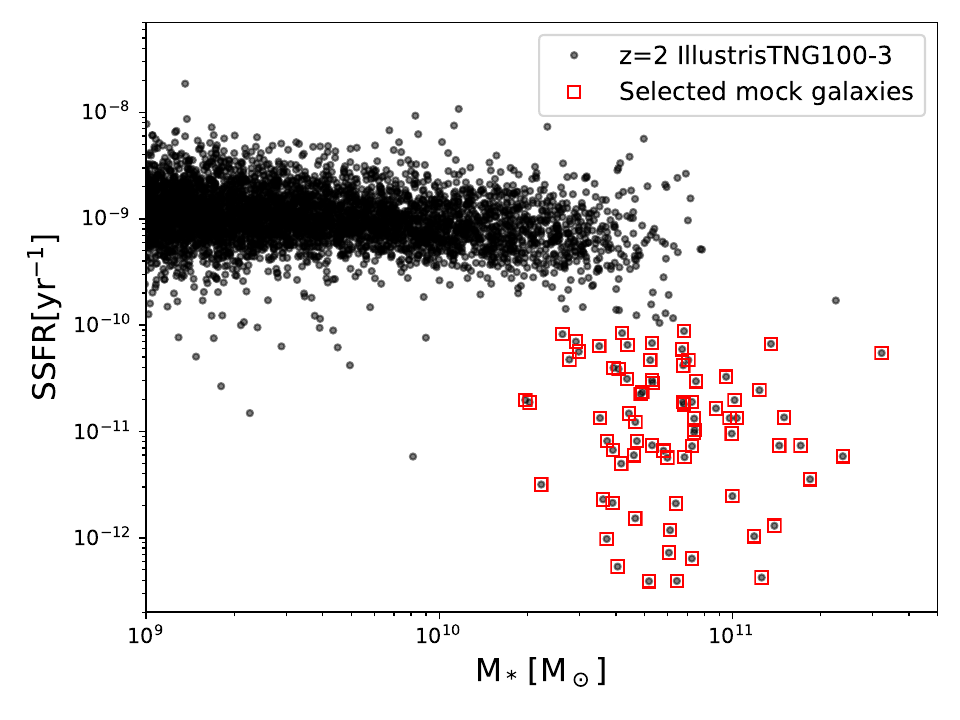}
    \includegraphics[width=0.527\textwidth]{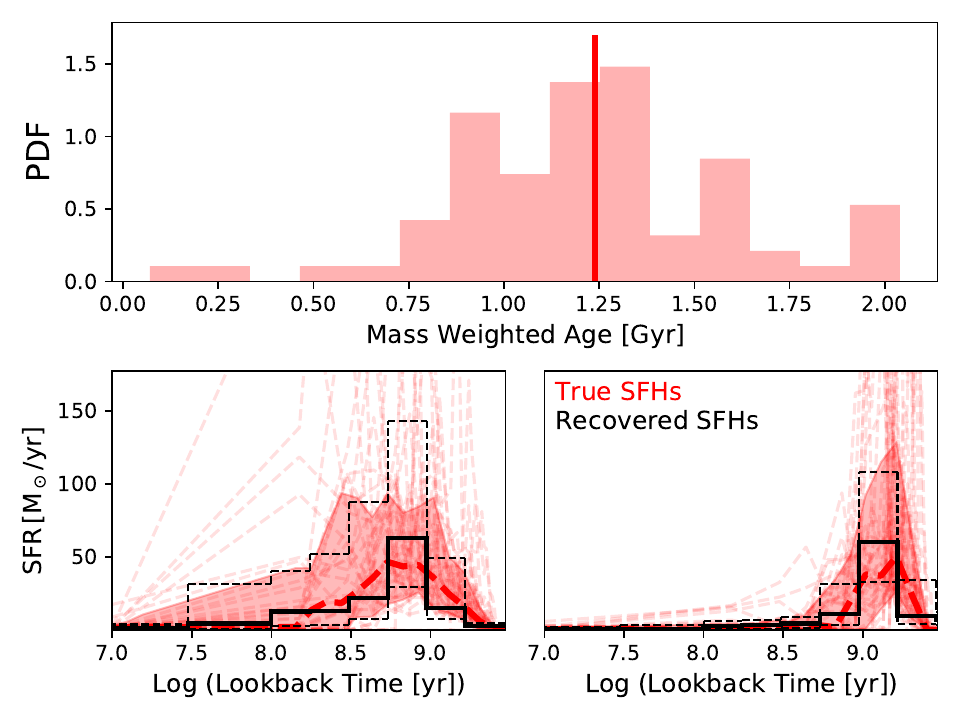}
    \caption{{\bf Left:} The sample selection of TNG QGs. The selection is done with the Illustris-TNG100-3 run and at its Snapshot\#33 corresponding to $z=2$. Galaxies marked with red squares are selected as our TNG QGs sample for the SED fitting robustness tests. {\bf Right:} The upper panel shows the distribution of mass-weighted age for the TNG QGs sample. The vertical red solid line marks the median. The whole sample is divided into two groups, the younger one with mass-weighted ages less than the median and the old one larger than the median. The bottom panels show the comparison between true SFHs and recovered SFHs using the Dirichlet prior, with the younger group shown on the left and the old one shown on the right. The thick red dashed line and the red shaded region show the median and 16th-84th percentile range of the true SFHs, while the black step plots with solid and dashed line styles show the recovered median and 16th-84th percentile range of SFHs.} 
    \label{fig:tng_sfms}
\end{figure*}

\begin{figure}
    \centering
    \includegraphics[width=0.87\textwidth]{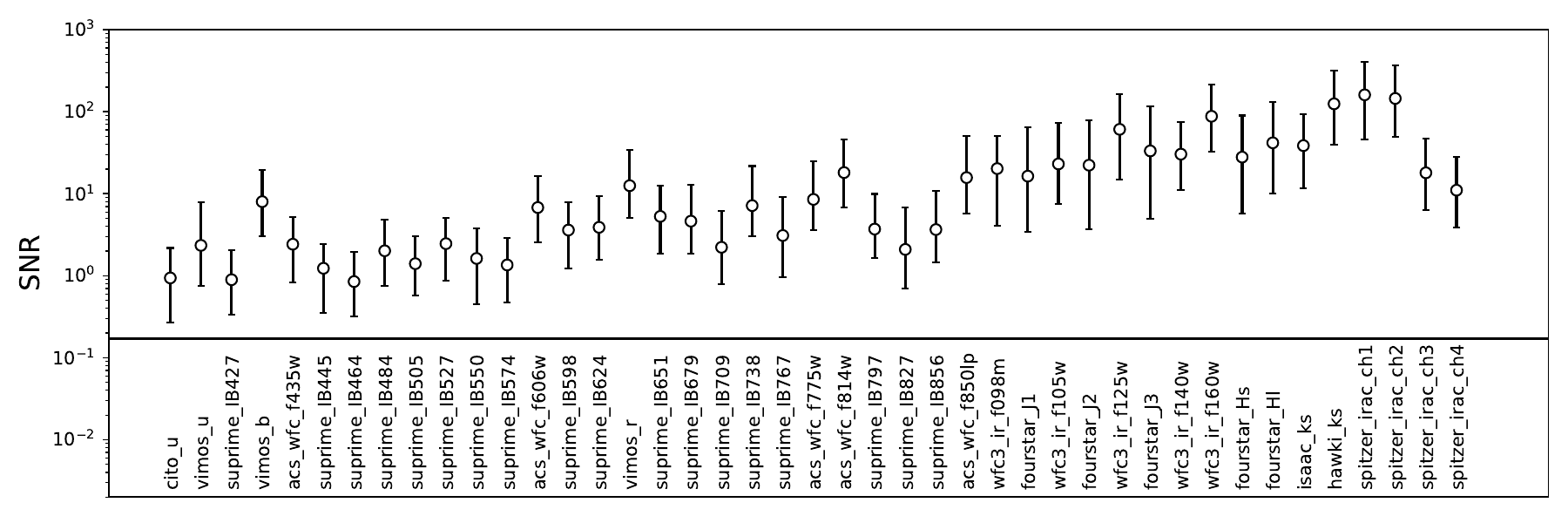}
    \caption{Observed signal-to-noise ratio (SNR) of each photometric band in the CANDELS/GOODS-South. From left to right show the bands from blue to red. Circles are the medians, while the error bars are the 16th-84th percentile ranges. } \label{fig:data_quality}
\end{figure}

We test the robustness of our {\sc Prospector} SED fitting procedure using the IllustrisTNG simulations. In particular, we use the results from the run TNG100-3 for this validation and choose the simulation snapshot (\#33) corresponding to $z=2$, i.e. about the median redshift ($\langle z\rangle=1.9$) of our QGs sample, for the TNG sample selection. As the left panel of Figure \ref{fig:tng_sfms} shows, we select the sample of TNG QGs using the ``main sequence" diagram, i.e. sSFR vs. $M_*$. Because the focus of this study is on massive galaxies, we only look at $M_*\ge 10^{10}M_\sun$ TNG galaxies and select those with $\rm{sSFR\le 10^{-10} yr^{-1}}$ as our TNG QGs sample (marked with red squares in the Figure). These give us a sample with 72 simulated QGs. We then use the TNG's ``SubLink'' halo merger tree to extract the full assembly histories for individual galaxies, which are then used to generate synthetic spectra with the stellar population synthetic code {\sc Fsps}-v3.0 \citep{Conroy2009,Conroy2010}. For each synthetic spectrum, we assume the \citealt{Kroupa2001} initial mass function, \citealt{Byler2017} nebular emission (line+continuum) model, \citealt{Calzetti2000} dust attenuation law with the fixed color excess $\rm{E(B-V)=0.2}$ that has been observed in typical early type galaxies \citep[e.g.][]{Dominguez2011}, and the stellar metallicity value that equals to the value of individual galaxies at the Snapshot\#33 (i.e. we ignore the evolution of metallicity during galaxy assemblies). In other words, we adopt the same assumptions we have made to run \prospector when fitting the SED of the real galaxies. With the synthetic spectra in hand, we then convolve them with the filter throughputs in the GOODS-South field (note that the CANDELS/COSMOS and GOODS-North have a very similar set). To model the data quality of observations, we assign the flux uncertainty in each filter according to the median observed SNR of our sample (Figure \ref{fig:data_quality}). 

\begin{figure*}
    \centering
    \includegraphics[width=0.97\textwidth]{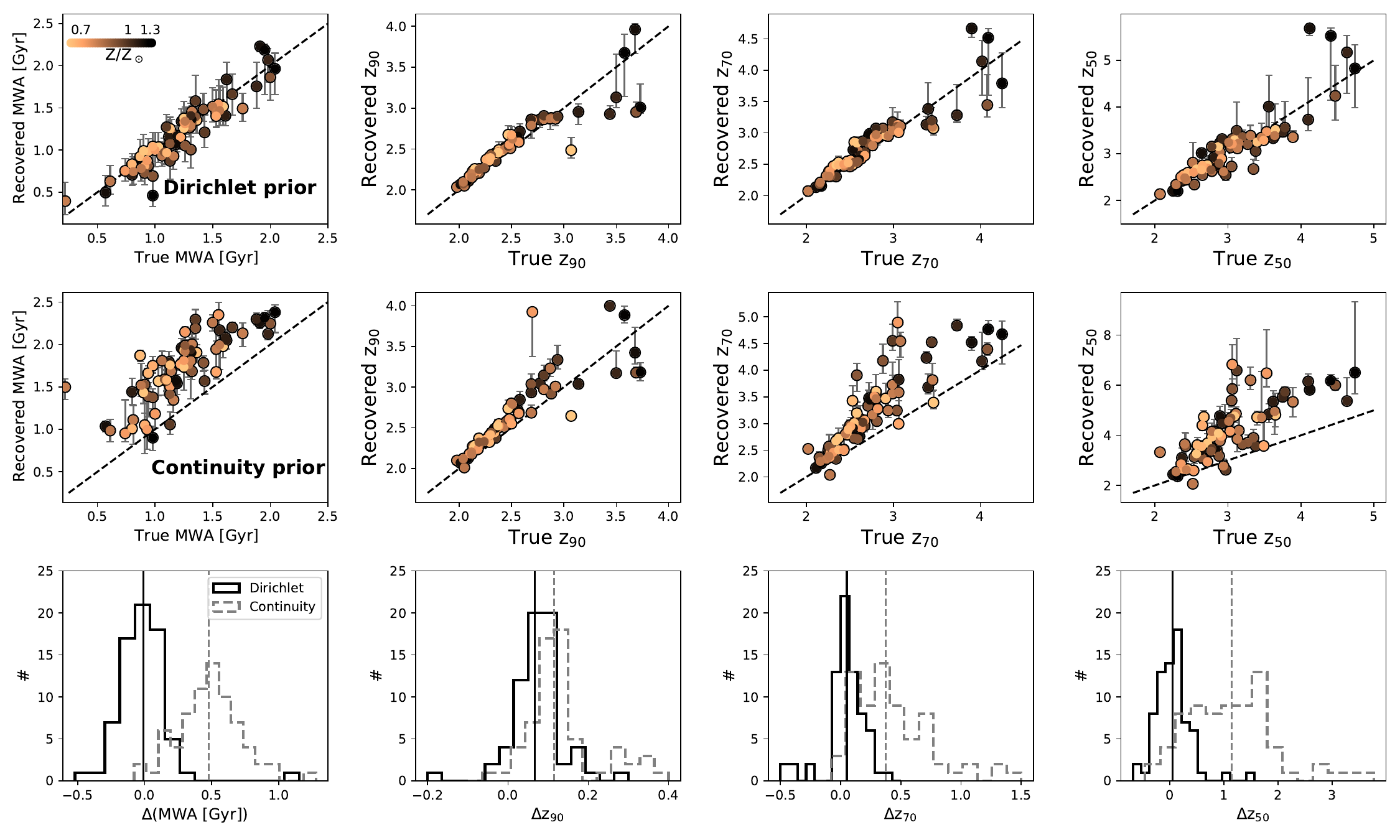}
    \caption{Comparisons between the recovered SFH-related parameters and the true values. {\bf The 1st row:} The comparisons for the measurements assuming the Dirichlet prior. From left to right show the results of mass-weighted age, $z_{90}$, $z_{70}$ and $z_{50}$. Each galaxy is color-coded with its metallicity. The black dashed line shows the one-to-one relation. {\bf The 2nd row:} The comparisons for the measurements assuming the Continuity prior. {\bf The 3rd row:} The distributions for the deviations of the measurements from the true values. Black histograms are for the Dirichlet prior, the grey ones are for the Continuity prior. Also marked in each histogram as vertical lines are the medians.} 
    \label{fig:tng_check}
\end{figure*}

Using the Dirichlet prior, we fit the synthetic data with the exactly same setup as described in section \ref{sec:SED}. The comparisons between the true SFHs and the {\sc Prospector}-recovered SFHs are shown in the right panels of Figure \ref{fig:tng_sfms}. Our fitting procedure is able to reconstruct galaxy assembly histories to a great extent. More quantitative comparisons are shown in the first row of Figure \ref{fig:tng_check}, where the recovered mass-weighted age, $z_{50}$, $z_{70}$ and $z_{90}$ are plotted against the true values. In all cases, using the Dirichlet prior results in robust measures of the SFHs for the TNG QGs. Also needed to point out is that, despite that we only have the photometric data hence we do not expect to have strong constrains on galaxy stellar metallicity, the deviation of the recovered SFH parameters from the true values does not correlate with the assumed metallicities, suggesting that the not well-contrained metallicity cannot lead to significant errors of the reconstructed SFHs.

In addition to the Dirichlet prior, we have also run {\sc Prospector} for the TNG QGs using the Continuity prior, which models changes of SFR in adjacent time bins by means of the variate $\rm{x=log(SFR_{t_i}/SFR_{t_{i+1}})}$ and, importantly, strongly {\it disfavours}  sharp changes of x. We refer readers to \citet{Leja2019} for the details of the model, but, briefly, the prior in the Continuity non-parametric SFH is assumed to be a Student's-t distribution $\rm{P(x,\nu,\sigma)}$\footnote{$\rm{P(x,\nu,\sigma)=\frac{\Gamma(\frac{\nu+1}{2})}{\sqrt{\nu\pi}\Gamma(\frac{1}{2}\nu)}(1+\frac{(x/\sigma)^2}{\nu})^{-\frac{\nu+1}{2}}}$, where $\Gamma$ is the complete Gamma function.}, where $\sigma$ controls the width of distribution and $\nu$ is the number of degrees of freedom that controls the tail of the distribution. Calibrated using $\approx30,000$ galaxies with $M_*\ge10^9M_\sun$ at $z=0$ (the vast majority are star-forming galaxies) from the Illustris simulations \citep{Vogelsberger2014}, \citealt{Leja2019} suggested $\sigma=0.3$ and $\nu=2$ which are used here. We used the Continuity prior to fit the synthetic galaxies and compared the recovered SFHs with the true ones, using the same procedures adopted for the Dirichlet prior. As the second row of Figure \ref{fig:tng_check} shows, while strong correlations between parameters related to the recovered and the true SFH are observed overall, systematic offsets exist. Specifically, compared with the Dirichlet prior (the third row of Figure \ref{fig:tng_check}), the Continuity prior in general over-estimates the mass-weighted age by $\approx0.5$ Gyr. It also over-estimates $z_{50}$, $z_{70}$ and $z_{90}$. These findings are consistent with what reported by \citet{Leja2019} for galaxies with the ``Sudden Quench'' SFHs (see the right-most column of their Figure 5). Also noticed from the third row of Figure \ref{fig:tng_check} is the fact that the dispersion of the measurements is larger when the Continuity prior is used, again showing that the Dirichlet prior generally works better for the TNG QGs.

\begin{figure}
    \centering
    \includegraphics[width=0.7\textwidth]{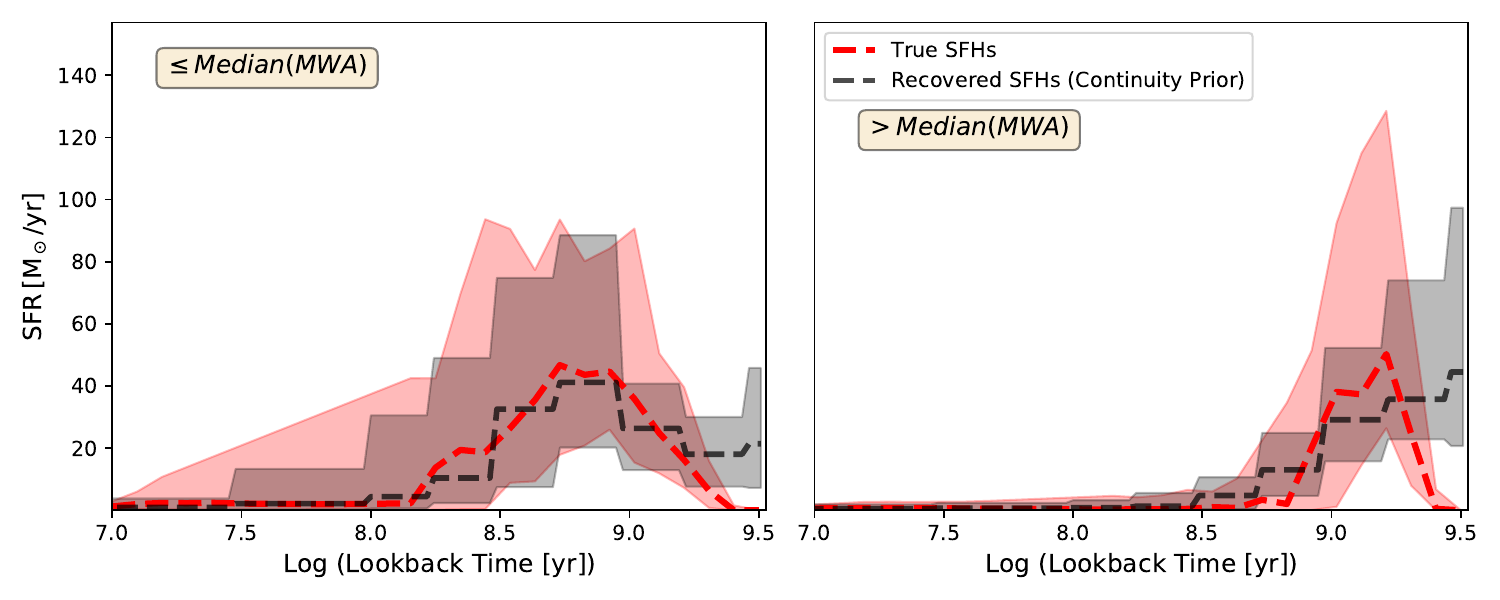}
    \caption{Similar to the bottom-right two panels of Figure \ref{fig:tng_sfms}, but we now compare the true SFHs with the recovered SFHs assuming the Continuity prior. The left panel shows the results for the younger TNG QGs, i.e. those have mass-weighted ages less than the median value of the whole TNG QGs sample, while the right panel shows the results for the older TNG QGs.} 
    \label{fig:tng_sfh_conti}
\end{figure}

The reason behind the over-estimation of stellar ages of the TNG QGs with the Continuity prior can be clearly seen in Figure \ref{fig:tng_sfh_conti}. Because the Continuity prior strongly opposes sharp changes in SFR, the reconstructed SFHs are forced to be smooth. Specifically, rather than putting more mass in fewer time bins, the Continuity prior tends to disperse the mass into more time bins. This is particularly seen in the old time bins because young and bright stars outshine older stars in the rest-frame optical-NIR, meaning that the goodness-of-fit does not change too much when slightly altering the SFH of the older stellar populations \citep{Papovich2001} given the data coverage. The similar was also found in \citet{Johnson2021} (see the bottom panel of their Figure 3). However, because the TNG QGs indeed show very sudden changes in SFRs at early epochs (i.e. old time bins), this leads to the obvious mismatch between the assumption of the Continuity prior and the true shape of SFHs of the TNG QGs, and consequentially the discrepancies between the recovered and the true SFHs.

We conclude that our {\sc Prospector} SED fitting procedure with the Dirichlet prior can robustly recover the assembly histories for the TNG QGs. Our tests also put in evidence the footprint of the assumed priors on the SFH reconstruction. While the Continuity prior may be a very good choice for star-forming galaxies, it is not as good as the Dirichlet prior for the QGs. However, we stress the limitations of our test with the TNG simulations. First and foremost, our conclusion on the best prior to use with non-parametric SFHs relies on the assumptions that 1) any difference between the SFHs of simulated TNG QGs and the real ones, regardless of how correctly the former capture the latter, does not affect the performance of \prospector in carrying out the reconstruction of the SFH itself, and 2) the systematic error from the mismatch between the real stellar populations at high redshift and the synthesis model ({\sc Fsps} here) used to generate synthetic photometry for the TNG QGs is minor. Second, we emphasize that we are {\it not} making the statement that the Continuity prior should not be used for the QGs. Recall that the Student's distribution used by the Continuity prior is governed by $\sigma$ and $\nu$. As already mentioned above, the values used here are from the calibration with the Illustris simulations at $z=0$ where the vast majority of the sample are star-forming galaxies. It is entirely plausible that, if we re-calibrate $\sigma$ and $\nu$ using QGs, the Continuity prior may still work well. Such calibration is beyond the scope of this work. The fact that using the Dirichlet prior can reproduce the overall shapes of true SFHs, and quantitatively recover the true values of parameters including mass-weighted age, $z_{50}$, $z_{70}$ and $z_{90}$, makes us believe our reconstructed SFHs of the QGs are statistically robust. 

\section{Physical parameters derived with the Continuity prior for the sample QGs} \label{app:conti}
\begin{figure}
    \centering
    \includegraphics[width=0.97\textwidth]{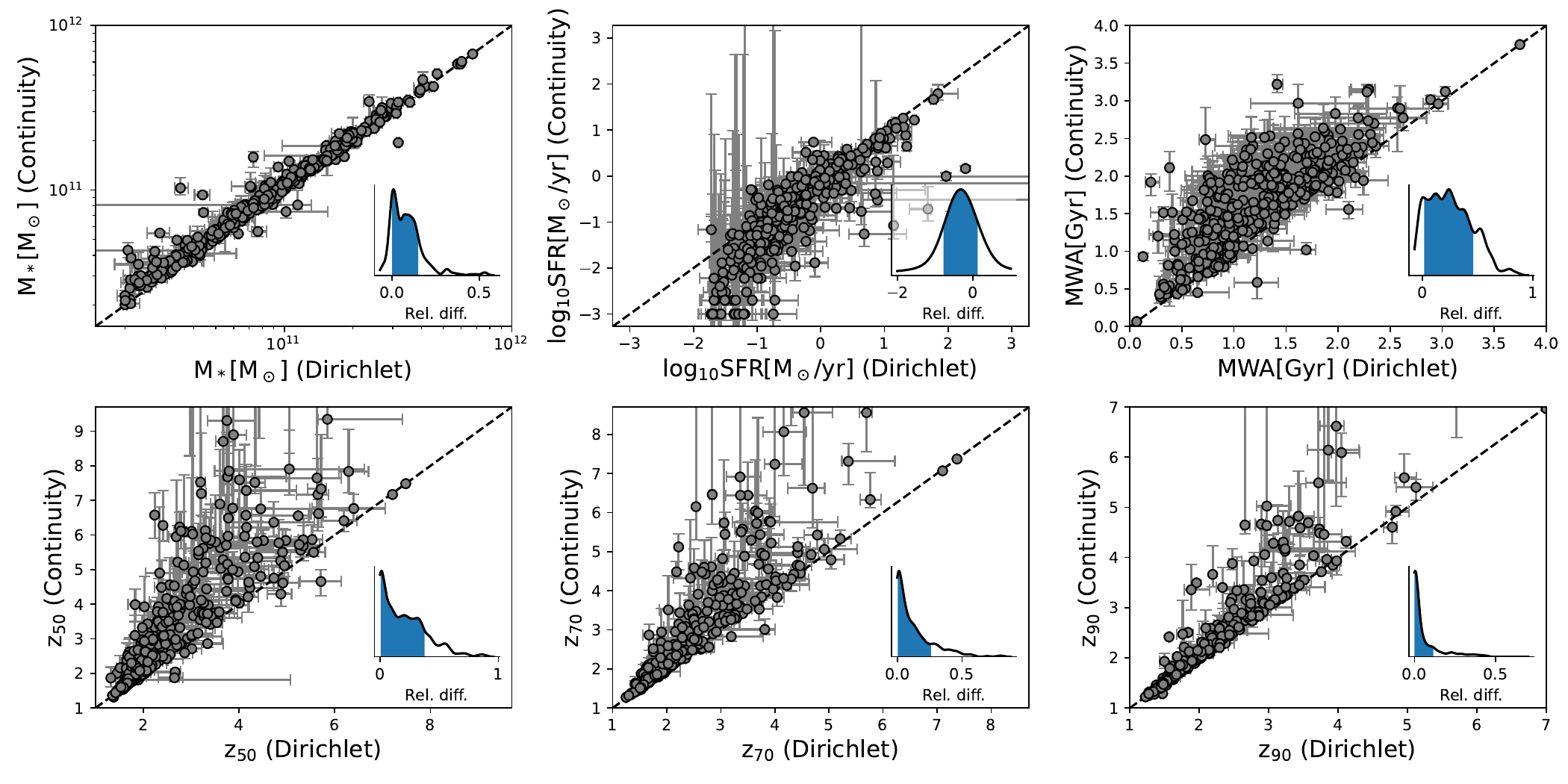}
    \caption{Comparisons of physical parameters derived using different priors of non-parametric SFHs for our sample QGs (section \ref{sec:sample}). The y-axis shows the measurements using the Continuity prior, and the x-axis shows the measurements using the Dirichlet prior. The black dashed line marks the one-to-one relation. The inset of each panel shows the distribution of relative differences, i.e. (y-x)/y, between the two measurements where the 1$\sigma$ range is filled with blue.} 
    \label{fig:diri_vs_cont}
\end{figure}

While tests with simulations provide very valuable guidance to assess the performance of the SED fitting procedure, considering the limitations that we detailed in the end of Appendix \ref{app:tng}, we decide to additionally do an empirical test, namely to run a new set of \prospector fittings with the Continuity prior for our sample QGs (section \ref{sec:sample}), to check if any significant systematic error introduced by the prior could exist in the measures. In any case, as the direct comparison of key parameters obtained using the two priors illustrates (see Figure \ref{fig:diri_vs_cont}), the strong correlations between the two sets of measurements are seen, meaning that switching from the Dirichlet prior to the Continuity one \textit{will not qualitatively} change any conclusion of this work.

\section{Physical parameters derived by fixing metallicity during the SED fitting}\label{app:metal}

\begin{figure}
    \centering
    \includegraphics[width=0.97\textwidth]{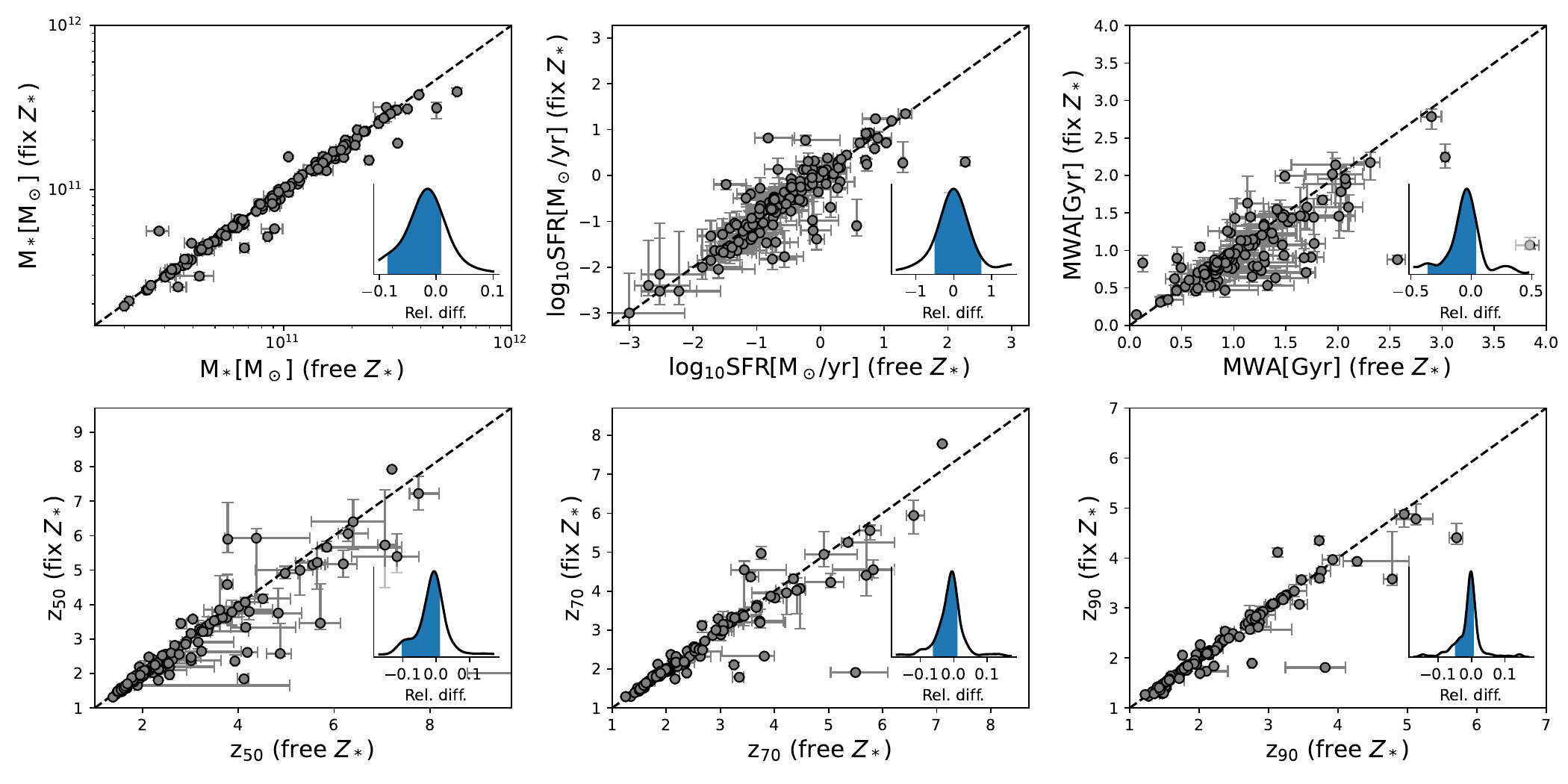}
    \caption{Similar to Figure \ref{fig:diri_vs_cont}, but for the comparisons of physical parameters derived by fixing (y-axis) or freeing (x-axis) metallicity during the SED fitting.} 
    \label{fig:freeZ_vs_fixZ}
\end{figure}

We also empirically test that if our results are sensitive to the assumption on metallicity during the SED fitting. Unlike Appendix \ref{app:conti} where the entire sample was used, here we only use the sample from the GOODS-South field for the test. We run a new set of SED fittings using the same assumptions as our fiducial model (Dirichlet) except that metallicity is fixed to the solar metallicity during the fittings. We compare the measurements in Figure \ref{fig:freeZ_vs_fixZ}. Compared to the systematic errors introduced by the priors (Figure \ref{fig:diri_vs_cont}), the errors introduced by the assumptions on metallicity are much smaller, and they cannot change any conclusion of this work.

\section{Comparing the Skewness of SFHs with \tausf/\tauq }\label{app:skew_ratio}

\begin{figure}
    \centering
    \includegraphics[width=0.97\textwidth]{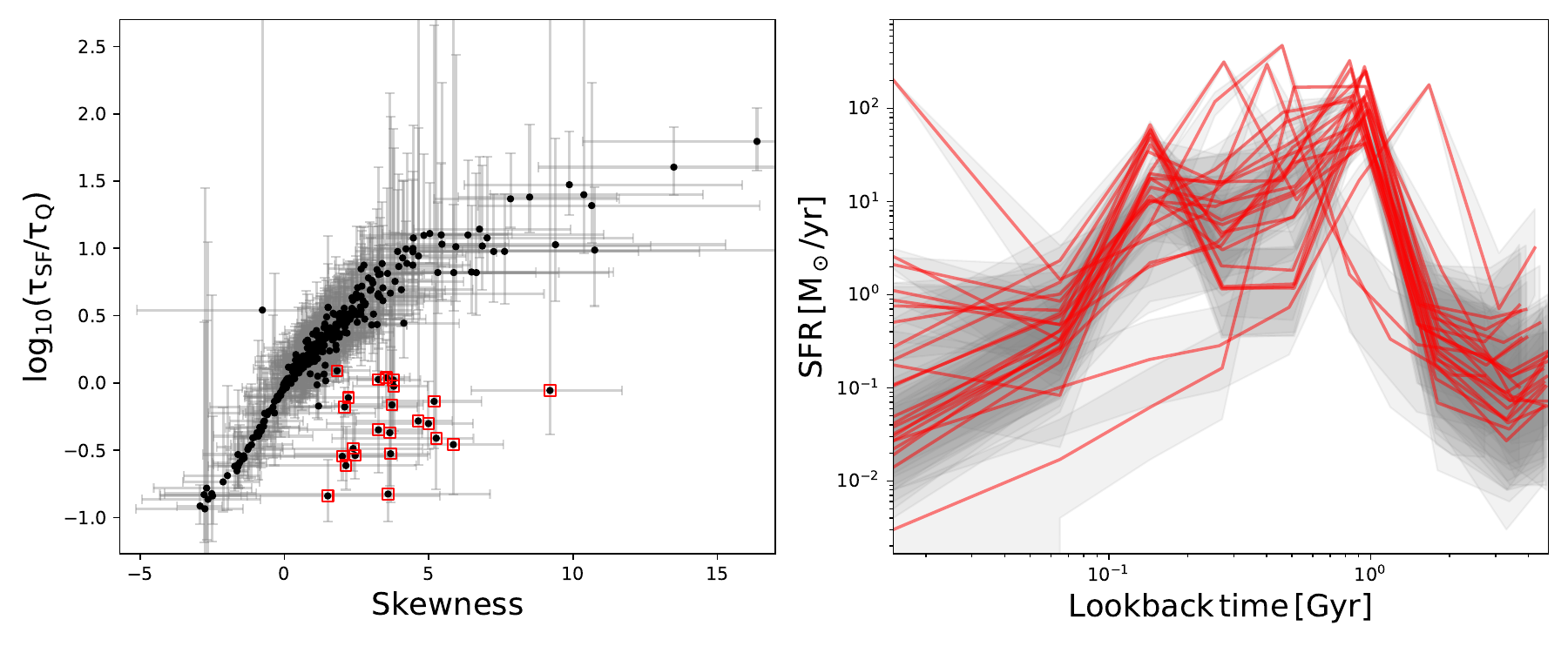}
    \caption{{\bf Left:} The comparison between the Skewness of SFHs and \tausf/\tauq (see the definitions in section \ref{sec:sfh_shape_def}). The two metrics describing the asymmetry of SFHs agree with other other very well, except some outliers which are labelled with red squares. {\bf Right:} SFHs of the individual outliers.} 
    \label{fig:skew_ratio}
\end{figure}

In section \ref{sec:sfh_shape_def}, we introduced two ways to characterize the asymmetry of SFHs, namely the Skewness and \tausf/\tauq. We compare them in the left panel of Figure \ref{fig:skew_ratio} where we see great agreement between the two metrics. We also find some outliers where clear discrepancies exist between the two metrics. In the right panel of the Figure, we plot SFHs of the individual outliers. We immediately see the complex shape of SFHs of the outliers, showing the reason of the discrepancies between the Skewness and \tausf/\tauq.

\section{The relationships between stellar mass, mass surface density and \zp} \label{app:results_w_zp}

\begin{figure*}
    \centering
    \includegraphics[width=1\textwidth]{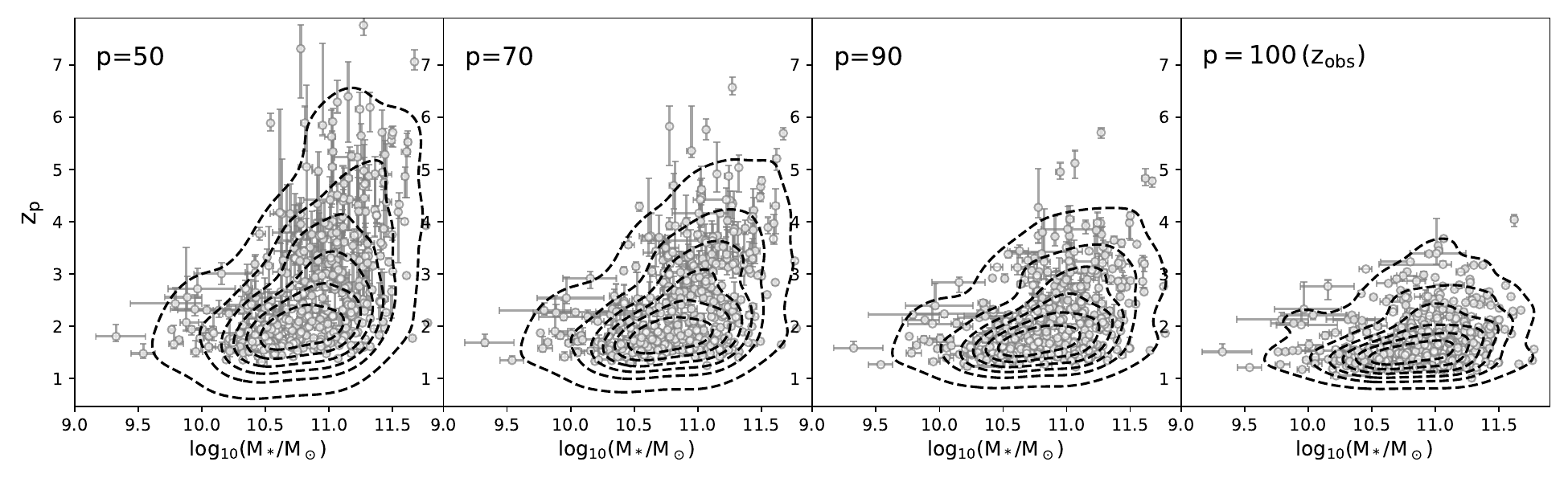}
    \caption{Similar to Figure \ref{fig:mass_zp}, but instead of using \zf, we show the relationships of $M_*$ with other characteristic redshifts \zp, i.e. the redshifts when the p-percentage of $M_*$ has already formed. The dashed contours show the density distributions estimated using a Gaussian kernel.} 
    \label{fig:mass_zp_app}
\end{figure*}

\begin{figure*}
    \centering
    \includegraphics[width=1\textwidth]{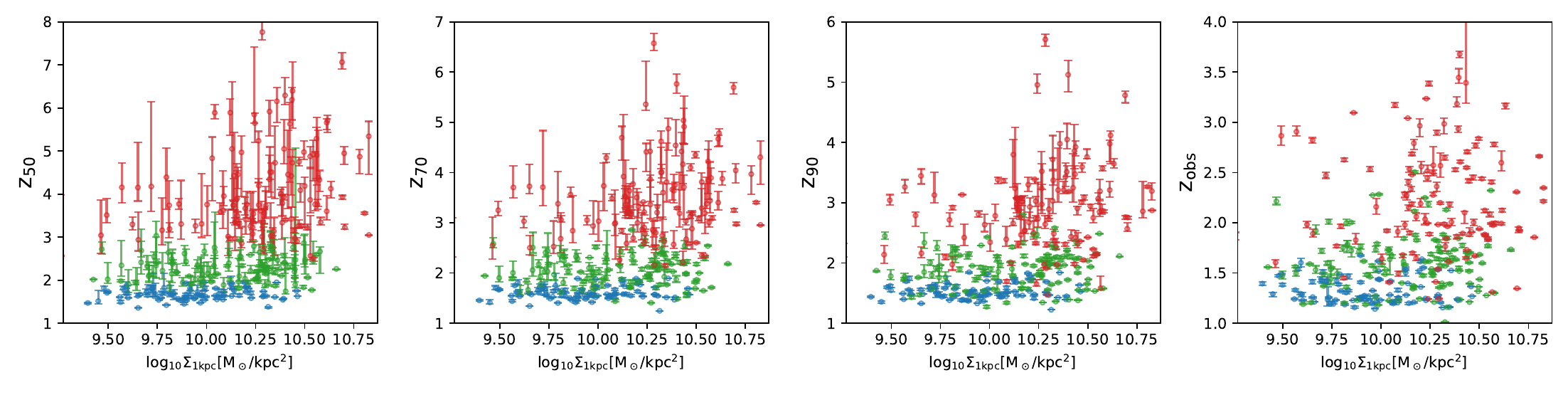}
    \includegraphics[width=1\textwidth]{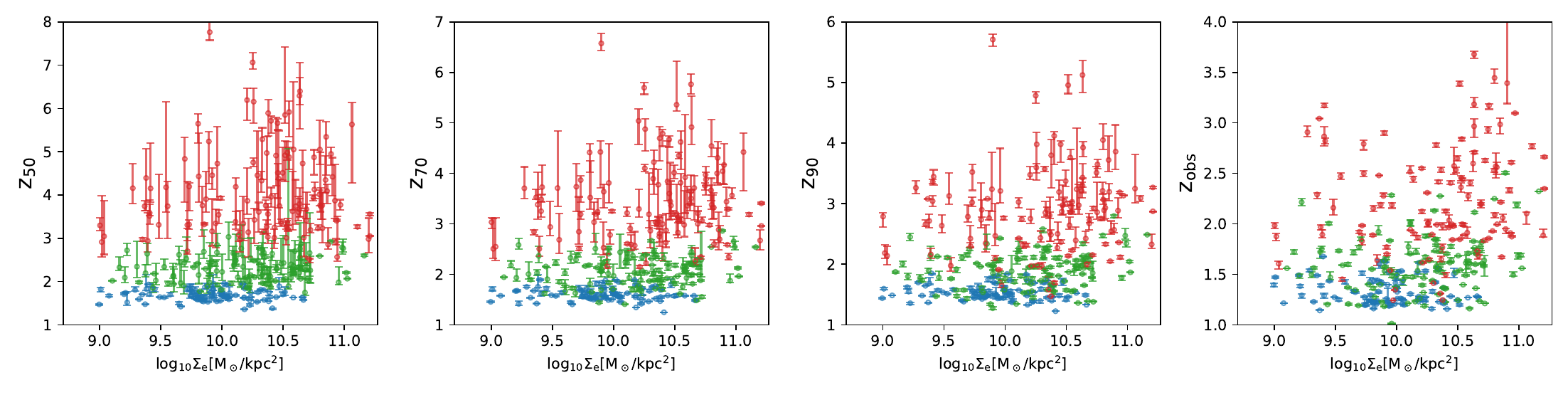}
    \includegraphics[width=1\textwidth]{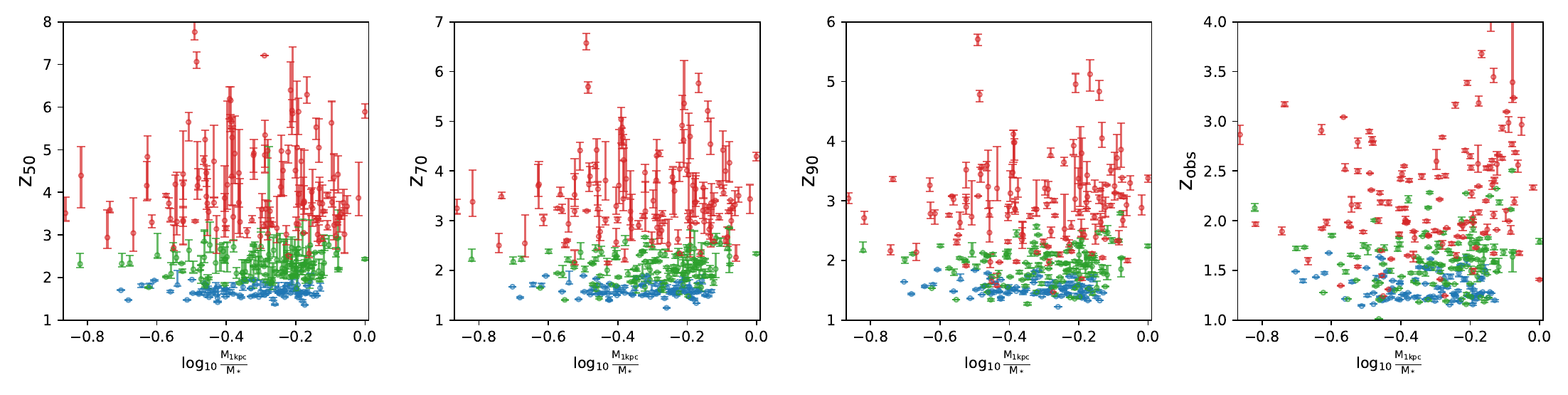}
    \caption{Similar to Figure \ref{fig:S1_zf}, \ref{fig:Se_zf} and \ref{fig:M1_zf}, but instead of using \zf, we show the relationships of \Sone (the first row), \Se (the second row) and \Mone (the third row) with other characteristic redshifts, namely $z_{50}$, $z_{70}$, $z_{90}$ and \zobs.} \label{fig:morph_zp_app}
\end{figure*}

In the main text we presented the relationships between stellar mass and \zf (section \ref{sec:mass_sfh}), and between galaxy compactness and \zf (section \ref{sec:zf_sigma}). Here, instead of using \zf, in Figure \ref{fig:mass_zp_app} and \ref{fig:morph_zp_app} we test the relationships using other characteristic \zp that we defined in section \ref{sec:sfh_shape_def}. 

\end{document}